\documentclass{aa}  
\usepackage{graphicx}
\usepackage{txfonts}
\usepackage{hyperref}
\usepackage{lscape}
\usepackage{threeparttable}
\usepackage{booktabs}
\usepackage{soul}
\usepackage{changepage}
\usepackage{adjustbox}

\defcitealias{2017MNRAS.469..800M}{Paper\,I}
\defcitealias{2018ApJ...853...86B}{Paper\,II}
\defcitealias{2018ApJ...854...45L}{Paper\,III}
\defcitealias{2021MNRAS.505.3549S}{Paper\,IV}
\defcitealias{2022ApJ...930...24G}{Paper\,V}

\begin{document}

\title{The HST Large Programme on $\omega$\,Centauri - VI.\\ The radial gradient of the stellar populations}


\author{M.\,Scalco\inst{1,2}\fnmsep\thanks{\email{michele.scalco@inaf.it}}
\and
L. Bedin\inst{2}
\and
E. Vesperini\inst{3}
}

\institute{Dipartimento di Fisica e Scienze della Terra, Università di Ferrara, Via Giuseppe Saragat 1, Ferrara I-44122, Italy
\and
Istituto Nazionale di Astrofisica, Osservatorio Astronomico di Padova, Vicolo dell’Osservatorio 5, Padova I-35122, Italy
\and
Department of Astronomy, Indiana University, Swain West, 727 E. 3rd Street, Bloomington, IN 47405, USA
}

\date{XXX,YYY,ZZZ}
 
\abstract
{In this paper we present the analysis of Hubble Space Telescope (HST) observations of the globular cluster Omega Centauri. Our analysis combines data obtained in this work with previously published HST data from an earlier article of this series and encompasses a broad portion of the cluster's radial extension. Our findings reveal a significant radial variation in the fraction of stars within the two most populous stellar populations showing that one of the main second-population groups (referred to as bMS) is more centrally concentrated than the first-population group (referred to as rMS). Additionally, we explore the spatial variations of the other less populous stellar populations (referred to as MSa and MSd) and find a qualitatively similar, but weaker, radial decrease in the fraction of stars in these populations at larger distances from the cluster centre. Only one of the populations identified (MSe) does not show any significant radial variation.}


\titlerunning{The HST Large Programme on $\omega$\,Centauri - VI.}
\authorrunning{M.\,Scalco et al.}
\maketitle

\section{Introduction}\label{Section1}
Photometric and spectroscopic studies over the last twenty years have revealed that stars within globular clusters (GCs) exhibit distinct chemical compositions, categorizing them into two main groups, each potentially containing sub-populations. The first population (1P) stars closely resemble the composition of stars in the halo field, whereas the second population (2P) stars display depletion in certain light elements like carbon, oxygen, and magnesium, along with enrichment in elements such as helium, nitrogen, aluminium, and sodium compared to 1P stars (see e.g. the reviews by \citep{1987PASP...99...67S,2019A&ARv..27....8G}.

Several scenarios have been proposed to explain the formation of multiple stellar populations (mPOPs) in GCs. However, all of these scenarios face significant challenges \citep[see][]{2015MNRAS.454.4197R,2018ARA&A..56...83B}. The spatial distributions of the different populations provide crucial information necessary to construct a comprehensive understanding of the formation and evolutionary history of GCs, as well as to constrain possible paths for theoretical investigations. According to various formation scenarios \citep[see e.g.][]{erc,2010ApJ...724L..99B,2019MNRAS.489.3269C}, 2P stars are expected to form more centrally concentrated in the cluster's inner regions and gradually mix with 1P stars during the cluster's evolution driven by two-body relaxation.

Omega Centauri (or NGC\,5139, hereafter $\omega$\,Cen), the most massive GC in the Milky Way, presents an intriguing case of mPOPs \citep{2004ApJ...605L.125B}. Characterized by low reddening \citep[$E(B-V)\sim$0.12;][]{1996AJ....112.1487H,2010arXiv1012.3224H} and located relatively close to the Sun ($\sim5$~kpc), $\omega$\,Cen is an ideal target for various photometric and spectroscopic investigations. It hosts a complex system of mPOPs, rendering it one of the most enigmatic stellar systems within the Galaxy. It is populated by at least two primary groups of stars, namely blue main sequence (bMS) and red main sequence (rMS) \citep[see][]{2004ApJ...605L.125B,2009A&A...493..959B}, exhibiting significant differences in their helium content \citep[Y$\sim$0.40 for the helium-rich component; see][]{2004ApJ...612L..25N,2012AJ....144....5K}. In a more recent paper, \citet{2017ApJ...844..164B}, found that in the core of $\omega$\,Cen, both bMS and rMS stars are split into three sub-components, and identified at least fifteen sub-populations. The origin story behind the intricate properties of the stellar populations in $\omega$\,Cen remains unclear. Several scenarios propose different explanations, suggesting that $\omega$\,Cen could be the nucleus of a dwarf galaxy absorbed by the Milky Way, or the outcome of the merger of two or more clusters (\citealt{1997ApJ...487L.187N,1998ApJ...506L.113J,2003MNRAS.346L..11B,2000ApJ...534L..83P,2006ApJ...637L.109B}; \citealt{2019NatAs...3..667I,2006A&A...445..513V}). $\omega$\,Cen represents an exceptional laboratory for unravelling many fundamental aspects of the origin of mPOPs. Its long relaxation time \citep[1.1~Gyr in the core and 10~Gyr at the half-mass radius;][]{1996AJ....112.1487H,2010arXiv1012.3224H} suggests that its present-day spatial and kinematic properties may retain some memory of those emerging at the end of the formation and early-evolutionary phases. Studies on $\omega$\,Cen showed that the bMS is more centrally concentrated than the rMS: within $\sim$2\,$r_c$ (where $r_{\rm c}$ = 2.37 arcmin is the core radius of the cluster; from \citealt{1996AJ....112.1487H,2010arXiv1012.3224H}), the fraction of bMS stars is similar to that of rMS stars but as the distance from the cluster centre increases, the relative abundance of bMS stars compared to rMS stars declines significantly. Beyond $\sim$8~arcmin, the relative proportions of bMS and rMS stars remain constant \citep{2007ApJ...654..915S,2009A&A...493..959B} although recent investigations suggest an increase in the bMS/rMS ratio at radii larger than $\sim$20~arcmin \citep{2020ApJ...891..167C}.

The Hubble Space Telescope (HST) GO-16247 programme (P.I.; Scalco), aims to expand upon the work conducted by \citet{2007ApJ...654..915S,2009A&A...493..959B}, which focused solely on the bMS and rMS groups, to include all fifteen sub-populations identified in the core of the cluster by \citet{2017ApJ...844..164B}. This program seeks to investigate, for the first time, the complete radial distribution of all identified mPOPs across the entire extension of the GC $\omega$\,Cen. 

\begin{centering} 
	\begin{figure}
		\centering
		\includegraphics[width=0.49\textwidth]{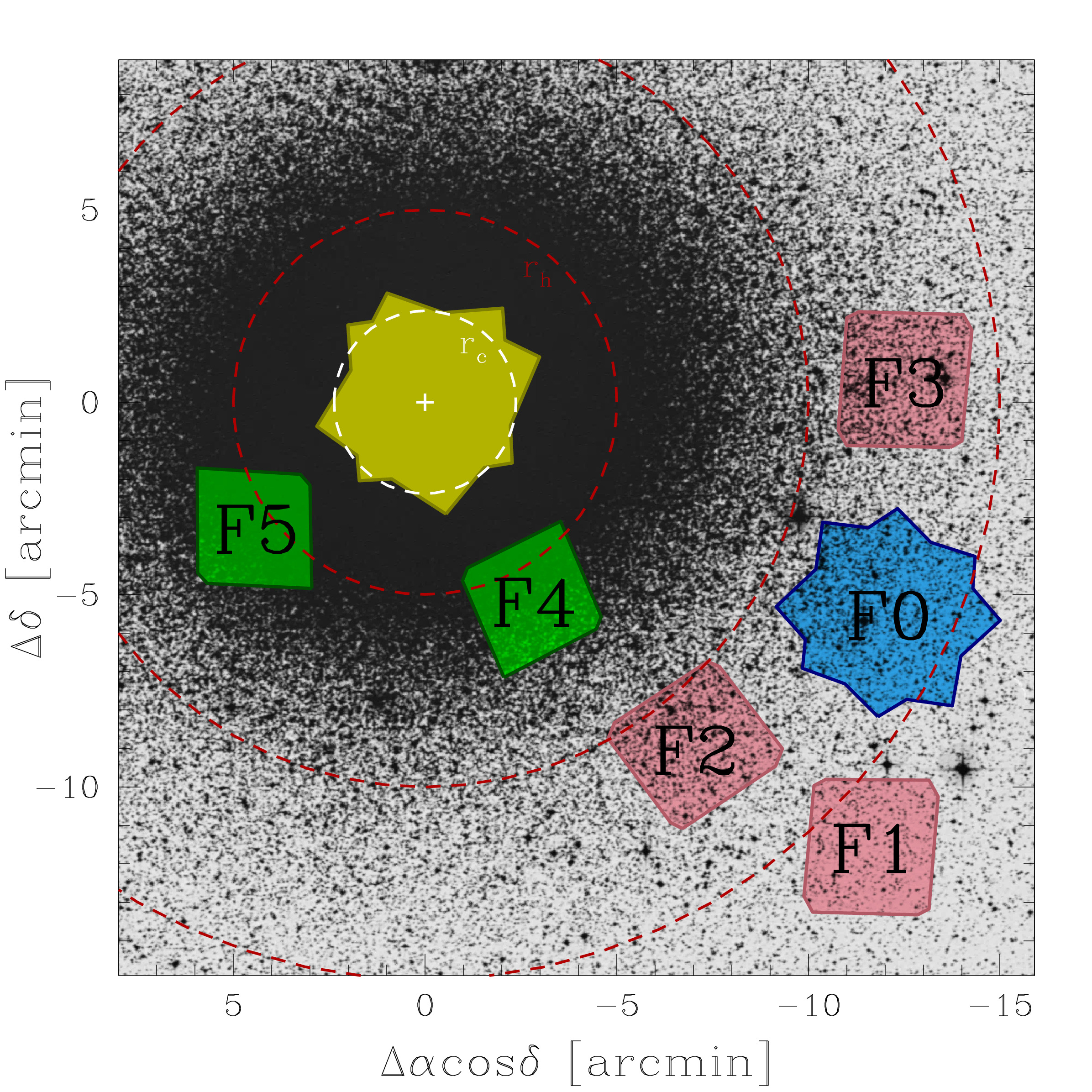}
		\caption{Outlines of the fields observed in HST programs GO-14118+14662 (P.I.: Bedin) and GO-16427 (P.I.: Scalco), superimposed on a DSS image of $\omega$\,Cen. The primary GO-14118+14662 field (F0) is in azure, while the three GO-14118+14662 parallel WFC3 fields are shown in pink. The two GO-16427 fields are shown in green. The data discussed in this paper come from fields F2, F3, F4 and F5. We also show, in yellow, the central field presented in \citet{2017ApJ...842....6B,2017ApJ...842....7B,2017ApJ...844..164B}. Units are in arcmin measured from the cluster centre. The white dashed circle marks the $r_{\rm c}$ while the red dashed circles mark the half-light radius \citep[$r_{\rm h}=5.00$ arcmin; from][]{1996AJ....112.1487H,2010arXiv1012.3224H}, and $2\,r_{\rm h}$, $3\,r_{\rm h}$, from the centre.}
	\label{Fields} 
	\end{figure} 
\end{centering}

Schematically, the existing HST radial coverage of the cluster, with filter coverage sufficient to effectively separate and identify the mPOPs, can be divided into three parts. Fig.\,\ref{Fields} shows the locations of the three parts, superimposed on an image from the Digital Sky Survey (DSS)\footnote{\href{https://archive.eso.org/dss/dss}{https://archive.eso.org/dss/dss}}. The part\,(i) of the radial coverage, maps the cluster from the centre out to $\sim$1\,$r_c$ and is represented in yellow in Fig.\,\ref{Fields}. The entire photometric catalogues for this field have been published and analysed by \citet{2017ApJ...842....6B,2017ApJ...842....7B,2017ApJ...844..164B}. 

The part\,(ii) of the radial coverage, maps the outskirts of the cluster and consists of three HST fields, mapping between $\sim$10~arcmin and $\sim$20~arcmin from the centre of $\omega$\,Cen, and collected under the multi-cycle programme GO-14118+14662 (P.I.: Bedin). Those fields are represented in pink in Fig.\,\ref{Fields} (fields F1, F2 and F3). We also show in azure the primary field of the GO-14118+14662 programme (field F0).

The exposures from the parallel field F1, F2 and F3 were reduced and presented in the five previous publications of this series: the mPOPs at very faint magnitudes in the field F1 were analysed by \citet[hereafter \citetalias{2017MNRAS.469..800M}]{2017MNRAS.469..800M}. \citet[hereafter \citetalias{2018ApJ...853...86B}]{2018ApJ...853...86B} analysed the internal kinematics of the mPOPs in the field F1, complementing the GO-14118+14662 data with archival images collected more than 10\,years earlier under HST programmes GO-9444 and GO-10101 (on both P.I.: King). The astrometric catalogue for the field F1 together with the photometry of some filters (F606W, F814W, F110W and F160W) were released in \citetalias{2018ApJ...853...86B}. \citet[hereafter \citetalias{2018ApJ...854...45L}]{2018ApJ...854...45L} presented the absolute proper motion estimate for $\omega$\,Cen in the field F1. \citet[hereafter \citetalias{2021MNRAS.505.3549S}]{2021MNRAS.505.3549S} released the astro-photometric catalogue for the fields F2 and F3. Finally, \citet[hereafter \citetalias{2022ApJ...930...24G}]{2022ApJ...930...24G} presented a set of stellar models designed to investigate low-mass stars and brown dwarfs in $\omega$\,Cen.

The part\,(iii) maps the radial distance between $\sim$3~arcmin and $\sim$10~arcmin from the cluster centre. Those fields are represented in green in Fig.\,\ref{Fields} (fields F4 and F5) and were observed under programmes GO-12580 (P.I.: Renzini) and GO-14759 (P.I.: Brown) and, more recently, under the GO-16247 programme (P.I.: Scalco). This part of the cluster covers a crucial radial range since it is where \citet{2009A&A...493..959B} found the strongest gradient in the ratio between bMS and rMS stars.

In this paper, we present the reduction of the two intermediate fields F4 and F5 observed under the GO-16247 programme (P.I.: Scalco) and the outer field F1 observed under the GO-14118+14662 programme (P.I.: Bedin), for which the photometry of some filters has not yet been released. Together with the analysis, we also release the catalogue and atlases of the analysed fields. We then combine the photometry obtained from fields F4, F5 and F1 with that from fields F2 and F3 (see \citetalias{2021MNRAS.505.3549S}) and examined the presence of the fifteen stellar populations identified by \citet{2017ApJ...844..164B} within these fields.

The paper is organized as follows: Section \ref{Section2} presents the data set and data reduction for fields F4, F5 and F1. Section \ref{Section3} describes the quality selection and differential-reddening correction applied to the analyzed fields F1, F2, F3, F4, and F5, while Section \ref{Section4} outlines the process used to identify and separate the mPOPs in these fields. Section \ref{Section5} discusses the radial variations observed among the identified populations. Finally, Section\,\ref{Section6} presents a brief summary of the results.

\section{Data set and reduction}\label{Section2}
Fields F4 and F5 were observed in 2012, 2016, 2017 and 2022 using the ultraviolet and visible (UVIS) channel of the Wide Field Camera 3 (WFC3). In each field, data were collected with five filters (F275W, F336W, F438W, F606W and F814W). Table\,\ref{Table1} reports the complete list of HST observations of fields F4 and F5. 

Field F1 was observed under the GO-14118+14662 programme in 2015 and 2017, utilizing the F275W, F336W, F438W, F606W, and F814W filters of the WFC3/UVIS channel and the F110W and F160W filters of the WFC3 Near Infrared (NIR) channel. While the astrometry and photometry of some filters were published in \citetalias{2018ApJ...853...86B}, the catalogue contains only data for which reliable proper motion (PM) measurements were possible. To recover the photometry and astrometry for all other stars, we reprocessed the entire F1 dataset from the GO-14118+14662 programme. We refer to \citetalias{2018ApJ...853...86B} (see Table\,1) for the complete description of the field F1 GO-14118+14662 data set.

\begin{table}
	\caption{List of HST observations of fields F4 and F5.}
        \label{Table1}
	\smallskip
	\centering
	\begin{tabular*}{0.49\textwidth}{@{\extracolsep{\fill}}l c @{\hspace{-.75mm}} c}
		& Field F4 &\\
		\hline
		\hline
		Filter & Exposures & Epoch\\
		\hline
		\hline
            & Epoch 1 (GO-12580; P.I.: Renzini) &\\            
		\hline
		& WFC3/UVIS &\\
		\hline
		F275W & 2$\times$909 s + 2$\times$914 s & 2012/03/09-04/29\\ 
                  & + 2$\times$1028 s + 2$\times$1030 s + 2$\times$1267 s & \\
		F336W & 2$\times$562 s + 2$\times$565 s & 2012/03/09-04/29\\ 
                  & + 1$\times$945 s + 1$\times$953 & \\
		F438W & 2$\times$200 s + 2$\times$210 s & 2012/03/09-04/29\\
		\hline
		& Epoch 2 (GO-14759; P.I.: Brown) &\\
		\hline
		& WFC3/UVIS & \\
		\hline
		F275W & 1$\times$765 s + 1$\times$850 s & 2017/04/13\\ 
            F336W & 1$\times$630 s + 1$\times$765 s & 2017/04/13\\ 
            F438W & 1$\times$630 s + 1$\times$1025 s & 2017/04/13\\
		\hline
		& Epoch 3 (GO-16247; P.I.: Scalco) &\\
		\hline
		& WFC3/UVIS & \\
		\hline
		F606W & 1$\times$325 s + 1$\times$348 s & 2022/03/12\\ 
		F814W & 1$\times$13 s + 4$\times$348 s & 2022/03/12\\ 
		\hline
	\end{tabular*}
	\newline
	\newline
        \newline
	\begin{tabular*}{0.49\textwidth}{@{\extracolsep{\fill}}c c c} 
		& Field F5 & \\
        \hline
        \hline
        Filter & Exposures & Epoch\\
        \hline
        \hline
        & Epoch 1 (GO-14759; P.I.: Brown) &\\
        \hline
        & WFC3/UVIS &\\
        \hline
        F275W & 2$\times$765 s + 2$\times$850 s & 2016/12/10-14\\ 
        F336W & 2$\times$630 s + 2$\times$765 s & 2016/12/10-14\\ 
        F438W & 2$\times$630 s + 2$\times$1025 s & 2016/12/10-14\\
        \hline
        & Epoch 2 (GO-16247; P.I.: Scalco) &\\
        \hline
        & WFC3/UVIS & \\
        \hline
        F606W & 1$\times$325 s + 1$\times$348 s & 2022/04/27\\ 
        F814W & 1$\times$13 s + 4$\times$348 s & 2022/04/27\\ 
        \hline
	\end{tabular*}
\end{table}

The data were reduced following the procedure outlined in \citetalias{2021MNRAS.505.3549S}. In summary, this procedure involves two main steps: the \textit{first-pass} and \textit{second-pass} photometry. During the \textit{first-pass} photometry, we perturbed a set of ``library'' WFC3/UVIS and WFC3/NIR effective Point Spread Functions \citep[ePSFs, see][]{2000PASP..112.1360A,2006acs..rept....1A} to determine the optimal spatially variable PSF for each image. Then, using these PSFs, we extracted the positions and fluxes of the stars within each image. This extraction was carried out using the \texttt{FORTRAN} code \texttt{hst1pass} \citep[see][]{2022wfc..rept....5A}. To account for geometric distortion, the stellar positions in each individual exposure catalogue were corrected using the publicly available WFC3/UVIS and WFC3/NIR correction \citep{2011PASP..123..622B,2016wfc..rept...12A}. For each filter, the positions and magnitudes were transformed to a common reference frame using six-parameter linear transformations and photometric zero points.

We performed the \textit{second-pass} photometry using the \texttt{FORTRAN} software package \texttt{KS2}, which is based on the software \texttt{kitchen\_sync} presented in \citet{2008AJ....135.2055A}.  This software routine makes use of the results obtained from the \textit{first-pass} stage to simultaneously identify and measure stars across all individual exposures and filters. By relying on multiple exposures, \texttt{KS2} effectively detects and measures faint stars that would be otherwise lost in the noise of individual exposures. The star-finding process is executed through a series of passes, gradually moving from the brightest to the faintest stars. In each iteration, the routine identifies stars that are fainter than those found in the previous iteration, subsequently measuring and subtracting them. This iterative approach ensures that progressively fainter stars are detected and accounted for, enhancing the overall accuracy of the photometric measurements. \texttt{KS2} employs three distinct methods for measuring stars, with each approach specifically tailored for different magnitude ranges. We refer to \citet{2017ApJ...842....6B,2018MNRAS.481.3382N}; \citetalias{2021MNRAS.505.3549S} for a detailed description of the methods and procedures. To make the catalogue as similar as possible to that of the F1, F2 and F3 fields released by \citetalias{2018ApJ...853...86B,2021MNRAS.505.3549S}, we performed the star-finding using the F606W and F814W filters. Our final photometric catalogue contains a total of 40\,397, 30\,929 and 4\,015 sources measured in all five filters for the fields F4, F5, and F1, respectively.

The photometry has been zero-pointed into the Vega magnitude system by following the recipe of \citet{2005MNRAS.357.1038B} and adopting the photometric zero-points provided by STScI web page for WFC3/UVIS and WFC3/NIR\footnote{\href{https://www.stsci.edu/hst/instrumentation/wfc3/data-analysis/photometric-calibration}{https://www.stsci.edu/hst/instrumentation/wfc3/data-analysis/photometric-calibration}}.

We cross-referenced the stars in our catalogue with the stars in the \textit{Gaia} Data Release 3 (\textit{Gaia} DR3, \citealt{2016A&A...595A...2G,2023A&A...674A...1G}). The sources found in common were used to anchor our positions (X, Y) to the \textit{Gaia} DR3 absolute astrometric system.

PMs were computed using the technique described in \citetalias{2021MNRAS.505.3549S} (see also \citealt{2014ApJ...797..115B}; \citetalias{2018ApJ...853...86B,2018ApJ...854...45L}; \citealt{2022ApJ...934..150L}). This iterative procedure treats each image as an independent epoch and can be summarised in two main steps: first, it transforms the stellar positions from each exposure into a common reference frame through a six-parameter linear transformation. Then, it fits these transformed positions as a function of the epoch using a least-square straight line. The slope of this line, determined after multiple outlier-rejection stages, provides a direct measurement of the PM. High-frequency-variation systematic effects were corrected as described in \citetalias{2018ApJ...853...86B}, i.e. according to the median value of the closest 100 likely cluster members (excluding the target star itself).

Finally, we computed the membership probability (MP) of each star, by following a method based on PMs described by \citet{1998A&AS..133..387B} (see also \citealt{2009A&A...493..959B,2018MNRAS.481.3382N}; \citetalias{2021MNRAS.505.3549S}).

As part of this publication, we are releasing publicly accessible astro-photometric catalogues and atlases derived from this study for fields F4, F5 and F1. These resources are provided in a format identical to the catalogues and atlases made available by \citetalias{2021MNRAS.505.3549S} for fields F2 and F3. For a comprehensive description of these resources, we refer to \citetalias{2021MNRAS.505.3549S}. When available, we also include the corresponding \textit{Gaia} DR3 identification numbers for sources in our catalogue. For field F1, we also provide the identification number of the corresponding source in the catalogue published in \citetalias{2018ApJ...853...86B} when the source is available. The supplementary electronic material for this journal will also be accessible through our website\footnote{\href{https://web.oapd.inaf.it/bedin/files/PAPERs_eMATERIALs/wCen_HST_LargeProgram/P06/}{https://web.oapd.inaf.it/bedin/files/PAPERs\_eMATERIALs/\\wCen\_HST\_LargeProgram/P06/}}.

\section{Sample selection and differential-redening correction}\label{Section3}
In addition to the catalogue obtained for fields F4, F5 and F1, we retrieved the catalogue for fields F2 and F3 as published in \citetalias{2021MNRAS.505.3549S}. Our combined catalogue encompasses a substantial number of sources across a wide range of radial distances from the cluster centre. The intermediate fields F4 and F5 exhibit higher population densities and a higher number of stars compared to the outer fields F1, F2 and F3, offering robust statistical significance. Notably, the photometry in field F4, with its extensive observations in F438W, F336W, and F275W, demonstrates greater precision and accuracy than that of field F5.

As outlined in the preceding section, the photometric measurements of stars within our catalogue were conducted using three distinct methods, each designed for specific magnitude ranges (refer to \citealt{2017ApJ...842....6B,2018MNRAS.481.3382N}; \citetalias{2021MNRAS.505.3549S} for details). For our analysis, we opted to utilize the photometry obtained through the first method, as it yields the most accurate photometry within the magnitude range under consideration.

In what follows we describe the selection process applied to identify a sample of well-measured stars and the subsequent differential reddening correction performed on the selected sample for each field.

\subsection{Sample selection}\label{Section3.1}
To ensure a well-measured sample of stars, we implemented a selection process using a set of quality parameters provided by \texttt{KS2}, following a similar approach as described in \citetalias{2021MNRAS.505.3549S}; \citet{2017ApJ...842....6B,2017ApJ...842....7B}. The quality parameters employed include the photometric error ($\sigma_{\rm PHO}$), the quality-of-fit (QFIT) parameter, which quantifies the PSF-fitting residuals, and the RADXS parameter, a shape parameter that allows for differentiation between stellar sources, galactic sources, and cosmic ray/hot pixels introduced in \citet{2008ApJ...678.1279B}. Further details regarding these parameters can be found in \citet{2017ApJ...842....6B,2018MNRAS.481.3382N}; \citetalias{2021MNRAS.505.3549S}. 

The selection process is described in the following and it has been applied to each field separately. For each filter, we divided the stars into 0.5 magnitude bins and evaluated for each bin the 2.5$\sigma$-clipped median value and dispersion ($\sigma$) of each photometric parameter. We then defined a series of points by adding (for $\sigma_{\rm PHO}$ and RADXS parameters) or subtracting (for the QFIT parameter) 2.5$\sigma$ from the median values of each magnitude bin and interpolated the points with a spline. Stars with $\sigma_{\rm PHO}$ or $\lvert$RADXS$\rvert$ values above or QFIT values below the interpolating spline are considered bad-measured stars and excluded from the analysis. However, we set two hard constraints: stars are always considered well-measured and included in the analysis if their QFIT values are above 0.95, and if their $\lvert$RADXS$\rvert$ values are below 0.1. Finally, we required selected stars to be cluster members by excluding all the sources with MP<90\%.

\subsection{Differential-reddening correction}\label{Section3.2}
We corrected our photometry for the effects of differential reddening on zero-point variations with location in the field following the procedure described by \citet{2017ApJ...842....7B} (see also \citealt{2007AJ....133.1658S,2012A&A...540A..16M}). The differential-reddening correction is been applied for each field separately. Briefly, for each field, we started by selecting a sample of reference stars by choosing all objects likely belonging to the most populated sequence in the $m_{\rm F814W}$ versus $m_{\rm F275W}-m_{\rm F814W}$ and $m_{\rm F814W}$ versus $m_{\rm F336W}-m_{\rm F438W}$ CMDs, in close analogy of what was done in \citet{2017ApJ...842....7B}. We limited our reference stars to be within the magnitude range $16.5<m_{\rm F814W}<18.8$. We evaluated a separate differential-reddening correction for each of the CMDs utilised in this article. For each CMD, we derived the fiducial line of our sample of reference stars in the CMD and measured the residual in colour between our sample of reference stars and the fiducial along the reddening directions. For each star, we considered the median of the residual values from the 75 or 50 (depending on the CMD) neighbouring reference stars as the best estimate of the differential reddening. 

\section{The main sequence multiple stellar populations}\label{Section4}
We employed the methodology outlined in \citet{2017ApJ...844..164B}, to discern and characterize the distinct stellar populations within our selected star sample. The procedure involves several steps and can be summarized as follows: we initiate the process with a preliminary selection of a specific population on the CMD where its features are most prominent. We then plot these preliminarily selected stars on various CMDs, in which outliers are easily discernible and can be excluded. We employed \textit{"two-pseudo-colour diagrams"} \citep[hereafter, TpCDs, for details, see][]{2015MNRAS.447..927M,2015ApJ...808...51M} to highlight finer population structures. Once stars belonging to a particular population are identified and isolated, they are removed from the sample, and the process is reiterated for other populations. We refer to \citet{2017ApJ...844..164B} for a complete description of the procedure.

For consistency and comparability with the findings presented in \citet{2017ApJ...844..164B}, we meticulously followed the same procedures, utilized identical CMDs, and employed the same fiducial lines (provided to us by Bellini via private communication) making minor zero-point adjustments in colour and magnitude to ensure alignment with our photometry. This approach ensured that our selections on the CMDs and the verticalization process remained consistent \citep[we refer to][for a comprehensive description of the CMDs and fiducials used]{2017ApJ...844..164B}. However, the envelopes used to define the sub-populations in TpCDs are not identical to those used in \citet{2017ApJ...844..164B}. Instead, they are defined manually for each field separately in this paper due to potential significant differences in TpCDs at various radial distances. The entire procedure was repeated independently for each field and it is briefly presented below for field F4. The corresponding figures for the other fields can be found in the appendix of this paper.

All CMDs feature $m_{\rm F438W}$ on the y-axis, with varying colours. Following the convention in \citet{2017ApJ...844..164B}, we will henceforth identify a CMD solely by its colour.

\subsection{MSa}\label{Section4.1}
\begin{figure*}
\centering
\includegraphics[width=\textwidth]{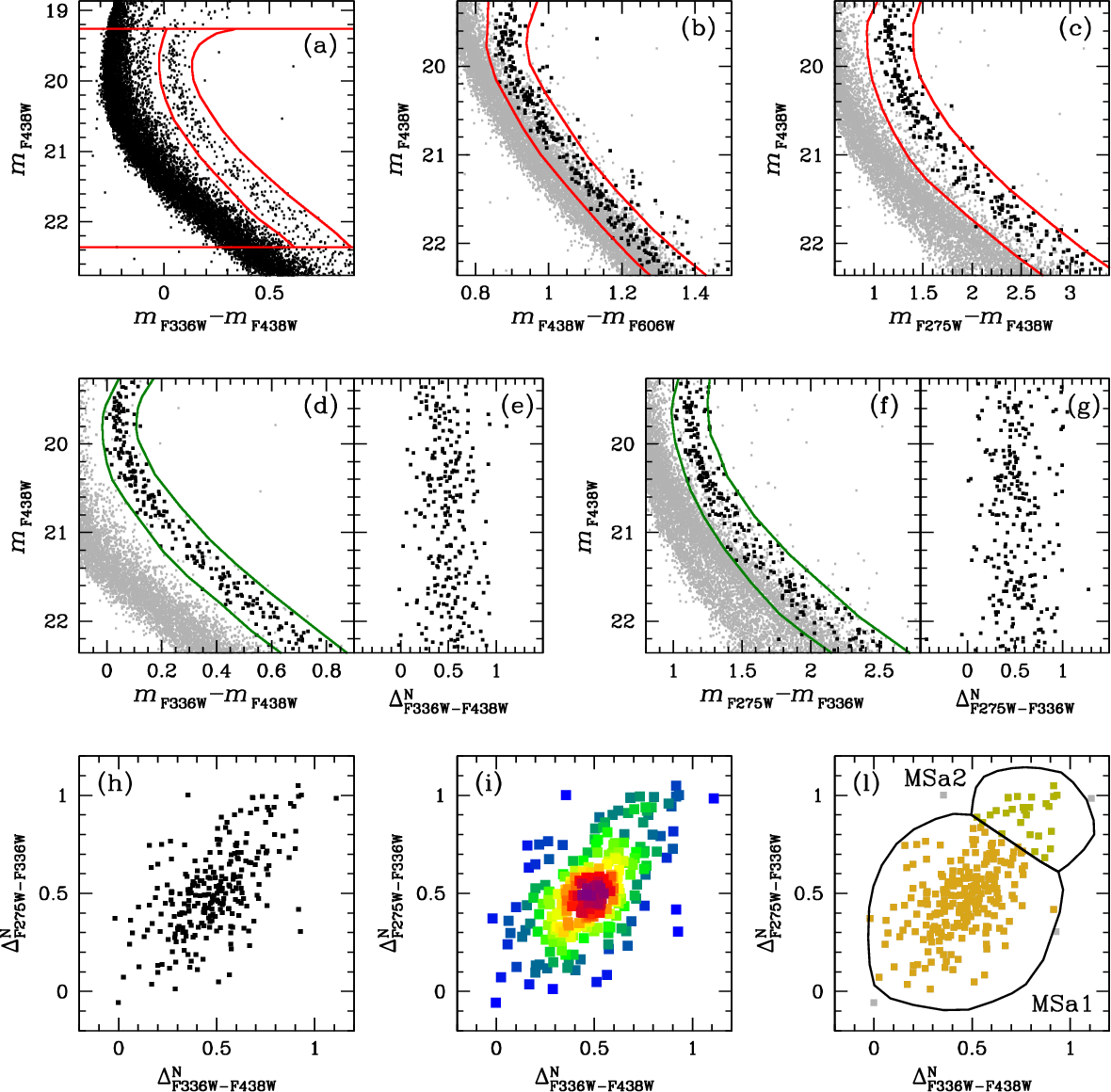}
\caption{Illustration of the selection procedures we applied to isolate MSa stars. (a) Preliminary selection of MSa candidates on the $m_{\rm F336W}-m_{\rm F438W}$ CMD (within the red lines). (b)-(c) Selection refinements using two CMDs of different colours. In black we show MSa stars selected from the previous panel, in grey the rest of the MS. Rejected stars are those outside the two red lines. (d) Fiducial lines (in green) used to verticalize the MSa in the $m_{\rm F336W}-m_{\rm F438W}$ CMD. Stars that survived the selections from panels (a)+(b)+(c) are represented in black, while other stars are in grey. (e) Verticalized $\Delta^{\rm N}_{m_{\rm F336W}-m_{\rm F438W}}$ CMD. (f)-(g) Same as panels\,(d) and (e) but for the $m_{\rm F275W}-m_{\rm F336W}$ CMD. (h) $\Delta^{\rm N}_{m_{\rm F275W}-m_{\rm F336W}}$ versus $\Delta^{\rm N}_{m_{\rm F336W}-m_{\rm F438W}}$ TpCD of MSa stars. (e) Hess diagram of the TpCD. (f) The two defined MSa sub-populations: MSa1 (in dark yellow) and MSa2 (in light yellow).}
\label{MSa}
\end{figure*}

One of the most distinguishable populations is the MSa \citep[where the \textit{a} stands for \textit{anomalous}, see][] {2010AJ....140..631B}, characterized by a narrow sequence that is notably redder and more curved compared to the majority of MS stars. This population corresponds to a group of stars rich in helium and with the largest Fe enrichment compared to the reference population (rMS, see below) of $\omega$\,Cen \citep[see Table\,1 of][see also \citealt{2021A&A...653L...8L}]{2017ApJ...844..164B}. Panel\,(a) of Fig.\,\ref{MSa} illustrates the $m_{\rm F336W}-m_{\rm F438W}$ CMD of stars within our selected sample for field F4. In this CMD, stars attributed to the MSa are distinctly separated from the remaining MS population, with the colour boundaries delineated by red lines. The magnitude range for selection is confined between $19.26<m_{\rm F438W}<22.36$, as indicated by two horizontal red lines in panel\,(a) (note that the bright limit will be set to a fainter level for the analysis of the other populations). The initially selected MSa stars are plotted on the $m_{\rm F438W}-m_{\rm F606W}$ CMD (black dots in panel\,(b)), while the remaining unidentified MS stars are shown in grey. Our MSa selection is restricted to stars within the two red lines (hereafter, red lines will always indicate our selection boundaries, black points will denote selected stars from the previous panel, while all other stars will be in grey). Panel\,(c) shows the $m_{\rm F275W}-m_{\rm F438W}$ CMD of the surviving MSa stars that passed both selections in panels\,(a) and (b). A few additional outliers from our MSa candidates (black points outside the two red lines) were subsequently removed. Finally, the $m_{\rm F336W}-m_{\rm F438W}$ and $m_{\rm F275W}-m_{\rm F336W}$ CMDs of the stars that passed all selections are shown in black in panels\,(d) and (f), respectively, while the excluded stars are shown in grey. In each panel, we used two fiducial lines (represented in green) enclosing the MSa population to rectify and parallelize the population sequence (hereafter, the fiducial used to verticalize the CMDs will be represented in green to differentiate it from the fiducials used for making selections). The verticalised $\Delta^{\rm N}_{\rm F336W-F438W}$ and $\Delta^{\rm N}_{\rm F275W-F336W}$ CMDs of only selected stars are represented in panels\,(e) and (g) (for clarity in the verticalized diagram, only the selected stars will be shown). The TpCD obtained from the combination of the two verticalised diagrams is shown in panel\,(h), while the Hess diagram of the TpCD is shown in panel\,(i). The colour mapping of this and the following Hess diagrams goes from blue (lowest density), green (average density), yellow, and red (highest density). The shape of the TpCDs shown in panels\,(h) and (i) closely resembles the one reported in \citet[see Fig.\,1]{2017ApJ...844..164B}, where a similar distribution was observed. \citet{2017ApJ...844..164B} identified a primary clump at the TpCD centre, extending into a tail towards the lower-left region, alongside a secondary, less populated clump in the upper-right area. This closely resembles what we observe here: there is a prominent main clump centred around coordinates (0.5, 0.5) with a tail extending towards (0.1, 0.1). A secondary, less populated clump positioned approximately at (0.75, 0.9) seems to emerge, although the statistical significance of this second clump is very low due to the limited number of sources available. Following \citet{2017ApJ...844..164B} two sub-populations of MSa are defined in panel\,(l): the MSa1 (dark yellow) and the MSa2 (light yellow), within the black envelopes. All the sources outside these envelopes are rejected.

The corresponding procedures for fields F5, F3, F2 are shown in Fig.\,\ref{MSa_5}, \ref{MSa_3}, \ref{MSa_2}, respectively. Because of the limited number of stars in field F1, the MSa sequence was not identifiable, and therefore, we have omitted the MSa analysis for this field. The TpCD obtained for field F5 closely resembles that of field F4, albeit with a slightly more blurred appearance due to the lower photometric quality of field F5. Similarly, the TpCDs obtained for fields F2 and F3 exhibit a similar pattern, but the limited number of stars in these fields, particularly belonging to the MSa populations, results in low statistical significance. As a result, a reliable separation of the two stellar populations is not feasible in these fields, and therefore, only the total number of MSa stars in these two fields was considered. The final sample of MSa stars in fields F2 and F3 consist of those falling inside the black envelope in panel\,(l) of Fig.\,\ref{MSa_2} and Fig.\,\ref{MSa_3}, respectively, represented by black dots. 

\subsection{bMS}\label{Section4.2}
\begin{figure*}
\centering
\includegraphics[width=\textwidth]{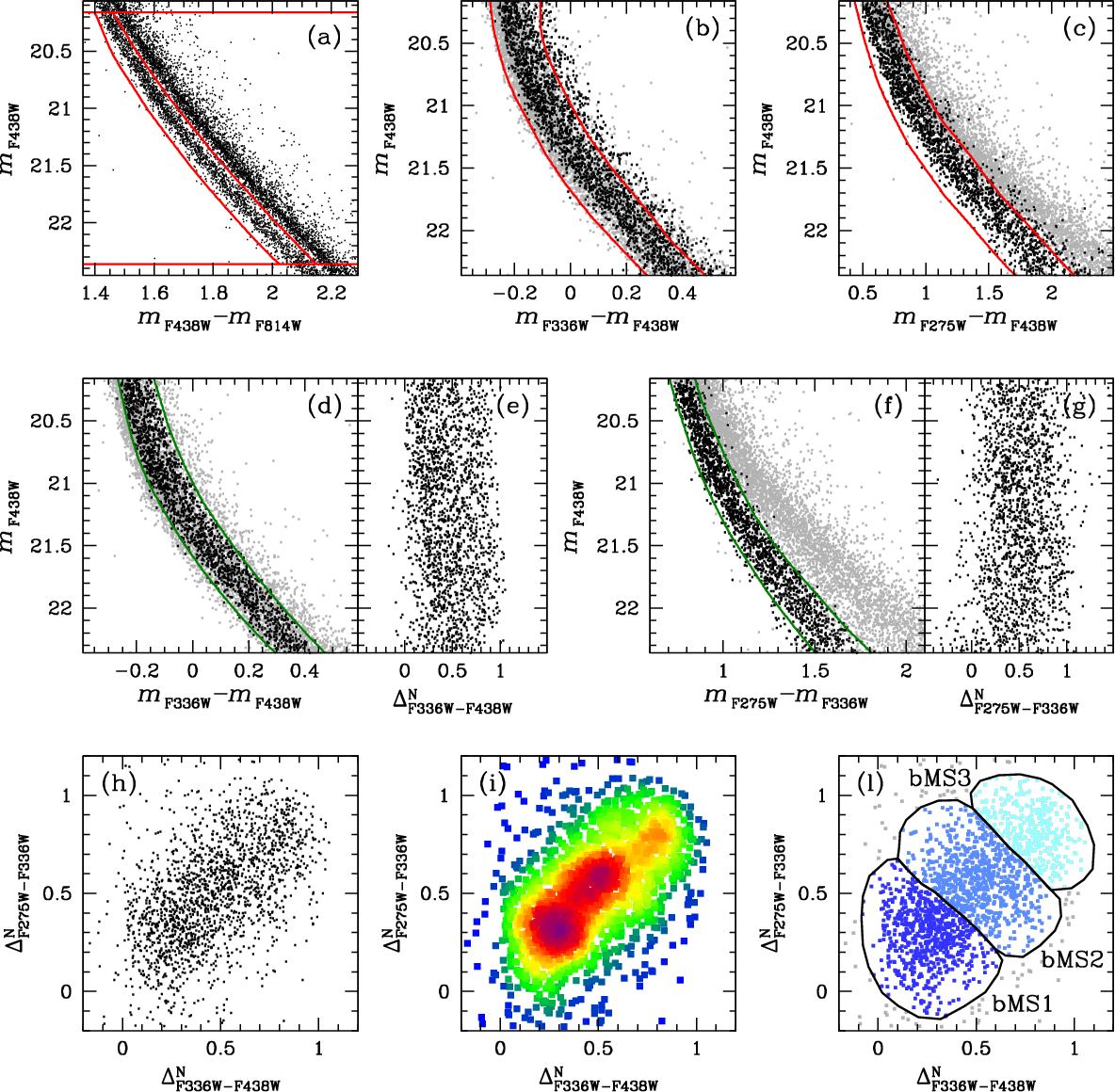}
\caption{Similar to Fig.\,\ref{MSa} but for the bMS stars. (a) Preliminary selection of bMS stars on the $m_{\rm F438W}-m_{\rm F814W}$ CMD (within the red lines). Already identified MSa1 and MSa2 stars have been removed from the CMD. (b)-(c) Preliminarily selected bMS stars are further refined using the $m_{\rm F336W}-m_{\rm F438W}$ and $m_{\rm F275W}-m_{\rm F438W}$ CMDs. As for Fig.\,\ref{MSa}, stars surviving from the previous panel are in black, while rejected stars are in grey. (d)-(f) Fiducial lines (in green) used to verticalize the MSa in the $m_{\rm F336W}-m_{\rm F438W}$ and $m_{\rm F275W}-m_{\rm F336W}$ CMDs. (e)-(g) Verticalized $\Delta^{\rm N}_{m_{\rm F336W}-m_{\rm F438W}}$ and $\Delta^{\rm N}_{m_{\rm F336W}-m_{\rm F438W}}$ CMDs. (h) $\Delta^{\rm N}_{m_{\rm F336W}-m_{\rm F438W}}$ versus $\Delta^{\rm N}_{m_{\rm F275W}-m_{\rm F814W}}$. (i) Hess diagram of the TpCD. (l) The three main subcomponents of the bMS, namely bMS1 (dark blue), bMS2 (azure), and bMS3 (light blue).}
\label{bMS}
\end{figure*}

The stands out distinctly on the blue side of the MS, rendering it easily distinguishable. This is evident in panel\,(a) of Fig.\,\ref{bMS}, which illustrates the $m_{\rm F438W}-m_{\rm F814W}$ CMD. This is the most populous of the chemical anomalous 2P groups \citep[see Table\,1 of][]{2017ApJ...844..164B} and corresponds to a population with significant enrichment in light elements such as helium and nitrogen. In this and the following panels of the figure, all previously identified stars (in this case, MSa1 and MSa2 stars) have been removed. The two red lines delineate the colour boundaries of our preliminarily selected bMS stars. The selection is confined to the magnitude range $20.16<m_{\rm F438W}<22.36$, as indicated by the two horizontal red lines. In panel\,(b) we show the $m_{\rm F336W}-m_{\rm F438W}$ CMD of preliminarily selected bMS stars from panel\,(a). Stars rejected in panel\,(a) are in grey. We removed a few outliers using the two red lines. A further selection refinement is applied on the $m_{\rm F275W}-m_{\rm F438W}$ CMD (panel\,(c)). The $m_{\rm F336W}-m_{\rm F438W}$ and $m_{\rm F275W}-m_{\rm F336W}$ CMDs of the stars surviving all the selections are represented in black in panel\,(d) and (f), respectively, while the excluded stars are shown in grey. The green fiducial lines are used to verticalise the sequences of bMS stars following the same procedure used above for the MSa. The two verticalised $\Delta^{\rm N}_{\rm F336W-F438W}$ and $\Delta^{\rm N}_{\rm F275W-F814W}$ CMDs are shown in panels\,(e) and (g). Panel\,(h) shows he $\Delta^{\rm N}_{\rm F336W-F438W}$ versus $\Delta^{\rm N}_{\rm F275W-F814W}$ TpCD of the bMS stars, while the Hess diagram of the TpCD is presented in panel\,(i). The shape of the TpCDs presented in panels\,(h) and (i) closely resemble those illustrated in \citet[see Fig.\,2]{2017ApJ...844..164B}. Consistent with the findings in \citet{2017ApJ...844..164B}, we can distinctly identify three stellar populations, each characterized by clumps located at coordinates (0.3, 0.3), (0.5, 0.6), and (0.8, 0.8). It's worth noting that the clump situated at (0.5, 0.6) appears to be more prominent and visually discernible compared to the findings reported in \citet{2017ApJ...844..164B}, suggesting a possible radial gradient within the bMS sub-populations. However, it is evident that all clumps display varying degrees of overlap and contamination with each other, and their structure appears highly fragmented. Following the same approach of \citet{2017ApJ...844..164B}, we defined three sub-populations of the bMS in panel\,(l): the bMS1 (blue), the bMS2 (azure), and the bMS3 (cyan), each defined as all stars within the respective black envelope. 

The procedures for identifying the bMS population in fields F5, F3, F2, and F1 are outlined in Fig.\,\ref{bMS_5}, \ref{bMS_3}, \ref{bMS_2}, and \ref{bMS_1}, respectively. Despite variations in photometric accuracy and statistical limitations, the general shape of the TpCDs remains consistent across all fields. In field F5, the TpCD resembles that of field F4, albeit slightly more blurred due to lower photometric accuracy. Given this limitation, we opted not to separate the three sub-populations but to consider only the total number of bMS stars for this field. 

In fields F1, F2 and F3, the visibility of the three clumps on the TpCD is reduced due to limited statistics. Nevertheless, in the TpCD in field F2, we can recognize a pattern similar to the TpCD in field F4, with three distinct clumps located around coordinates (0.3, 0.3), (0.55, 0.55), and (0.85, 0.85). Notably, the central clump appears even more prominent compared to field F4, which aligns with the earlier anticipation of a potential radial gradient within the bMS sub-populations. 

Similar to field F2, in field F1, we can also barely discern three clumps, which might correspond to the three clumps identified in field F4. These clumps have coordinates (0.15, 0.25), (0.55, 0.45), and (0.95, 0.95), with the central clump appearing more prominent once again.

The field F3 presents a challenging scenario in the TpCD analysis. The bMS1 populations appear to be absent, while there is a possible indication of an increase in the populations of bMS2 and bMS3. However, the positions of the three subpopulations in the TpCD are not clearly defined and a reliable separation is not feasible. In spite of these considerations, we attempted to identify the three bMS sub-populations in field F1 and F2 (see panel\,(l) of Fig.\,\ref{bMS_2} and \ref{bMS_1}, respectively), while we only considered the total number of bMS stars for field F3. The final sample of bMS stars in fields F5 and F3 consist of those falling inside the black envelope in panel\,(l) of Fig.\,\ref{bMS_5} and Fig.\,\ref{bMS_3}, respectively, represented by black dots. 

\subsection{rMS}\label{Section4.3}
\begin{figure*}
\centering
\includegraphics[width=\textwidth]{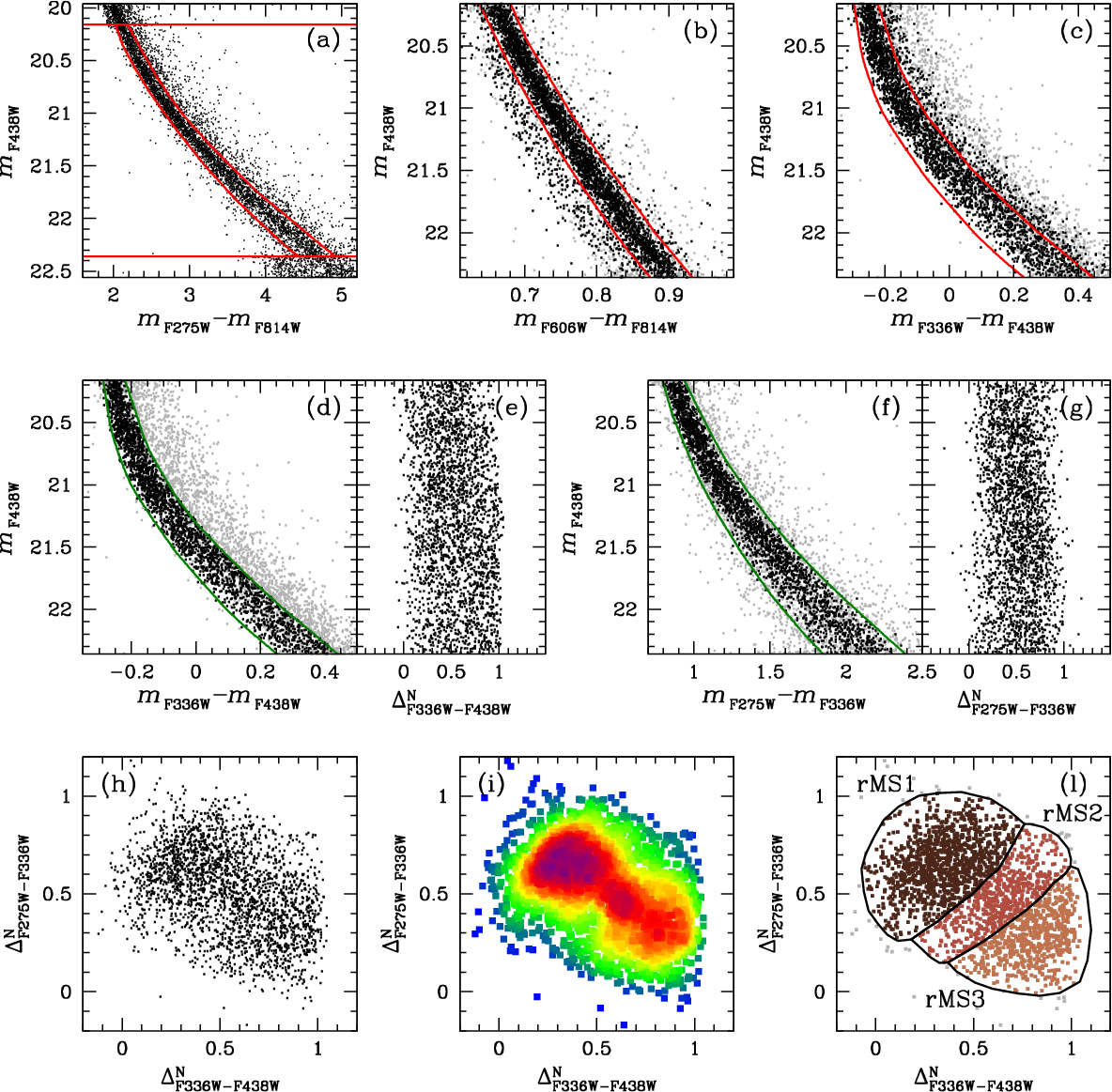}
\caption{Similar to Fig.\,\ref{MSa} and \ref{bMS} but for the rMs stars. The Hess diagram in panel (i) reveals three main subcomponents, labelled as rMS1 (brown), rMS2 (red), and rMS3 (orange) in panel\,(l).}
\label{rMS}
\end{figure*}

Another straightforward population to isolate is the rMS. Following \citet{2017ApJ...844..164B}, we consider this our reference 1P group \citep[although it might include two subgroups of stars with a mild nitrogen enhancement; see Table 1 of][]{2017ApJ...844..164B}. This population becomes apparent in the $m_{\rm F275W}-m_{\rm F814W}$ CMD shown in panel\,(a) of Fig.\,\ref{rMS}, after the removal of MSa and bMS stars. We kept the same magnitude limits as for the bMS, and preliminarily selected rMS stars on this panel by means of the two red lines. Selected stars are then plotted in black in the $m_{\rm F606W}-m_{\rm F814W}$ CMD of panel\,(b), where we remove a few outliers. An additional rejection of likely outliers is performed on the $m_{\rm F336W}-m_{\rm F438W}$ CMD of panel\,(c). We verticalised the sequencies of rMS stars on the $m_{\rm F336W}-m_{\rm F438W}$ and $m_{\rm F275W}-m_{\rm F336W}$ CMDs by means of the green iducials shown in panels\,(d) and (f). The two verticalised $\Delta^{\rm N}_{\rm F336W-F438W}$ and $\Delta^{\rm N}_{\rm F275W-F814W}$ CMDs are shown in panels\,(e) and (g). Panel\,(h) shows the $\Delta^{\rm N}_{\rm F275W-F336W}$ versus $\Delta^{\rm N}_{\rm F336W-F438W}$ TpCD of selected rMS stars (in black). The corresponding Hess diagram is in panel\,(i). The shape of the TpCDs presented in panels\,(h) and (i) resembles those reported by \citet[see Fig.\,3]{2017ApJ...844..164B}. Similar to the findings in \citet{2017ApJ...844..164B}, we can distinguish three distinct stellar populations, characterized by clumps located at coordinates (0.35, 0.65), (0.65, 0.45), and (0.8, 0.3). However, it is important to note that all clumps exhibit a higher degree of overlap and contamination compared to the results reported by \citet{2017ApJ...844..164B}. The shape of these clumps and the relative abundance of their populations appear slightly different from the findings in \citet{2017ApJ...844..164B}. Specifically, the leftmost clumps appear more prominent than those observed in the previous study. This variation could be attributed to contamination by other populations, such as the populations MSe (discussed in Section \ref{Section4.5}), or it could indicate intrinsic variations in the number of stars within the rMS sub-populations as a function of radial distance from the cluster centre. Following the approach of \citet{2017ApJ...844..164B} we defined three rMS sub-populations in panel\,(l): rMS1 (brown), rMS2 (red), and rMS3 (orange).

Corresponding figures for the other fields are presented in Fig.\,\ref{rMS_5}, \ref{rMS_3}, \ref{rMS_2}, and \ref{rMS_1}. Across all fields, the general shape of the TpCDs remains consistent. In field F5, the TpCD resembles that of field F4, yet it is significantly blurred, hindering accurate identification of the three sub-populations. In fields F1, F2 and F3, only the leftmost clump is clearly visible, appearing more prominent compared to fields F4 and F5, particularly in field F3. The other two clumps are barely discernible, posing challenges for identification. Similar to fields F4 and F5, the dominance of the leftmost clump in fields F1, F2 and F3 might stem from contamination by other populations or intrinsic radial variations in the number of stars within the three rMS sub-populations. As a result, we opted not to estimate any sub-populations for these fields but we only consider the total number of rMS stars. The final sample of rMS stars in these fields consists of those falling inside the black envelope in panel\,(l) of Fig.\,\ref{rMS_5}, \ref{rMS_3} \ref{rMS_2}, and \ref{rMS_1} represented by black dots. 

\subsection{MSd}\label{Section4.4}
\begin{figure*}
\centering
\includegraphics[width=\textwidth]{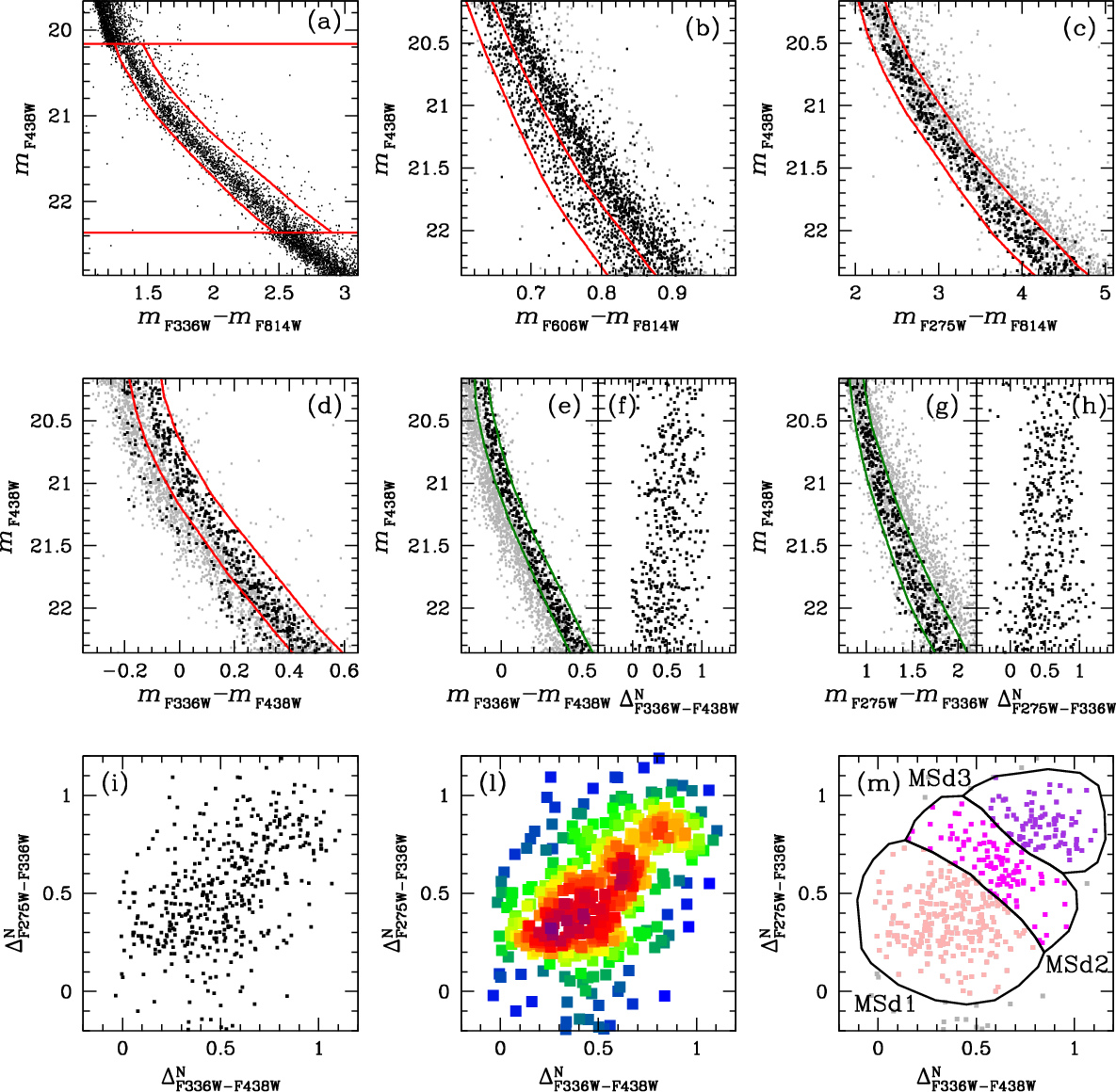}
\caption{(a) $m_{\rm F336W}-m_{\rm F814W}$ CMD of the MS stars not belonging to MSa, bMs and rMS. We selected all stars falling between the two red diagonal lines. (b) $m_{\rm F606W}-m_{\rm F814W}$ CMD of the stars selected in (a), where they clearly split into two components. The MSd stars, situated in the blue component, are preliminarily selected. (c)-(d) Selection refinements for MSd stars. (e)-(g) Fiducials (in green) used to verticalise the MSd sequence in the $m_{\rm F336W}-m_{\rm F438W}$ and $m_{\rm F275W}-m_{\rm F336W}$ CMDs. (f)-(h) Verticalised $\Delta^{\rm N}_{\rm F336W-F438W}$ and $\Delta^{\rm N}_{\rm F275W-F814W}$ CMDs. The TpCD (i) and Hess diagram (l) of MSd stars reveal three main sub-populations, which we define in panel\,(m) as MSd1 (pink), MSd2 (magenta), and MSd3 (purple).}
\label{MSd}
\end{figure*}

Here, we turn our focus to the MSd population \citep[see][]{2017ApJ...844..164B}. According to the study of \citet{2017ApJ...844..164B}, the chemical properties of this 2P subgroup would be characterized by Fe and helium enhancement (less extreme than the MSa population discussed in Section\,\ref{Section4.1}) and a modest nitrogen enhancement compared to the reference rMS population. In panel\,(a) of Fig.\,\ref{MSd}, we present the $m_{\rm F336W}-m_{\rm F814W}$ CMD for stars not categorized under populations MSa, bMS, and rMS. We selected all stars falling between the two red diagonal lines and kept the same magnitude limits used for the bMS and rMS stars (red horizontal lines). Panel\,(b) shows the $m_{\rm F606W}-m_{\rm F814W}$ CMD of these selected stars in black, where two distinct sequences are evident. The MSd stars, situated in the blue component, are preliminarily selected (red lines) in panel\,(b). Further refinement of the MSd sample involved removing a few outliers using the $m_{\rm F275W}-m_{\rm F814W}$ and $m_{\rm F336W}-m_{\rm F438W}$ CMDs (panels\,(c) and (d), respectively). The fiducials used to verticalised the MSd sequence in the $m_{\rm F336W}-m_{\rm F438W}$ and $m_{\rm F275W}-m_{\rm F336W}$ CMDs are shown in green in panels\,(e) and (g), while the verticalised $\Delta^{\rm N}_{\rm F336W-F438W}$ and $\Delta^{\rm N}_{\rm F275W-F814W}$ CMDs are shown in panels\,(f) and (h). The TpCD and Hess diagrams of selected MSd stars are shown in panels\,(i) and (l), respectively. The TpCD shape resembles that presented in \citet[see Fig.\,5]{2017ApJ...844..164B}, where three clumps were identified: two primary clumps situated in the lower-left and centre sections of the plot, and a less populated clump positioned in the upper-right section. Similarly, we observe two main clumps at coordinates (0.25, 0.3) and (0.6, 0.65), along with a less populated clump at (0.8, 0.85). However, the two main clumps exhibit significant overlap and contamination, posing challenges in clear sub-population separation. The border between these clumps appears higher and more towards the right than what was presented in \citet{2017ApJ...844..164B}, with the (0.25, 0.3) clump appearing more prominent compared to the findings reported therein. The three sub-populations are defined in panel\,(m), labelled as MSd1 (pink), MSd2 (magenta), and MSd3 (purple).

The TpCD for field F5 (see Fig.\,\ref{MSd_5}) presents a challenging scenario for the identification of the three sub-populations, displaying a different structure compared to what was obtained for field F4. Additionally, the TpCDs for fields F3, F2, and F1 (see Fig.\,\ref{MSd_3}, \ref{MSd_2}, and \ref{MSd_1}, respectively) are characterized by a low number of stars and poor statistics, making it very difficult to identify the three clumps. Due to these limitations, we decided not to apply any sub-population selection for these fields, but to only consider the total number of MSd stars. The final sample of MSd stars in these fields consists of those falling inside the black envelope in panel\,(m) of Fig.\,\ref{MSd_5}, \ref{MSd_3}, \ref{MSd_2}, and \ref{MSd_1} represented by black dots.

\subsection{MSe}\label{Section4.5}
\begin{figure*}
\centering
\includegraphics[width=\textwidth]{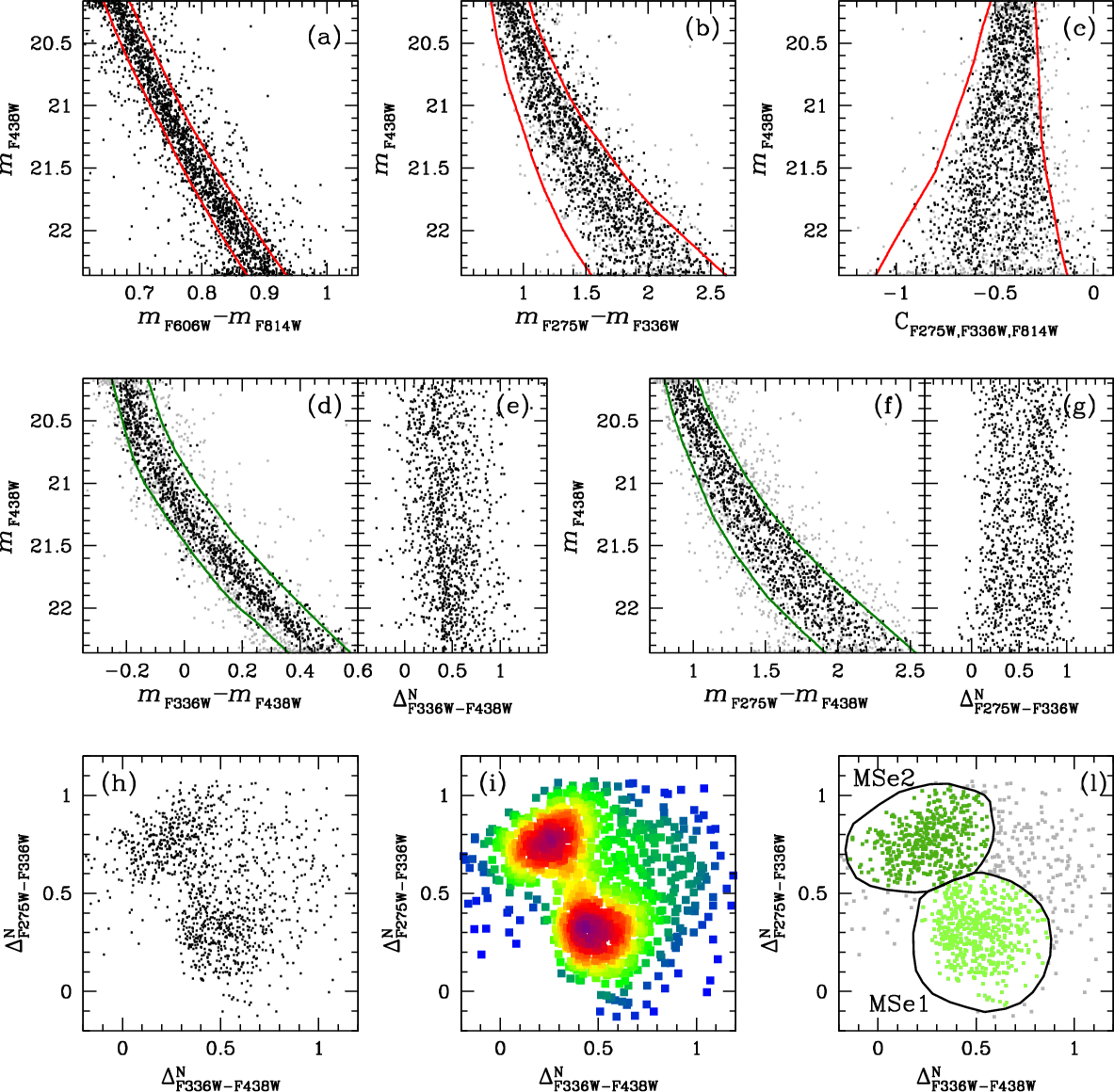}
\caption{(a) Same as panel\,(b) of Fig.\,\ref{MSd}, in which we also removed MSd stars. The remaining stars, constituting the MSe population, form a well-defined sequence on this plane, which we select and further refine in panels\,(b) and (c). (d)-(f) fiducials (in green) used to verticalise the MSe sequences. (e)-(g) verticalised CMDs. (h)-(i) TpCD and the Hess diagram of MSe stars. (l) We identified the two MSe populations as MSe1 (lime) and MSe2 (green).}
\label{MSe}
\end{figure*}

Here, we focus our attention on the red component, constituting the MSe population \citep[see][]{2017ApJ...844..164B}, which was excluded in panel\,(b) of Fig.\,\ref{MSd}. This population is composed of various sub-populations with chemical abundances slightly enhanced in Fe and/or nitrogen relative to the rMS population \citep[see][]{2017ApJ...844..164B}. Panel\,(a) of Fig.\,\ref{MSe} is similar to panel\,(b) of Fig.\,\ref{MSd}, but without MSd stars. Within this panel, we initially selected MSe stars as those enclosed by the two red lines. We refined the MSe sample as shown in panels\,(b) and (c). Panels\,(d) and (f) of Fig.\,\ref{MSe} show the fiducials utilized to verticalize the MSe sequence, while panels\,(e) and (g) exhibit the verticalized CMDs. Panels\,(h) and (i) shows the TpCD of the selected MSe stars and the corresponding Hess diagram. The TpCD presented here closely resembles that illustrated in \citet[see Fig.\,6]{2017ApJ...844..164B}: two prominent clumps are clearly discernible, situated approximately at (0.45, 0.3) and (0.25, 0.75). However, due to lower statistics, the identification of the additional two less populated clumps introduced by \citet{2017ApJ...844..164B} is not feasible. Therefore, we concentrate our attention solely on the two main clumps. In panel\,(l), we consequently delineate the following two MSe sub-populations: MSe1 (lime) and MSe2 (green).

The corresponding figures for fields F5, F3, F2, and F1 are presented in Fig.\,\ref{MSe_5}, \ref{MSe_3}, \ref{MSe_2}, and \ref{MSe_1} respectively. For field F5, the obtained TpCD closely resembles that of field F4, albeit more blurred. In fields F2 and F3, despite lower statistics, we are still able to identify the two sub-populations. However, in field F1, the two sub-populations are not discernible, so we only consider the total number of MSe stars in this field.

\section{Radial variation among stellar populations}\label{Section5}
\begin{figure}
\centering
\includegraphics[width=\columnwidth]{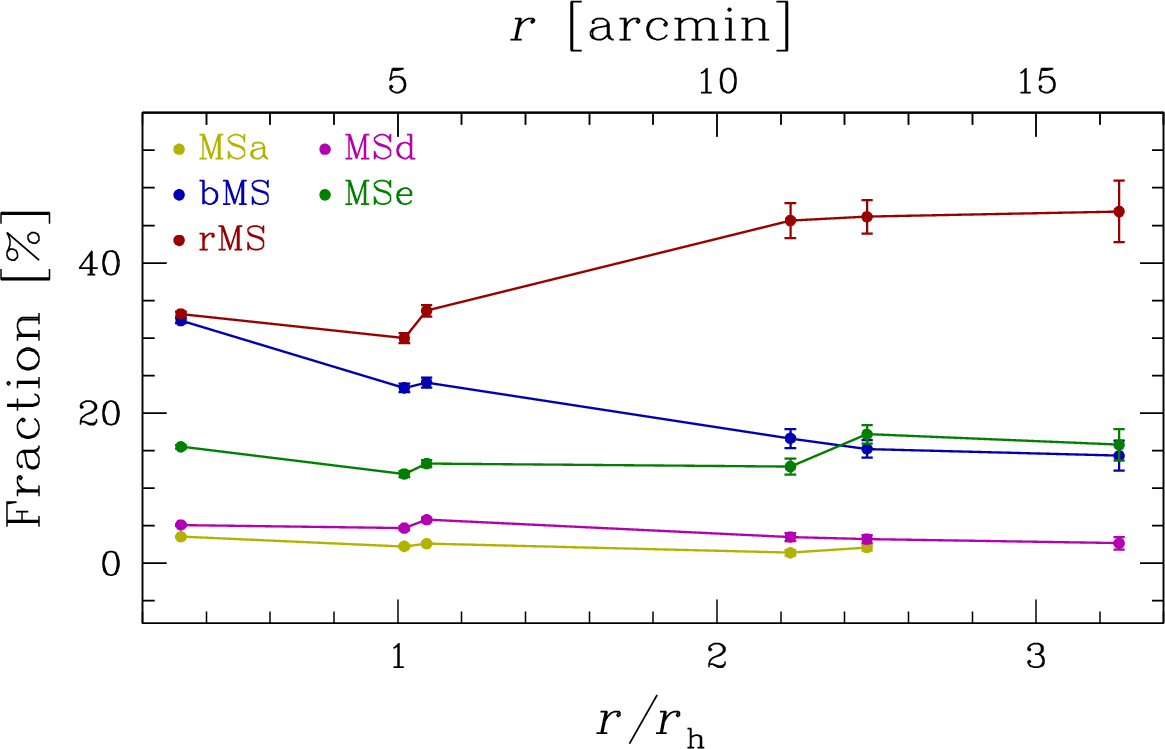}
\caption{Fractions of stars in each main population for each field as a function of radial distance from the cluster centre, with each population colour-coded as indicated in the top-left corner of the plot.}
\label{Figure2}
\end{figure}

Table\,\ref{Table2} presents the count and relative percentage of stars in the magnitude range $20.16<m_{\rm F438W}<22.36$, across the five identified mPOPs and their sub-populations, alongside the estimated number of unidentified stars\footnote{The count of unidentified stars in our analysis also includes binaries. In contrast to \citet{2017ApJ...844..164B}, this study does not provide an estimate of the number of binaries. The decision to neglect the presence of binaries stems from uncertainties and their relatively low abundance in this cluster, as noted in previous studies \citep[see][]{2022A&A...659A..96M,2023tmib.confE..12W}.} in fields F1, F2, F3, F4, and F5. Errors are estimated using Poisson errors and propagating the uncertainties. For convenience, the corresponding values from the central field reported by \citet{2017ApJ...844..164B} are also included in Table\,\ref{Table2}. Each field's distance interval from the cluster centre is provided, with the fields listed in order from the closest to the farthest. The fractions of stars in each main population for each field are also illustrated in Fig.\,\ref{Figure2} as a function of radial distance from the cluster centre.

The separation and identification of mPOPs, particularly their sub-populations, have proven to be more challenging than outlined in \citet{2017ApJ...844..164B}. This difficulty stems from lower photometric accuracy and precision, attributed to fewer exposures (especially in field F5) and limited star counts and statistics, particularly evident in the three outer fields, F1, F2 and F3. Notably, the proportion of unidentified stars in our fields exceeds that reported for the central field (10.34\% $\pm$ 0.17\%), with the highest proportion observed in field F5 (27.84\% $\pm$ 0.59\%). This discrepancy is primarily due to the reduced photometric accuracy in this field. To account for the different photometric quality among the analyzed fields and the varying numbers of unidentified stars in each, we also provide in Table\,\ref{Table2}, within parentheses, ratios obtained solely from the counts of classified stars.

In all fields analyzed, the two most populous stellar populations observed are the rMS and the bMS populations, except for fields F1 and F3, where the MSe populations outnumber the bMS population.

The fraction of MSa stars is approximately constant over the range of radial distances from the cluster centre we have explored, although there is a possible subtle indication of a slight decrease in the outermost regions: the fraction of MSa stars varies from 3.53\% $\pm$ 0.10\% to $\sim$2.4\% in the intermediate (F4 and F5) and $\sim$1.7\% in the outer (F2 and F3) fields. A similar trend is observed when considering only classified stars. The two MSa sub-populations (MSa1 and MSa2) can be separated only in the central and the two intermediate fields. In that radial range, the two populations follow a trend similar to that found for the entire MSa population with no strong indication of a variation in the relative number of stars in each of the two groups.

A clear trend is observed in the bMS population with radial distance from the cluster centre, with bMS populations decreasing from the centre to the outskirts. Specifically, there is a central value of 32.32\% $\pm$ 0.33\%, dropping to $\sim$23-24\% in the two intermediate fields (F4 and F5), and further to $\sim$14-16\% in the three outer fields (F1, F2 and F3).
Similar considerations are valid also when considering only classified stars, although the radial variation appears less steep. A reliable division of the three bMS sub-populations was feasible only in the central field and fields F4, F2 and F1. The results obtained from the analysis of these fields show that the fraction of stars in each of these sub-populations follows a radial gradient similar to that found in the total bMS population and decreases at larger distances from the cluster centre. Such a trend, however, appears milder in the bMS2 sub-population. 

The fraction of the total number of stars in the rMS group follows a variation with the distance from the cluster centre characterized by an approximately constant fraction in the inner regions and increasing in the outer regions, a behaviour complementary to that of the other dominant population (the bMS population). This increase becomes more pronounced when considering only classified stars. We note that, as anticipated in Section\,\ref{Section4.3}, the leftmost clump in the TpCDs for the fields F5, F4, F3, F2 and F1 present a prominence more evident with respect to what was found in the central field \citep[see Fig.\,3 of][]{2017ApJ...844..164B}. This can be due to contamination from other populations, especially the MSe populations, which partially overlap the rMS population in most CMDs, or can indicate a radial gradient within the rMS sub-populations. The only field in which we were able to separate the rMS sub-populations with sufficient accuracy was the F4 field. The identification of the three rMS subgroups was possible only in the central field and in the F4 field making the study of the radial variation of the fraction of stars in the three groups difficult. The data available suggest that the fraction of the total number of rMS belonging to the rMS1 (rMS3) group increases (decreases) with the distance from the cluster centre and it is approximately constant for the rMS2 group. 

For the MSd population, the number of MSd stars in the two intermediate fields (4.68\% $\pm$ 0.24\% and 5.80\% $\pm$ 0.29\% for F5 and F4, respectively) aligns closely with the central field value (5.10\% $\pm$ 0.12\%) but slightly decreases in the two outer fields (3.50\% $\pm$ 0.54\%, 3.23\% $\pm$ 0.50\% and 2.67\% $\pm$ 0.82\% for F2, F3 and F1, respectively). However, when considering only classified stars, we observe a slight increase in the two intermediate fields (6.49\% $\pm$ 0.33\% and 7.31\% $\pm$ 0.37\% for F5 and F4, respectively) with respect to the central value (5.69\% $\pm$ 0.13\%), followed by a decrease in the outer fields (4.38\% $\pm$ 0.68\%, 3.85\% $\pm$ 0.60\% and 3.35\% $\pm$ 1.03\% for F2, F3, and F1, respectively). Among these fields, only field F4 allowed for the separation of the MSd sub-populations, although with considerable uncertainties due to the envelope positions in the TpCD. Nonetheless, the number counts suggest a potential trend among the MSd sub-populations. While the fraction of the total number of MSd stars in the MSd3 group remains relatively constant in the central field and field F4 ($\sim$0.20-0.22), the fraction of MSd stars in the MSd1 and MSd2 sub-populations exhibit a radial gradient. Specifically, from the central field to field F4, the MSd1/MSd ratio increases from MSd1/MSd$\sim$0.38 $\pm$ 0.02 to MSd1/MSd$\sim$0.56 $\pm$ 0.05 (an increase of $\sim$68\%), while the MSd2/MSd ratio decreases from MSd2/MSd$\sim$0.41 $\pm$ 0.02 to MSd2/MSd$\sim$0.24 $\pm$ 0.03 (a decrease of $\sim$60\%). However, it is important to note that these numbers are affected by uncertainties stemming from the envelope positions on the TpCD, as mentioned earlier.

For the MSe population, the number of stars remains relatively stable across varying radial distances from the cluster centre, ranging between $\sim$13.8\% to 17.2\%. When considering only classified stars, we observe a consistent ratio in the central field and fields F5, F4, and F2 ($\sim$16-17\%) and then a slight increase in the most outer fields, F3 and F1 ($\sim$20\%). Although we were unable to separate the smallest MSe sub-populations (MSe3 and MSe4) due to low statistics, we achieved accurate separation of the two main MSe sub-populations (MSe1 and MSe2) in fields F2, F3, F4 and F5. The relative number of the two sub-populations, MSe1 and MSe2, maintains a constant ratio in the central field and in the intermediate fields (F4 and F5), with MSe2/MSe1 $\sim$1, while it shows a notable increase in the two outer fields (F2 and F3), with a value of MSe2/MSe1 $\sim$2.2.

\begin{figure}
\centering
\includegraphics[width=\columnwidth]{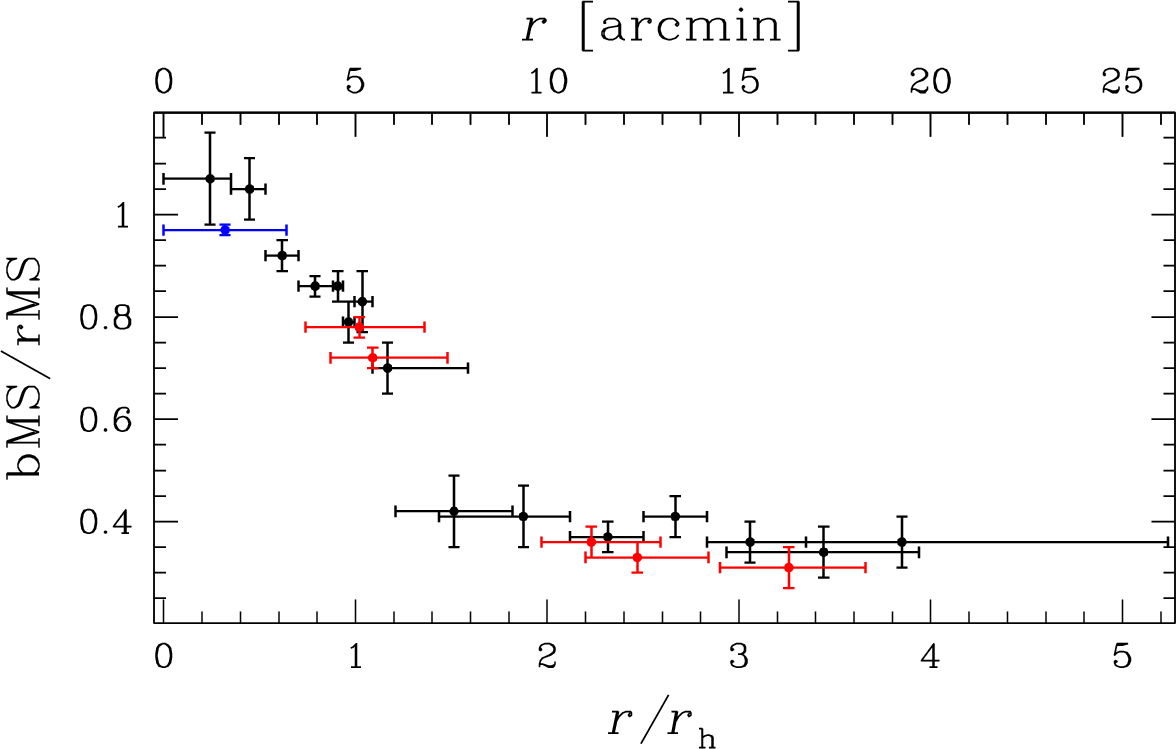}
\caption{bMS/rMS as a function of the radial distance. Black points represent data from \citet{2009A&A...493..959B}, the blue point from \citet{2017ApJ...844..164B}, and red points from our current study.}
\label{Figure1}
\end{figure}

As outlined in Section\,\ref{Section1}, \citet{2009A&A...493..959B} conducted an extensive examination of the radial distribution of the two primary stellar populations in $\omega$\,Cen (rMs and bMS), identifying a radial gradient in their counts-number fraction (bMS/rMS). In Fig.\,\ref{Figure1}, we show a comparison with the radial profile presented in \citet[black points]{2009A&A...493..959B} alongside the ratios derived from the data presented in this study (red points) and in \citet[blue point][]{2017ApJ...844..164B}, revealing a notable agreement between the datasets.

\section{Summary}\label{Section6}
In this paper, we presented the reduction process of HST data acquired through the GO-16247 (P.I.: Scalco) and GO-14118+14662 (P.I.: Bedin) programmes. We provided an overview of the dataset and the process of data reduction. We have combined the newly obtained data with previously processed data \citep[see][]{2021MNRAS.505.3549S} to extend the analysis of \citet{2017ApJ...844..164B}, thus presenting, for the first time, a comprehensive investigation of the radial gradient of the mPOPs within $\omega$\,Cen spanning a significant portion of the cluster.

Our analysis revealed significant radial variations in the fraction of stars in the stellar populations of $\omega$\,Cen and their sub-populations. For the two dominant populations, the bMS and the rMS, our results show that the fraction of bMS (rMS) decreases (increases) with the distance from the cluster centre. Our findings are consistent with those reported by \citet{2009A&A...493..959B}. 

As discussed in Section\,\ref{Section1}, according to various formation scenarios \citep[see e.g.][]{erc,2010ApJ...724L..99B,2011MNRAS.412.2241B,2019MNRAS.489.3269C}, 2P stars are expected to form more centrally concentrated in the cluster’s inner regions and gradually mix with 1P stars during the cluster’s evolution. $\omega$\,Cen has a long relaxation time \citep[$\sim$1.1~Gyr in the core and $\sim$10~Gyr at the half-mass radius,][]{1996AJ....112.1487H,2010arXiv1012.3224H}, suggesting that the effects of two-body relaxation should not have completely erased the spatial differences imprinted by the formation processes and some memory of those differences should still be present in the current properties. This is consistent with our observations finding that the main 2P population (represented by the bMS population) is more centrally concentrated compared to the 1P population (represented by the rMS population).

It is important to emphasize that the variety of stellar populations and sub-populations hosted by $\omega$\,Cen are a clear indication of a complex formation history. An interpretation of the differences in the spatial variations requires particular attention and caution. Spectroscopic follow-ups --of the \textit{already} characterized stars in this work-- are essential for an accurate classification of the chemical properties of these stellar populations to gain a deeper understanding of their characteristics. Additionally, the complex formation history and intricate system of stellar populations in $\omega$\,Cen further underscore the need for future theoretical simulations involving mPOPs, beyond just 1P and 2P, to enhance our understanding of the presented results. Some initial theoretical studies in this direction have been presented by \citet{2019ApJ...886..121B,2021MNRAS.506.5951L,2022MNRAS.517.1171L} and have shown that more extreme 2P stars (i.e., those with higher helium content) may form initially more concentrated than the 2P with lower helium content \citep[see also][for a similar trend in NGC\,2808]{2016MNRAS.463..449S}. The much weaker radial variation of the MSd and MSe populations might be consistent with those predictions, but we reiterate the importance of caution in this comparison and recognize the necessity for further observational efforts to characterize the chemical properties of all the populations and further theoretical studies before drawing more definitive conclusions.

Finally, we point out that wide-field missions recently approved by \textit{NASA} (such as \textit{UVEX}) and ongoing (such as \textit{Euclid}) will contribute to statistical analysis and spatial distribution of these stellar populations.

\begin{acknowledgements}
Michele Scalco and Luigi Rolly Bedin acknowledge support by MIUR under the PRIN-2017 programme \#2017Z2HSMF, and by INAF under the PRIN-2019 programme \#10-Bedin.
\end{acknowledgements}

\bibliographystyle{aa}
\bibliography{main.bib}

\begin{thebibliography}{57}
\expandafter\ifx\csname natexlab\endcsname\relax\def\natexlab#1{#1}\fi

\bibitem[{{Anderson}(2016)}]{2016wfc..rept...12A}
{Anderson}, J. 2016, {Empirical Models for the WFC3/IR PSF}, Instrument Science Report WFC3 2016-12, 42 pages

\bibitem[{{Anderson}(2022)}]{2022wfc..rept....5A}
{Anderson}, J. 2022, {One-Pass HST Photometry with hst1pass}, Instrument Science Report WFC3 2022-5, 55 pages

\bibitem[{{Anderson} \& {King}(2000)}]{2000PASP..112.1360A}
{Anderson}, J. \& {King}, I.~R. 2000, \pasp, 112, 1360

\bibitem[{{Anderson} \& {King}(2006)}]{2006acs..rept....1A}
{Anderson}, J. \& {King}, I.~R. 2006, {PSFs, Photometry, and Astronomy for the ACS/WFC}, Instrument Science Report ACS 2006-01, 34 pages

\bibitem[{{Anderson} {et~al.}(2008){Anderson}, {Sarajedini}, {Bedin}, {King}, {Piotto}, {Reid}, {Siegel}, {Majewski}, {Paust}, {Aparicio}, {Milone}, {Chaboyer}, \& {Rosenberg}}]{2008AJ....135.2055A}
{Anderson}, J., {Sarajedini}, A., {Bedin}, L.~R., {et~al.} 2008, \aj, 135, 2055

\bibitem[{{Balaguer-N{\'u}nez} {et~al.}(1998){Balaguer-N{\'u}nez}, {Tian}, \& {Zhao}}]{1998A&AS..133..387B}
{Balaguer-N{\'u}nez}, L., {Tian}, K.~P., \& {Zhao}, J.~L. 1998, \aaps, 133, 387

\bibitem[{{Bastian} \& {Lardo}(2018)}]{2018ARA&A..56...83B}
{Bastian}, N. \& {Lardo}, C. 2018, ARA\&A, 56, 83

\bibitem[{{Bedin} {et~al.}(2005){Bedin}, {Cassisi}, {Castelli}, {Piotto}, {Anderson}, {Salaris}, {Momany}, \& {Pietrinferni}}]{2005MNRAS.357.1038B}
{Bedin}, L.~R., {Cassisi}, S., {Castelli}, F., {et~al.} 2005, \mnras, 357, 1038

\bibitem[{{Bedin} {et~al.}(2008){Bedin}, {King}, {Anderson}, {Piotto}, {Salaris}, {Cassisi}, \& {Serenelli}}]{2008ApJ...678.1279B}
{Bedin}, L.~R., {King}, I.~R., {Anderson}, J., {et~al.} 2008, \apj, 678, 1279

\bibitem[{{Bedin} {et~al.}(2004){Bedin}, {Piotto}, {Anderson}, {Cassisi}, {King}, {Momany}, \& {Carraro}}]{2004ApJ...605L.125B}
{Bedin}, L.~R., {Piotto}, G., {Anderson}, J., {et~al.} 2004, \apjl, 605, L125

\bibitem[{{Bekki}(2010)}]{2010ApJ...724L..99B}
{Bekki}, K. 2010, \apjl, 724, L99

\bibitem[{{Bekki}(2011)}]{2011MNRAS.412.2241B}
{Bekki}, K. 2011, \mnras, 412, 2241

\bibitem[{{Bekki} \& {Freeman}(2003)}]{2003MNRAS.346L..11B}
{Bekki}, K. \& {Freeman}, K.~C. 2003, \mnras, 346, L11

\bibitem[{{Bekki} \& {Norris}(2006)}]{2006ApJ...637L.109B}
{Bekki}, K. \& {Norris}, J.~E. 2006, \apjl, 637, L109

\bibitem[{{Bekki} \& {Tsujimoto}(2019)}]{2019ApJ...886..121B}
{Bekki}, K. \& {Tsujimoto}, T. 2019, ApJ, 886, 121

\bibitem[{{Bellini} {et~al.}(2011){Bellini}, {Anderson}, \& {Bedin}}]{2011PASP..123..622B}
{Bellini}, A., {Anderson}, J., \& {Bedin}, L.~R. 2011, \pasp, 123, 622

\bibitem[{{Bellini} {et~al.}(2017{\natexlab{a}}){Bellini}, {Anderson}, {Bedin}, {King}, {van der Marel}, {Piotto}, \& {Cool}}]{2017ApJ...842....6B}
{Bellini}, A., {Anderson}, J., {Bedin}, L.~R., {et~al.} 2017{\natexlab{a}}, \apj, 842, 6

\bibitem[{{Bellini} {et~al.}(2017{\natexlab{b}}){Bellini}, {Anderson}, {van der Marel}, {King}, {Piotto}, \& {Bedin}}]{2017ApJ...842....7B}
{Bellini}, A., {Anderson}, J., {van der Marel}, R.~P., {et~al.} 2017{\natexlab{b}}, \apj, 842, 7

\bibitem[{{Bellini} {et~al.}(2014){Bellini}, {Anderson}, {van der Marel}, {Watkins}, {King}, {Bianchini}, {Chanam{\'e}}, {Chandar}, {Cool}, {Ferraro}, {Ford}, \& {Massari}}]{2014ApJ...797..115B}
{Bellini}, A., {Anderson}, J., {van der Marel}, R.~P., {et~al.} 2014, \apj, 797, 115

\bibitem[{{Bellini} {et~al.}(2010){Bellini}, {Bedin}, {Piotto}, {Milone}, {Marino}, \& {Villanova}}]{2010AJ....140..631B}
{Bellini}, A., {Bedin}, L.~R., {Piotto}, G., {et~al.} 2010, \aj, 140, 631

\bibitem[{{Bellini} {et~al.}(2018){Bellini}, {Libralato}, {Bedin}, {Milone}, {van der Marel}, {Anderson}, {Apai}, {Burgasser}, {Marino}, \& {Rees}}]{2018ApJ...853...86B}
{Bellini}, A., {Libralato}, M., {Bedin}, L.~R., {et~al.} 2018, \apj, 853, 86

\bibitem[{{Bellini} {et~al.}(2017{\natexlab{c}}){Bellini}, {Milone}, {Anderson}, {Marino}, {Piotto}, {van der Marel}, {Bedin}, \& {King}}]{2017ApJ...844..164B}
{Bellini}, A., {Milone}, A.~P., {Anderson}, J., {et~al.} 2017{\natexlab{c}}, \apj, 844, 164

\bibitem[{{Bellini} {et~al.}(2009){Bellini}, {Piotto}, {Bedin}, {Anderson}, {Platais}, {Momany}, {Moretti}, {Milone}, \& {Ortolani}}]{2009A&A...493..959B}
{Bellini}, A., {Piotto}, G., {Bedin}, L.~R., {et~al.} 2009, \aap, 493, 959

\bibitem[{{Calamida} {et~al.}(2020){Calamida}, {Zocchi}, {Bono}, {Ferraro}, {Mastrobuono-Battisti}, {Saha}, {Iannicola}, {Rest}, {Strampelli}, \& {Zenteno}}]{2020ApJ...891..167C}
{Calamida}, A., {Zocchi}, A., {Bono}, G., {et~al.} 2020, \apj, 891, 167

\bibitem[{{Calura} {et~al.}(2019){Calura}, {D'Ercole}, {Vesperini}, {Vanzella}, \& {Sollima}}]{2019MNRAS.489.3269C}
{Calura}, F., {D'Ercole}, A., {Vesperini}, E., {Vanzella}, E., \& {Sollima}, A. 2019, MNRAS, 489, 3269

\bibitem[{{D'Ercole} {et~al.}(2008){D'Ercole}, {Vesperini}, {D'Antona}, {McMillan}, \& {Recchi}}]{erc}
{D'Ercole}, A., {Vesperini}, E., {D'Antona}, F., {McMillan}, S. L.~W., \& {Recchi}, S. 2008, MNRAS, 391, 825

\bibitem[{{Gaia Collaboration} {et~al.}(2016){Gaia Collaboration}, {Brown}, {Vallenari}, {Prusti}, {de Bruijne}, {Mignard}, {Drimmel}, {Babusiaux}, {Bailer-Jones}, {Bastian}, {Biermann}, {Evans}, {Eyer}, {Jansen}, {Jordi}, {Katz}, {Klioner}, {Lammers}, {Lindegren}, {Luri}, {O'Mullane}, {Panem}, {Pourbaix}, {Randich}, {Sartoretti}, {Siddiqui}, {Soubiran}, {Valette}, {van Leeuwen}, {Walton}, {Aerts}, {Arenou}, {Cropper}, {H{\o}g}, {Lattanzi}, {Grebel}, {Holland}, {Huc}, {Passot}, {Perryman}, {Bramante}, {Cacciari}, {Casta{\~n}eda}, {Chaoul}, {Cheek}, {De Angeli}, {Fabricius}, {Guerra}, {Hern{\'a}ndez}, {Jean-Antoine-Piccolo}, {Masana}, {Messineo}, {Mowlavi}, {Nienartowicz}, {Ord{\'o}{\~n}ez-Blanco}, {Panuzzo}, {Portell}, {Richards}, {Riello}, {Seabroke}, {Tanga}, {Th{\'e}venin}, {Torra}, {Els}, {Gracia-Abril}, {Comoretto}, {Garcia-Reinaldos}, {Lock}, {Mercier}, {Altmann}, {Andrae}, {Astraatmadja}, {Bellas-Velidis}, {Benson}, {Berthier}, {Blomme}, {Busso}, {Carry}, {Cellino}, {Clementini}, {Cowell}, {Creevey},
  {Cuypers}, {Davidson}, {De Ridder}, {de Torres}, {Delchambre}, {Dell'Oro}, {Ducourant}, {Fr{\'e}mat}, {Garc{\'\i}a-Torres}, {Gosset}, {Halbwachs}, {Hambly}, {Harrison}, {Hauser}, {Hestroffer}, {Hodgkin}, {Huckle}, {Hutton}, {Jasniewicz}, {Jordan}, {Kontizas}, {Korn}, {Lanzafame}, {Manteiga}, {Moitinho}, {Muinonen}, {Osinde}, {Pancino}, {Pauwels}, {Petit}, {Recio-Blanco}, {Robin}, {Sarro}, {Siopis}, {Smith}, {Smith}, {Sozzetti}, {Thuillot}, {van Reeven}, {Viala}, {Abbas}, {Abreu Aramburu}, {Accart}, {Aguado}, {Allan}, {Allasia}, {Altavilla}, {{\'A}lvarez}, {Alves}, {Anderson}, {Andrei}, {Anglada Varela}, {Antiche}, {Antoja}, {Ant{\'o}n}, {Arcay}, {Bach}, {Baker}, {Balaguer-N{\'u}{\~n}ez}, {Barache}, {Barata}, {Barbier}, {Barblan}, {Barrado y Navascu{\'e}s}, {Barros}, {Barstow}, {Becciani}, {Bellazzini}, {Bello Garc{\'\i}a}, {Belokurov}, {Bendjoya}, {Berihuete}, {Bianchi}, {Bienaym{\'e}}, {Billebaud}, {Blagorodnova}, {Blanco-Cuaresma}, {Boch}, {Bombrun}, {Borrachero}, {Bouquillon}, {Bourda}, {Bouy},
  {Bragaglia}, {Breddels}, {Brouillet}, {Br{\"u}semeister}, {Bucciarelli}, {Burgess}, {Burgon}, {Burlacu}, {Busonero}, {Buzzi}, {Caffau}, {Cambras}, {Campbell}, {Cancelliere}, {Cantat-Gaudin}, {Carlucci}, {Carrasco}, {Castellani}, {Charlot}, {Charnas}, {Chiavassa}, {Clotet}, {Cocozza}, {Collins}, {Costigan}, {Crifo}, {Cross}, {Crosta}, {Crowley}, {Dafonte}, {Damerdji}, {Dapergolas}, {David}, {David}, {De Cat}, {de Felice}, {de Laverny}, {De Luise}, {De March}, {de Martino}, {de Souza}, {Debosscher}, {del Pozo}, {Delbo}, {Delgado}, {Delgado}, {Di Matteo}, {Diakite}, {Distefano}, {Dolding}, {Dos Anjos}, {Drazinos}, {Duran}, {Dzigan}, {Edvardsson}, {Enke}, {Evans}, {Eynard Bontemps}, {Fabre}, {Fabrizio}, {Faigler}, {Falc{\~a}o}, {Farr{\`a}s Casas}, {Federici}, {Fedorets}, {Fern{\'a}ndez-Hern{\'a}ndez}, {Fernique}, {Fienga}, {Figueras}, {Filippi}, {Findeisen}, {Fonti}, {Fouesneau}, {Fraile}, {Fraser}, {Fuchs}, {Gai}, {Galleti}, {Galluccio}, {Garabato}, {Garc{\'\i}a-Sedano}, {Garofalo}, {Garralda}, {Gavras},
  {Gerssen}, {Geyer}, {Gilmore}, {Girona}, {Giuffrida}, {Gomes}, {Gonz{\'a}lez-Marcos}, {Gonz{\'a}lez-N{\'u}{\~n}ez}, {Gonz{\'a}lez-Vidal}, {Granvik}, {Guerrier}, {Guillout}, {Guiraud}, {G{\'u}rpide}, {Guti{\'e}rrez-S{\'a}nchez}, {Guy}, {Haigron}, {Hatzidimitriou}, {Haywood}, {Heiter}, {Helmi}, {Hobbs}, {Hofmann}, {Holl}, {Holland}, {Hunt}, {Hypki}, {Icardi}, {Irwin}, {Jevardat de Fombelle}, {Jofr{\'e}}, {Jonker}, {Jorissen}, {Julbe}, {Karampelas}, {Kochoska}, {Kohley}, {Kolenberg}, {Kontizas}, {Koposov}, {Kordopatis}, {Koubsky}, {Krone-Martins}, {Kudryashova}, {Kull}, {Bachchan}, {Lacoste-Seris}, {Lanza}, {Lavigne}, {Le Poncin-Lafitte}, {Lebreton}, {Lebzelter}, {Leccia}, {Leclerc}, {Lecoeur-Taibi}, {Lemaitre}, {Lenhardt}, {Leroux}, {Liao}, {Licata}, {Lindstr{\o}m}, {Lister}, {Livanou}, {Lobel}, {L{\"o}ffler}, {L{\'o}pez}, {Lorenz}, {MacDonald}, {Magalh{\~a}es Fernandes}, {Managau}, {Mann}, {Mantelet}, {Marchal}, {Marchant}, {Marconi}, {Marinoni}, {Marrese}, {Marschalk{\'o}}, {Marshall},
  {Mart{\'\i}n-Fleitas}, {Martino}, {Mary}, {Matijevi{\v{c}}}, {Mazeh}, {McMillan}, {Messina}, {Michalik}, {Millar}, {Miranda}, {Molina}, {Molinaro}, {Molinaro}, {Moln{\'a}r}, {Moniez}, {Montegriffo}, {Mor}, {Mora}, {Morbidelli}, {Morel}, {Morgenthaler}, {Morris}, {Mulone}, {Muraveva}, {Musella}, {Narbonne}, {Nelemans}, {Nicastro}, {Noval}, {Ord{\'e}novic}, {Ordieres-Mer{\'e}}, {Osborne}, {Pagani}, {Pagano}, {Pailler}, {Palacin}, {Palaversa}, {Parsons}, {Pecoraro}, {Pedrosa}, {Pentik{\"a}inen}, {Pichon}, {Piersimoni}, {Pineau}, {Plachy}, {Plum}, {Poujoulet}, {Pr{\v{s}}a}, {Pulone}, {Ragaini}, {Rago}, {Rambaux}, {Ramos-Lerate}, {Ranalli}, {Rauw}, {Read}, {Regibo}, {Reyl{\'e}}, {Ribeiro}, {Rimoldini}, {Ripepi}, {Riva}, {Rixon}, {Roelens}, {Romero-G{\'o}mez}, {Rowell}, {Royer}, {Ruiz-Dern}, {Sadowski}, {Sagrist{\`a} Sell{\'e}s}, {Sahlmann}, {Salgado}, {Salguero}, {Sarasso}, {Savietto}, {Schultheis}, {Sciacca}, {Segol}, {Segovia}, {Segransan}, {Shih}, {Smareglia}, {Smart}, {Solano}, {Solitro}, {Sordo}, {Soria
  Nieto}, {Souchay}, {Spagna}, {Spoto}, {Stampa}, {Steele}, {Steidelm{\"u}ller}, {Stephenson}, {Stoev}, {Suess}, {S{\"u}veges}, {Surdej}, {Szabados}, {Szegedi-Elek}, {Tapiador}, {Taris}, {Tauran}, {Taylor}, {Teixeira}, {Terrett}, {Tingley}, {Trager}, {Turon}, {Ulla}, {Utrilla}, {Valentini}, {van Elteren}, {Van Hemelryck}, {van Leeuwen}, {Varadi}, {Vecchiato}, {Veljanoski}, {Via}, {Vicente}, {Vogt}, {Voss}, {Votruba}, {Voutsinas}, {Walmsley}, {Weiler}, {Weingrill}, {Wevers}, {Wyrzykowski}, {Yoldas}, {{\v{Z}}erjal}, {Zucker}, {Zurbach}, {Zwitter}, {Alecu}, {Allen}, {Allende Prieto}, {Amorim}, {Anglada-Escud{\'e}}, {Arsenijevic}, {Azaz}, {Balm}, {Beck}, {Bernstein}, {Bigot}, {Bijaoui}, {Blasco}, {Bonfigli}, {Bono}, {Boudreault}, {Bressan}, {Brown}, {Brunet}, {Bunclark}, {Buonanno}, {Butkevich}, {Carret}, {Carrion}, {Chemin}, {Ch{\'e}reau}, {Corcione}, {Darmigny}, {de Boer}, {de Teodoro}, {de Zeeuw}, {Delle Luche}, {Domingues}, {Dubath}, {Fodor}, {Fr{\'e}zouls}, {Fries}, {Fustes}, {Fyfe}, {Gallardo}, {Gallegos},
  {Gardiol}, {Gebran}, {Gomboc}, {G{\'o}mez}, {Grux}, {Gueguen}, {Heyrovsky}, {Hoar}, {Iannicola}, {Isasi Parache}, {Janotto}, {Joliet}, {Jonckheere}, {Keil}, {Kim}, {Klagyivik}, {Klar}, {Knude}, {Kochukhov}, {Kolka}, {Kos}, {Kutka}, {Lainey}, {LeBouquin}, {Liu}, {Loreggia}, {Makarov}, {Marseille}, {Martayan}, {Martinez-Rubi}, {Massart}, {Meynadier}, {Mignot}, {Munari}, {Nguyen}, {Nordlander}, {Ocvirk}, {O'Flaherty}, {Olias Sanz}, {Ortiz}, {Osorio}, {Oszkiewicz}, {Ouzounis}, {Palmer}, {Park}, {Pasquato}, {Peltzer}, {Peralta}, {P{\'e}turaud}, {Pieniluoma}, {Pigozzi}, {Poels}, {Prat}, {Prod'homme}, {Raison}, {Rebordao}, {Risquez}, {Rocca-Volmerange}, {Rosen}, {Ruiz-Fuertes}, {Russo}, {Sembay}, {Serraller Vizcaino}, {Short}, {Siebert}, {Silva}, {Sinachopoulos}, {Slezak}, {Soffel}, {Sosnowska}, {Strai{\v{z}}ys}, {ter Linden}, {Terrell}, {Theil}, {Tiede}, {Troisi}, {Tsalmantza}, {Tur}, {Vaccari}, {Vachier}, {Valles}, {Van Hamme}, {Veltz}, {Virtanen}, {Wallut}, {Wichmann}, {Wilkinson}, {Ziaeepour}, \&
  {Zschocke}}]{2016A&A...595A...2G}
{Gaia Collaboration}, {Brown}, A.~G.~A., {Vallenari}, A., {et~al.} 2016, \aap, 595, A2

\bibitem[{{Gaia Collaboration} {et~al.}(2023){Gaia Collaboration}, {Vallenari}, {Brown}, {Prusti}, {de Bruijne}, {Arenou}, {Babusiaux}, {Biermann}, {Creevey}, {Ducourant}, {Evans}, {Eyer}, {Guerra}, {Hutton}, {Jordi}, {Klioner}, {Lammers}, {Lindegren}, {Luri}, {Mignard}, {Panem}, {Pourbaix}, {Randich}, {Sartoretti}, {Soubiran}, {Tanga}, {Walton}, {Bailer-Jones}, {Bastian}, {Drimmel}, {Jansen}, {Katz}, {Lattanzi}, {van Leeuwen}, {Bakker}, {Cacciari}, {Casta{\~n}eda}, {De Angeli}, {Fabricius}, {Fouesneau}, {Fr{\'e}mat}, {Galluccio}, {Guerrier}, {Heiter}, {Masana}, {Messineo}, {Mowlavi}, {Nicolas}, {Nienartowicz}, {Pailler}, {Panuzzo}, {Riclet}, {Roux}, {Seabroke}, {Sordo}, {Th{\'e}venin}, {Gracia-Abril}, {Portell}, {Teyssier}, {Altmann}, {Andrae}, {Audard}, {Bellas-Velidis}, {Benson}, {Berthier}, {Blomme}, {Burgess}, {Busonero}, {Busso}, {C{\'a}novas}, {Carry}, {Cellino}, {Cheek}, {Clementini}, {Damerdji}, {Davidson}, {de Teodoro}, {Nu{\~n}ez Campos}, {Delchambre}, {Dell'Oro}, {Esquej},
  {Fern{\'a}ndez-Hern{\'a}ndez}, {Fraile}, {Garabato}, {Garc{\'\i}a-Lario}, {Gosset}, {Haigron}, {Halbwachs}, {Hambly}, {Harrison}, {Hern{\'a}ndez}, {Hestroffer}, {Hodgkin}, {Holl}, {Jan{\ss}en}, {Jevardat de Fombelle}, {Jordan}, {Krone-Martins}, {Lanzafame}, {L{\"o}ffler}, {Marchal}, {Marrese}, {Moitinho}, {Muinonen}, {Osborne}, {Pancino}, {Pauwels}, {Recio-Blanco}, {Reyl{\'e}}, {Riello}, {Rimoldini}, {Roegiers}, {Rybizki}, {Sarro}, {Siopis}, {Smith}, {Sozzetti}, {Utrilla}, {van Leeuwen}, {Abbas}, {{\'A}brah{\'a}m}, {Abreu Aramburu}, {Aerts}, {Aguado}, {Ajaj}, {Aldea-Montero}, {Altavilla}, {{\'A}lvarez}, {Alves}, {Anders}, {Anderson}, {Anglada Varela}, {Antoja}, {Baines}, {Baker}, {Balaguer-N{\'u}{\~n}ez}, {Balbinot}, {Balog}, {Barache}, {Barbato}, {Barros}, {Barstow}, {Bartolom{\'e}}, {Bassilana}, {Bauchet}, {Becciani}, {Bellazzini}, {Berihuete}, {Bernet}, {Bertone}, {Bianchi}, {Binnenfeld}, {Blanco-Cuaresma}, {Blazere}, {Boch}, {Bombrun}, {Bossini}, {Bouquillon}, {Bragaglia}, {Bramante}, {Breedt},
  {Bressan}, {Brouillet}, {Brugaletta}, {Bucciarelli}, {Burlacu}, {Butkevich}, {Buzzi}, {Caffau}, {Cancelliere}, {Cantat-Gaudin}, {Carballo}, {Carlucci}, {Carnerero}, {Carrasco}, {Casamiquela}, {Castellani}, {Castro-Ginard}, {Chaoul}, {Charlot}, {Chemin}, {Chiaramida}, {Chiavassa}, {Chornay}, {Comoretto}, {Contursi}, {Cooper}, {Cornez}, {Cowell}, {Crifo}, {Cropper}, {Crosta}, {Crowley}, {Dafonte}, {Dapergolas}, {David}, {David}, {de Laverny}, {De Luise}, {De March}, {De Ridder}, {de Souza}, {de Torres}, {del Peloso}, {del Pozo}, {Delbo}, {Delgado}, {Delisle}, {Demouchy}, {Dharmawardena}, {Di Matteo}, {Diakite}, {Diener}, {Distefano}, {Dolding}, {Edvardsson}, {Enke}, {Fabre}, {Fabrizio}, {Faigler}, {Fedorets}, {Fernique}, {Fienga}, {Figueras}, {Fournier}, {Fouron}, {Fragkoudi}, {Gai}, {Garcia-Gutierrez}, {Garcia-Reinaldos}, {Garc{\'\i}a-Torres}, {Garofalo}, {Gavel}, {Gavras}, {Gerlach}, {Geyer}, {Giacobbe}, {Gilmore}, {Girona}, {Giuffrida}, {Gomel}, {Gomez}, {Gonz{\'a}lez-N{\'u}{\~n}ez},
  {Gonz{\'a}lez-Santamar{\'\i}a}, {Gonz{\'a}lez-Vidal}, {Granvik}, {Guillout}, {Guiraud}, {Guti{\'e}rrez-S{\'a}nchez}, {Guy}, {Hatzidimitriou}, {Hauser}, {Haywood}, {Helmer}, {Helmi}, {Sarmiento}, {Hidalgo}, {Hilger}, {H{\l}adczuk}, {Hobbs}, {Holland}, {Huckle}, {Jardine}, {Jasniewicz}, {Jean-Antoine Piccolo}, {Jim{\'e}nez-Arranz}, {Jorissen}, {Juaristi Campillo}, {Julbe}, {Karbevska}, {Kervella}, {Khanna}, {Kontizas}, {Kordopatis}, {Korn}, {K{\'o}sp{\'a}l}, {Kostrzewa-Rutkowska}, {Kruszy{\'n}ska}, {Kun}, {Laizeau}, {Lambert}, {Lanza}, {Lasne}, {Le Campion}, {Lebreton}, {Lebzelter}, {Leccia}, {Leclerc}, {Lecoeur-Taibi}, {Liao}, {Licata}, {Lindstr{\o}m}, {Lister}, {Livanou}, {Lobel}, {Lorca}, {Loup}, {Madrero Pardo}, {Magdaleno Romeo}, {Managau}, {Mann}, {Manteiga}, {Marchant}, {Marconi}, {Marcos}, {Marcos Santos}, {Mar{\'\i}n Pina}, {Marinoni}, {Marocco}, {Marshall}, {Martin Polo}, {Mart{\'\i}n-Fleitas}, {Marton}, {Mary}, {Masip}, {Massari}, {Mastrobuono-Battisti}, {Mazeh}, {McMillan}, {Messina}, {Michalik},
  {Millar}, {Mints}, {Molina}, {Molinaro}, {Moln{\'a}r}, {Monari}, {Mongui{\'o}}, {Montegriffo}, {Montero}, {Mor}, {Mora}, {Morbidelli}, {Morel}, {Morris}, {Muraveva}, {Murphy}, {Musella}, {Nagy}, {Noval}, {Oca{\~n}a}, {Ogden}, {Ordenovic}, {Osinde}, {Pagani}, {Pagano}, {Palaversa}, {Palicio}, {Pallas-Quintela}, {Panahi}, {Payne-Wardenaar}, {Pe{\~n}alosa Esteller}, {Penttil{\"a}}, {Pichon}, {Piersimoni}, {Pineau}, {Plachy}, {Plum}, {Poggio}, {Pr{\v{s}}a}, {Pulone}, {Racero}, {Ragaini}, {Rainer}, {Raiteri}, {Rambaux}, {Ramos}, {Ramos-Lerate}, {Re Fiorentin}, {Regibo}, {Richards}, {Rios Diaz}, {Ripepi}, {Riva}, {Rix}, {Rixon}, {Robichon}, {Robin}, {Robin}, {Roelens}, {Rogues}, {Rohrbasser}, {Romero-G{\'o}mez}, {Rowell}, {Royer}, {Ruz Mieres}, {Rybicki}, {Sadowski}, {S{\'a}ez N{\'u}{\~n}ez}, {Sagrist{\`a} Sell{\'e}s}, {Sahlmann}, {Salguero}, {Samaras}, {Sanchez Gimenez}, {Sanna}, {Santove{\~n}a}, {Sarasso}, {Schultheis}, {Sciacca}, {Segol}, {Segovia}, {S{\'e}gransan}, {Semeux}, {Shahaf}, {Siddiqui}, {Siebert},
  {Siltala}, {Silvelo}, {Slezak}, {Slezak}, {Smart}, {Snaith}, {Solano}, {Solitro}, {Souami}, {Souchay}, {Spagna}, {Spina}, {Spoto}, {Steele}, {Steidelm{\"u}ller}, {Stephenson}, {S{\"u}veges}, {Surdej}, {Szabados}, {Szegedi-Elek}, {Taris}, {Taylor}, {Teixeira}, {Tolomei}, {Tonello}, {Torra}, {Torra}, {Torralba Elipe}, {Trabucchi}, {Tsounis}, {Turon}, {Ulla}, {Unger}, {Vaillant}, {van Dillen}, {van Reeven}, {Vanel}, {Vecchiato}, {Viala}, {Vicente}, {Voutsinas}, {Weiler}, {Wevers}, {Wyrzykowski}, {Yoldas}, {Yvard}, {Zhao}, {Zorec}, {Zucker}, \& {Zwitter}}]{2023A&A...674A...1G}
{Gaia Collaboration}, {Vallenari}, A., {Brown}, A.~G.~A., {et~al.} 2023, \aap, 674, A1

\bibitem[{{Gerasimov} {et~al.}(2022){Gerasimov}, {Burgasser}, {Homeier}, {Bedin}, {Rees}, {Scalco}, {Anderson}, \& {Salaris}}]{2022ApJ...930...24G}
{Gerasimov}, R., {Burgasser}, A.~J., {Homeier}, D., {et~al.} 2022, \apj, 930, 24

\bibitem[{{Gratton} {et~al.}(2019){Gratton}, {Bragaglia}, {Carretta}, {D'Orazi}, {Lucatello}, \& {Sollima}}]{2019A&ARv..27....8G}
{Gratton}, R., {Bragaglia}, A., {Carretta}, E., {et~al.} 2019, A\&AR, 27, 8

\bibitem[{{Harris}(1996)}]{1996AJ....112.1487H}
{Harris}, W.~E. 1996, \aj, 112, 1487

\bibitem[{{Harris}(2010)}]{2010arXiv1012.3224H}
{Harris}, W.~E. 2010, arXiv e-prints, arXiv:1012.3224

\bibitem[{{Ibata} {et~al.}(2019){Ibata}, {Bellazzini}, {Malhan}, {Martin}, \& {Bianchini}}]{2019NatAs...3..667I}
{Ibata}, R.~A., {Bellazzini}, M., {Malhan}, K., {Martin}, N., \& {Bianchini}, P. 2019, Nature Astronomy, 3, 667

\bibitem[{{Jurcsik}(1998)}]{1998ApJ...506L.113J}
{Jurcsik}, J. 1998, \apjl, 506, L113

\bibitem[{{King} {et~al.}(2012){King}, {Bedin}, {Cassisi}, {Milone}, {Bellini}, {Piotto}, {Anderson}, {Pietrinferni}, \& {Cordier}}]{2012AJ....144....5K}
{King}, I.~R., {Bedin}, L.~R., {Cassisi}, S., {et~al.} 2012, AJ, 144, 5

\bibitem[{{Lacchin} {et~al.}(2021){Lacchin}, {Calura}, \& {Vesperini}}]{2021MNRAS.506.5951L}
{Lacchin}, E., {Calura}, F., \& {Vesperini}, E. 2021, MNRAS, 506, 5951

\bibitem[{{Lacchin} {et~al.}(2022){Lacchin}, {Calura}, {Vesperini}, \& {Mastrobuono-Battisti}}]{2022MNRAS.517.1171L}
{Lacchin}, E., {Calura}, F., {Vesperini}, E., \& {Mastrobuono-Battisti}, A. 2022, \mnras, 517, 1171

\bibitem[{{Latour} {et~al.}(2021){Latour}, {Calamida}, {Husser}, {Kamann}, {Dreizler}, \& {Brinchmann}}]{2021A&A...653L...8L}
{Latour}, M., {Calamida}, A., {Husser}, T.~O., {et~al.} 2021, \aap, 653, L8

\bibitem[{{Libralato} {et~al.}(2018){Libralato}, {Bellini}, {Bedin}, {Moreno D.}, {Fern{\'a}ndez-Trincado}, {Pichardo}, {van der Marel}, {Anderson}, {Apai}, {Burgasser}, {Fabiola Marino}, {Milone}, {Rees}, \& {Watkins}}]{2018ApJ...854...45L}
{Libralato}, M., {Bellini}, A., {Bedin}, L.~R., {et~al.} 2018, \apj, 854, 45

\bibitem[{{Libralato} {et~al.}(2022){Libralato}, {Bellini}, {Vesperini}, {Piotto}, {Milone}, {van der Marel}, {Anderson}, {Aparicio}, {Barbuy}, {Bedin}, {Borsato}, {Cassisi}, {Dalessandro}, {Ferraro}, {King}, {Lanzoni}, {Nardiello}, {Ortolani}, {Sarajedini}, \& {Sohn}}]{2022ApJ...934..150L}
{Libralato}, M., {Bellini}, A., {Vesperini}, E., {et~al.} 2022, \apj, 934, 150

\bibitem[{{Marks} {et~al.}(2022){Marks}, {Kroupa}, \& {Dabringhausen}}]{2022A&A...659A..96M}
{Marks}, M., {Kroupa}, P., \& {Dabringhausen}, J. 2022, \aap, 659, A96

\bibitem[{{Milone} {et~al.}(2017){Milone}, {Marino}, {Bedin}, {Anderson}, {Apai}, {Bellini}, {Bergeron}, {Burgasser}, {Dotter}, \& {Rees}}]{2017MNRAS.469..800M}
{Milone}, A.~P., {Marino}, A.~F., {Bedin}, L.~R., {et~al.} 2017, \mnras, 469, 800

\bibitem[{{Milone} {et~al.}(2015{\natexlab{a}}){Milone}, {Marino}, {Piotto}, {Bedin}, {Anderson}, {Renzini}, {King}, {Bellini}, {Brown}, {Cassisi}, {D'Antona}, {Jerjen}, {Nardiello}, {Salaris}, {Marel}, {Vesperini}, {Yong}, {Aparicio}, {Sarajedini}, \& {Zoccali}}]{2015MNRAS.447..927M}
{Milone}, A.~P., {Marino}, A.~F., {Piotto}, G., {et~al.} 2015{\natexlab{a}}, \mnras, 447, 927

\bibitem[{{Milone} {et~al.}(2015{\natexlab{b}}){Milone}, {Marino}, {Piotto}, {Renzini}, {Bedin}, {Anderson}, {Cassisi}, {D'Antona}, {Bellini}, {Jerjen}, {Pietrinferni}, \& {Ventura}}]{2015ApJ...808...51M}
{Milone}, A.~P., {Marino}, A.~F., {Piotto}, G., {et~al.} 2015{\natexlab{b}}, \apj, 808, 51

\bibitem[{{Milone} {et~al.}(2012){Milone}, {Piotto}, {Bedin}, {Aparicio}, {Anderson}, {Sarajedini}, {Marino}, {Moretti}, {Davies}, {Chaboyer}, {Dotter}, {Hempel}, {Mar{\'\i}n-Franch}, {Majewski}, {Paust}, {Reid}, {Rosenberg}, \& {Siegel}}]{2012A&A...540A..16M}
{Milone}, A.~P., {Piotto}, G., {Bedin}, L.~R., {et~al.} 2012, \aap, 540, A16

\bibitem[{{Nardiello} {et~al.}(2018){Nardiello}, {Libralato}, {Piotto}, {Anderson}, {Bellini}, {Aparicio}, {Bedin}, {Cassisi}, {Granata}, {King}, {Lucertini}, {Marino}, {Milone}, {Ortolani}, {Platais}, \& {van der Marel}}]{2018MNRAS.481.3382N}
{Nardiello}, D., {Libralato}, M., {Piotto}, G., {et~al.} 2018, \mnras, 481, 3382

\bibitem[{{Norris}(2004)}]{2004ApJ...612L..25N}
{Norris}, J.~E. 2004, ApJl, 612, L25

\bibitem[{{Norris} {et~al.}(1997){Norris}, {Freeman}, {Mayor}, \& {Seitzer}}]{1997ApJ...487L.187N}
{Norris}, J.~E., {Freeman}, K.~C., {Mayor}, M., \& {Seitzer}, P. 1997, \apjl, 487, L187

\bibitem[{{Pancino} {et~al.}(2000){Pancino}, {Ferraro}, {Bellazzini}, {Piotto}, \& {Zoccali}}]{2000ApJ...534L..83P}
{Pancino}, E., {Ferraro}, F.~R., {Bellazzini}, M., {Piotto}, G., \& {Zoccali}, M. 2000, \apjl, 534, L83

\bibitem[{{Renzini} {et~al.}(2015){Renzini}, {D'Antona}, {Cassisi}, {King}, {Milone}, {Ventura}, {Anderson}, {Bedin}, {Bellini}, {Brown}, {Piotto}, {van der Marel}, {Barbuy}, {Dalessandro}, {Hidalgo}, {Marino}, {Ortolani}, {Salaris}, \& {Sarajedini}}]{2015MNRAS.454.4197R}
{Renzini}, A., {D'Antona}, F., {Cassisi}, S., {et~al.} 2015, \mnras, 454, 4197

\bibitem[{{Sarajedini} {et~al.}(2007){Sarajedini}, {Bedin}, {Chaboyer}, {Dotter}, {Siegel}, {Anderson}, {Aparicio}, {King}, {Majewski}, {Mar{\'\i}n-Franch}, {Piotto}, {Reid}, \& {Rosenberg}}]{2007AJ....133.1658S}
{Sarajedini}, A., {Bedin}, L.~R., {Chaboyer}, B., {et~al.} 2007, \aj, 133, 1658

\bibitem[{{Scalco} {et~al.}(2021){Scalco}, {Bellini}, {Bedin}, {Anderson}, {Rosati}, {Libralato}, {Salaris}, {Vesperini}, {Nardiello}, {Apai}, {Burgasser}, \& {Gerasimov}}]{2021MNRAS.505.3549S}
{Scalco}, M., {Bellini}, A., {Bedin}, L.~R., {et~al.} 2021, \mnras, 505, 3549

\bibitem[{{Simioni} {et~al.}(2016){Simioni}, {Milone}, {Bedin}, {Aparicio}, {Piotto}, {Vesperini}, \& {Hong}}]{2016MNRAS.463..449S}
{Simioni}, M., {Milone}, A.~P., {Bedin}, L.~R., {et~al.} 2016, \mnras, 463, 449

\bibitem[{{Smith}(1987)}]{1987PASP...99...67S}
{Smith}, G.~H. 1987, PASP, 99, 67

\bibitem[{{Sollima} {et~al.}(2007){Sollima}, {Ferraro}, {Bellazzini}, {Origlia}, {Straniero}, \& {Pancino}}]{2007ApJ...654..915S}
{Sollima}, A., {Ferraro}, F.~R., {Bellazzini}, M., {et~al.} 2007, \apj, 654, 915

\bibitem[{{van de Ven} {et~al.}(2006){van de Ven}, {van den Bosch}, {Verolme}, \& {de Zeeuw}}]{2006A&A...445..513V}
{van de Ven}, G., {van den Bosch}, R.~C.~E., {Verolme}, E.~K., \& {de Zeeuw}, P.~T. 2006, \aap, 445, 513

\bibitem[{{Wragg}(2023)}]{2023tmib.confE..12W}
{Wragg}, F. 2023, in Two in a Million - The Interplay Between Binaries and Star Clusters, 12

\end{thebibliography}

\begin{appendix}

\renewcommand{\thetable}{\arabic{table}}
\setcounter{table}{1}

\begin{landscape}
\begin{table}
\centering
\begin{adjustbox}{width=\linewidth,totalheight=\textheight,keepaspectratio}
\begin{threeparttable}
    \caption{Distribution of stars among various populations in the analysed fields.}
    \label{Table2}
    \begin{tabular}{l c c c c c c c}
        \hline
            \hline
            \multicolumn{1}{l}{} & \multicolumn{1}{c}{} & \multicolumn{2}{c}{Central field\tnote{1}} & \multicolumn{2}{c}{Field F5} & \multicolumn{2}{c}{Field F4}\\
        \multicolumn{1}{l}{} & \multicolumn{1}{c}{} & \multicolumn{2}{c}{$r \lesssim$3.21\,arcmin ($r/r_{\rm h}\lesssim$0.64)} & \multicolumn{2}{c}{3.70$\lesssim r \lesssim$6.80\,arcmin (0.74$\lesssim r/r_{\rm h}\lesssim$1.36)} & \multicolumn{2}{c}{4.36$\lesssim r \lesssim$7.40\,arcmin (0.87$\lesssim r/r_{\rm h}\lesssim$1.48)}\\
            \hline 
            Main Group & Subgroup & N$_{\rm Stars}$ & Fraction & N$_{\rm Stars}$ & Fraction & N$_{\rm Stars}$ & Fraction\\
            \hline 
            Entire MS & & 39\,526 & 100\% & 8635 & 100\% & 7174 & 100\%\\
            \hline
            \textbf{MSa} & & 1394 & 3.53\% $\pm$ 0.10\% (3.93\% $\pm$ 0.11\%) & 194 & 2.25\% $\pm$ 0.16\% (3.11\% $\pm$ 0.23\%) & 187 & 2.61\% $\pm$ 0.19\% (3.28\% $\pm$ 0.24\%)\\
            & MSa1 & 1283 & 3.25\% $\pm$ 0.09\% (3.62\% $\pm$ 0.10\%) & 180 & 2.09\% $\pm$ 0.16\% (2.89\% $\pm$ 0.22\%) & 166 & 2.32\% $\pm$ 0.18\% (2.91\% $\pm$ 0.23\%)\\
            & MSa2 & 111 & 0.28\% $\pm$ 0.03\% (0.31\% $\pm$ 0.03\%) & 14 & 0.16\% $\pm$ 0.04\% (0.22\% $\pm$ 0.06\%) & 21 & 0.29\% $\pm$ 0.06\% (0.37\% $\pm$ 0.08\%)\\
            \hline
            \textbf{bMS} & & 12\,776 & 32.32\% $\pm$ 0.33\% (36.05\% $\pm$ 0.37\%) & 2017 & 23.36\% $\pm$ 0.58\% (32.37\% $\pm$ 0.83\%) & 1727 & 24.07\% $\pm$ 0.65\% (30.33\% $\pm$ 0.83\%)\\
		& bMS1 & 5141 & 13.01\% $\pm$ 0.19\% (14.51\% $\pm$ 0.22\%) & - & - & 669 & 9.32\% $\pm$ 0.38\% (11.75\% $\pm$ 0.48\%)\\
		& bMS2 & 3683 & 9.32\% $\pm$ 0.16\% (10.39\% $\pm$ 0.18\%) & - & - & 721 & 10.05\% $\pm$ 0.39\% (12.66\% $\pm$ 0.50\%)\\
		& bMS3 & 3952 & 10.00\% $\pm$ 0.17\% (11.15\% $\pm$ 0.19\%) & - & - & 337 & 4.70\% $\pm$ 0.26\% (5.92\% $\pm$ 0.33\%)\\	
		\hline
		\textbf{rMS} & & 13\,124 & 33.20\% $\pm$ 0.33\% (37.03\% $\pm$ 0.38\%) & 2589 & 29.98\% $\pm$ 0.67\% (41.55\% $\pm$ 0.97\%) & 2412 & 33.62\% $\pm$ 0.79\% (42.36\% $\pm$ 1.03\%)\\
		& rMS1 & 3739 & 9.46\% $\pm$ 0.16\% (10.55\% $\pm$ 0.18\%) & - & - & 1171 & 16.32\% $\pm$ 0.51\% (20.57\% $\pm$ 0.66\%)\\
		& rMS2 & 3838 & 9.71\% $\pm$ 0.16\% (10.83\% $\pm$ 0.18\%) & - & - & 602 & 8.39\% $\pm$ 0.36\% (10.57\% $\pm$ 0.45\%)\\
		& rMS3 & 5547 & 14.03\% $\pm$ 0.20\% (15.65\% $\pm$ 0.23\%) & - & - & 639 & 8.91\% $\pm$ 0.37\% (11.22\% $\pm$ 0.47\%)\\	
		\hline
		\textbf{MSd} & & 2016 & 5.10\% $\pm$ 0.12\% (5.69\% $\pm$ 0.13\%) & 404 & 4.68\% $\pm$ 0.24\% (6.49\% $\pm$ 0.33\%) & 416 & 5.80\%$ \pm$ 0.29\% (7.31\% $\pm$ 0.37\%)\\
		& MSd1 & 757 & 1.92\% $\pm$ 0.07\% (2.14\% $\pm$ 0.08\%) & - & - & 231 & 3.22\% $\pm$ 0.22\% (4.06\% $\pm$ 0.27\%)\\
		& MSd2 & 819 & 2.07\% $\pm$ 0.07\% (2.31\% $\pm$ 0.08\%) & - & - & 101 & 1.41\% $\pm$ 0.14\% (1.77\% $\pm$ 0.18\%)\\
		& MSd3 & 440 & 1.11\% $\pm$ 0.05\% (1.24\% $\pm$ 0.06\%) & - & - & 84 & 1.17\% $\pm$ 0.13\% (1.48\% $\pm$ 0.16\%)\\	
		\hline
		\textbf{MSe} & & 6129 & 15.51\% $\pm$ 0.21\% (17.30\% $\pm$ 0.24\%) & 1027 & 11.89\% $\pm$ 0.39\% (16.48\% $\pm$ 0.56\%) & 952 & 13.27\% $\pm$ 0.46\% (16.72\% $\pm$ 0.59\%)\\
		& MSe1 & 2555 & 6.46\% $\pm$ 0.13\% (7.21\% $\pm$ 0.15\%) & 519 & 6.01\% $\pm$ 0.27\% (8.33\% $\pm$ 0.38\%) & 501 & 6.98\% $\pm$ 0.32\% (8.80\% $\pm$ 0.41\%)\\
		& MSe2 & 2591 & 6.56\% $\pm$ 0.13\% (7.31\% $\pm$ 0.15\%) & 508 & 5.88\% $\pm$ 0.27\% (8.15\% $\pm$ 0.38\%) & 451 & 6.29\% $\pm$ 0.31\% (7.92\% $\pm$ 0.39\%)\\
		& MSe3 & 463 & 1.17\% $\pm$ 0.05\% (1.31\% $\pm$ 0.06\%) & - & - & - & -\\	
		& MSe4 & 520 & 1.32\% $\pm$ 0.06\% (1.47\% $\pm$ 0.06\%) & - & - & - & -\\
		\hline
            Unidentified+Binaries & & 4087 & 10.34\% $\pm$ 0.17\% & 2404 & 27.84\% $\pm$ 0.64\% & 1480 & 20.63\% $\pm$ 0.59\%\\
            \hline
            \addlinespace
		\hline
            \hline
            \multicolumn{1}{l}{} & \multicolumn{1}{c}{} & \multicolumn{2}{c}{Field F2} & \multicolumn{2}{c}{Field F3} & \multicolumn{2}{c}{Field F1}\\
		\multicolumn{1}{l}{} & \multicolumn{1}{c}{} & \multicolumn{2}{c}{9.83$\lesssim r/ \lesssim$12.94\,arcmin (1.97$\lesssim r/r_{\rm h}\lesssim$2.59)} & \multicolumn{2}{c}{10.98$\lesssim r/ \lesssim$14.21\,arcmin (2.20$\lesssim r/r_{\rm h}\lesssim$2.84)} & \multicolumn{2}{c}{14.48$\lesssim r \lesssim$18.28\,arcmin (2.90$\lesssim r/r_{\rm h}\lesssim$3.66)}\\      
            \hline 
            Main Group & Subgroup & N$_{\rm Stars}$ & Fraction & N$_{\rm Stars}$ & Fraction & N$_{\rm Stars}$ & Fraction\\
            \hline 
            Entire MS & & 1227 & 100\% & 1333 & 100\% & 412 & 100\%\\
		\hline
		\textbf{MSa} & & 17 & 1.39\% $\pm$ 0.34\% (1.73\% $\pm$ 0.42\%) & 28 & 2.10\% $\pm$ 0.40\% (2.51\% $\pm$ 0.48\%)& - & -\\
		& MSa1 &  - & - & - & - & - & -\\
		& MSa2 &  - & - & - & - & - & -\\
		\hline
		\textbf{bMS} & & 204 & 16.63\% $\pm$ 1.26\% (20.77\% $\pm$ 1.60\%) & 203 & 15.23\% $\pm$ 1.15\% (18.16\% $\pm$ 1.39\%) & 59 & 14.32\% $\pm$ 1.99\% (17.99\% $\pm$ 2.54\%)\\
		& bMS1 & 79 & 6.44\% $\pm$ 0.75\% (8.04\% $\pm$ 0.94\%) & - & - & 20 & 4.86\% $\pm$ 1.11\% (6.10\% $\pm$ 1.40\%)\\
		& bMS2 & 97 & 7.91\% $\pm$ 0.83\% (9.88\% $\pm$ 1.05\%) & - & - & 31 & 7.52\% $\pm$ 1.40\% (9.45\% $\pm$ 1.78\%)\\
		& bMS3 & 28 & 2.28\% $\pm$ 0.52\% (2.85\% $\pm$ 0.55\%) & - & - & 8 & 1.94\% $\pm$ 0.69\% (2.44\% $\pm$ 0.87\%)\\	
		\hline
		\textbf{rMS} & & 560 & 45.64\% $\pm$ 2.33\% (57.03\% $\pm$ 3.02\%) & 615 & 46.14\% $\pm$ 2.25\% (55.01\% $\pm$ 2.76\%) & 193 & 46.84\% $\pm$ 4.09\% (58.84\% $\pm$ 5.34\%)\\
		& rMS1 & - & - & - & - & - & -\\
		& rMS2 & - & - & - & - & - & -\\
		& rMS3 & - & - & - & - & - & -\\	
		\hline
		\textbf{MSd} & & 43 & 3.50\% $\pm$ 0.54\% (4.38\% $\pm$ 0.68\%) & 43 & 3.22\% $\pm$ 0.50\% (3.85\% $\pm$ 0.60\%) & 11 & 2.67\% $\pm$ 0.82\% (3.35\% $\pm$ 1.03\%)\\
		& MSd1 & - & - & - & - & - & -\\
		& MSd2 & - & - & - & - & - & -\\
		& MSd3 & - & - & - & - & - & -\\	
		\hline
		\textbf{MSe} & & 158 & 12.87\% $\pm$ 1.09\% (16.09\% $\pm$ 1.38\%) & 229 & 17.18\% $\pm$ 1.23\% (20.48\% $\pm$ 1.49\%) & 65 & 15.78\% $\pm$ 2.11\% (19.82\% $\pm$ 2.69\%)\\
		& MSe1 & 48 & 3.91\% $\pm$ 0.58\% (4.89\% $\pm$ 0.72\%) & 72 & 5.40\% $\pm$ 0.65\% (6.44\% $\pm$ 0.78\%) & - & -\\
		& MSe2 & 110 & 8.96\% $\pm$ 0.89\% (11.2\% $\pm$ 1.13\%) & 157 & 11.78\% $\pm$ 0.99\% (14.04\% $\pm$ 1.20\%) & - & -\\
		& MSe3 & - & - & - & - & - & -\\	
		& MSe4 & - & - & - & - & - & -\\
		\hline
            Unidentified+Binaries & & 245 & 19.97\% $\pm$ 1.4\% & 215 & 16.13\% $\pm$ 1.19\% & 84 & 20.39\% $\pm$ 2.44\%\\
            \hline
    \end{tabular}
    \tablefoot{The table includes data for stars in the magnitude range $20.16<m_{\rm F438W}<22.36$ from the two intermediate fields (F4 and F5) and the three outer fields (F1, F2, and F3) examined in this paper, along with the central field analyzed in \citet{2017ApJ...844..164B}. Ratios obtained solely from the counts of classified stars are reported within parentheses.}
     \begin{tablenotes}
      \small
      \item [1] values from \citet{2017ApJ...844..164B}.
    \end{tablenotes}
  \end{threeparttable}
  \end{adjustbox}
\end{table}
\end{landscape}

\renewcommand{\thefigure}{A.\arabic{figure}}
\setcounter{figure}{0}

\begin{figure*}
\centering
\includegraphics[width=\textwidth]{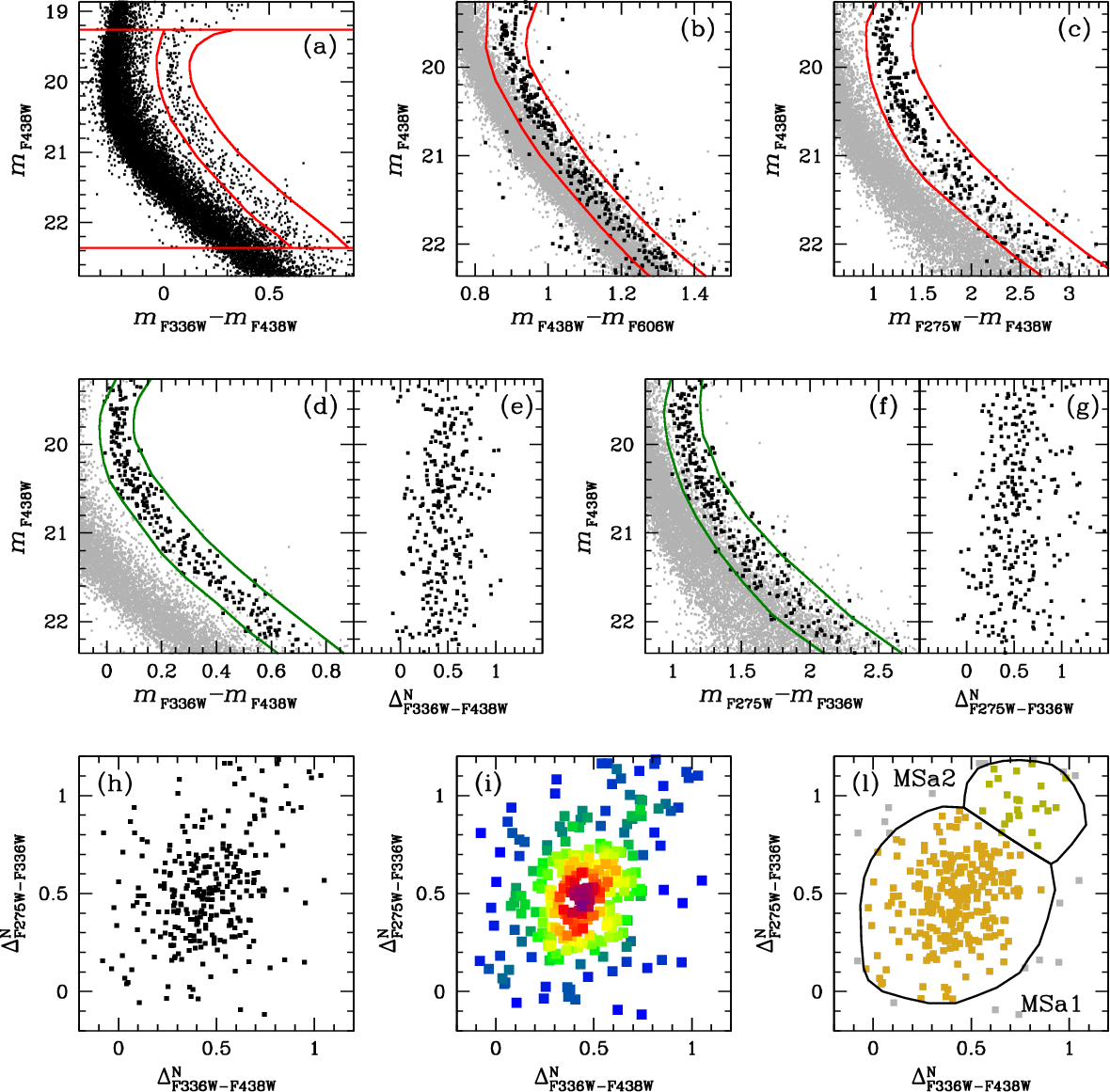}
\caption{Same as Fig.\,\ref{MSa} but for the F5 fields.}
\label{MSa_5}
\end{figure*}

\begin{figure*}
\centering
\includegraphics[width=\textwidth]{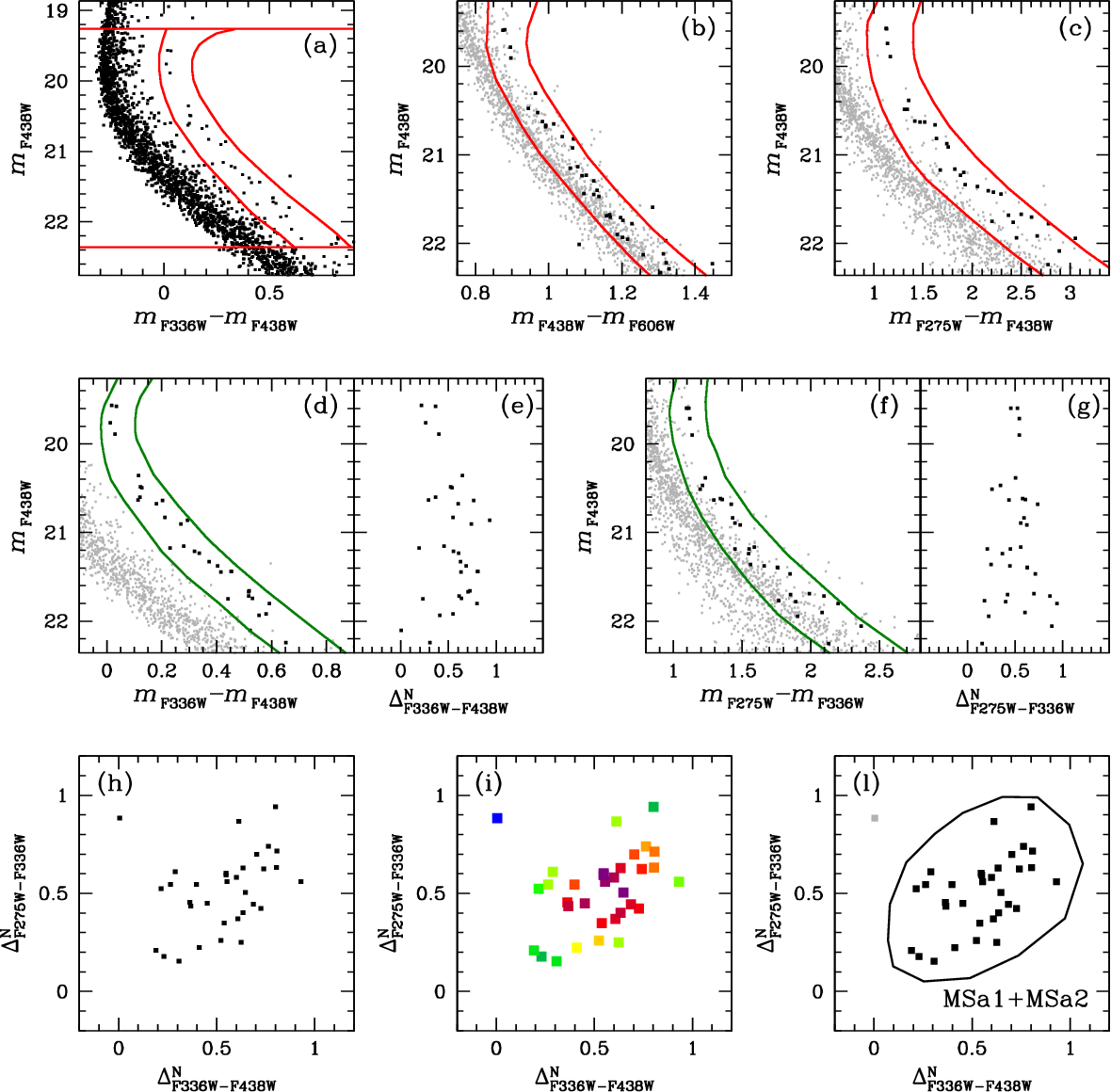}
\caption{Same as Fig.\,\ref{MSa} but for the F3 fields.}
\label{MSa_3}
\end{figure*}

\begin{figure*}
\centering
\includegraphics[width=\textwidth]{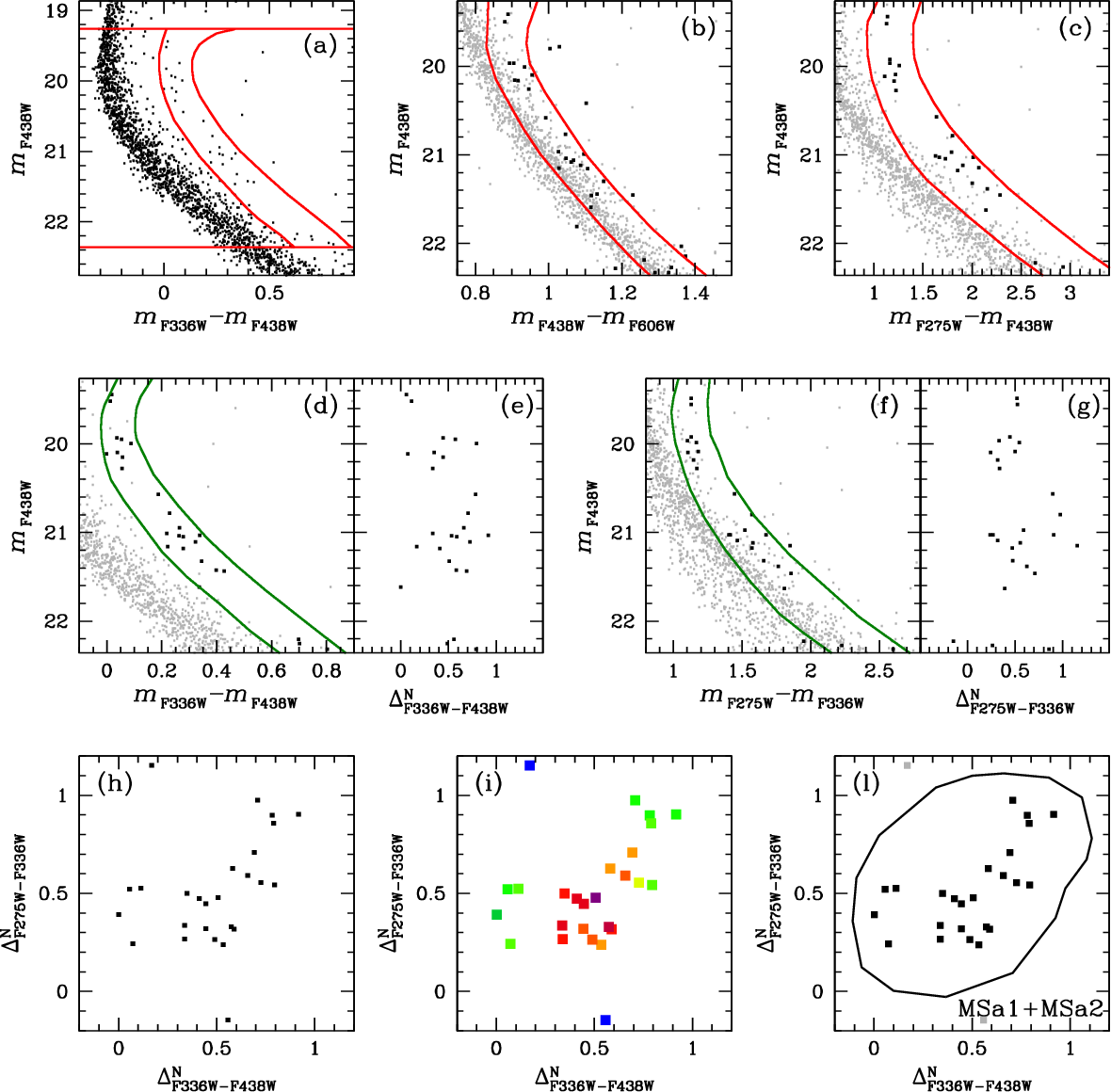}
\caption{Same as Fig.\,\ref{MSa} but for the F2 fields.}
\label{MSa_2}
\end{figure*}

\begin{figure*}
\centering
\includegraphics[width=\textwidth]{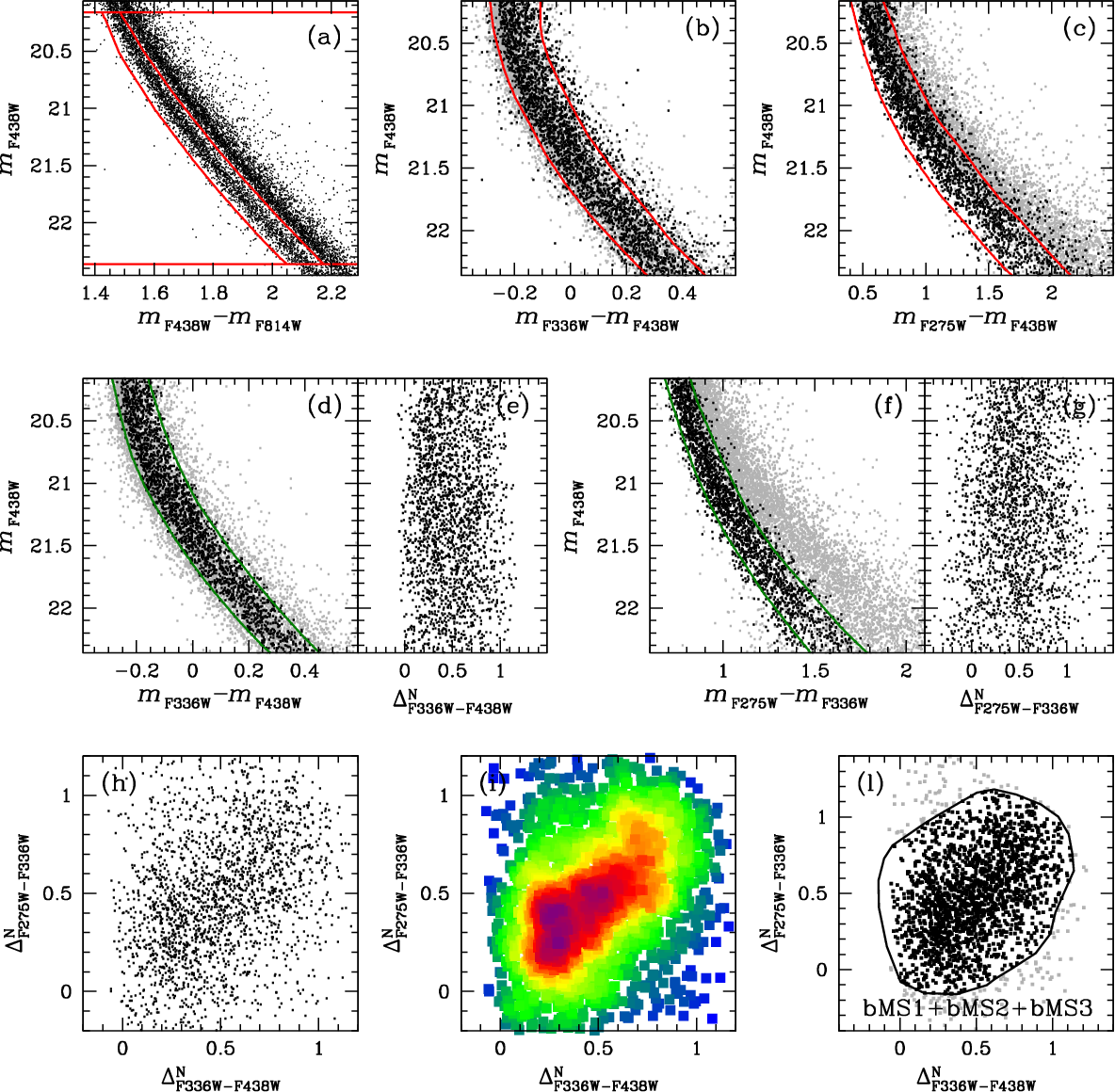}
\caption{Same as Fig.\,\ref{bMS} but for the F5 fields.}
\label{bMS_5}
\end{figure*}

\begin{figure*}
\centering
\includegraphics[width=\textwidth]{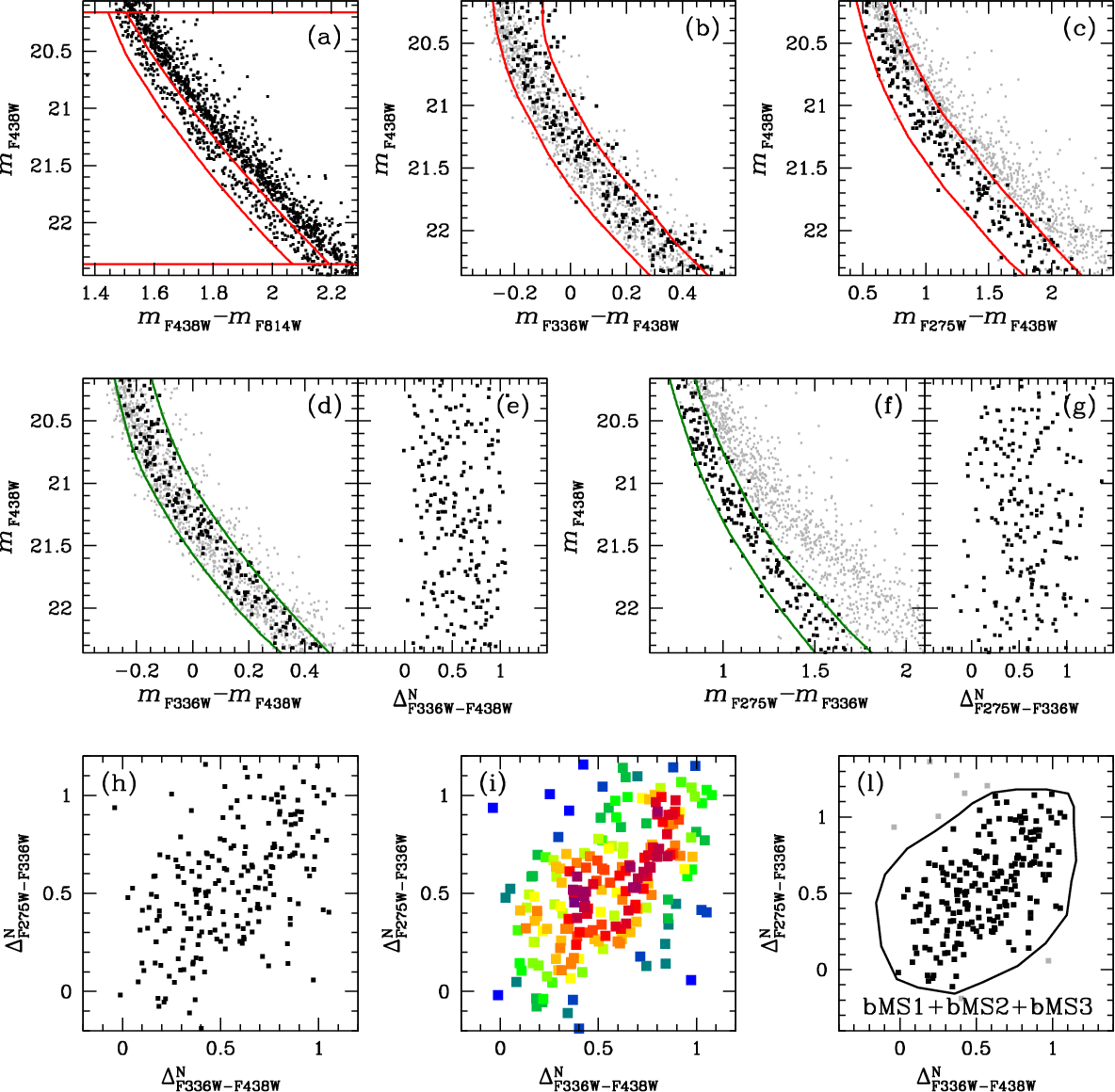}
\caption{Same as Fig.\,\ref{bMS} but for the F3 fields.}
\label{bMS_3}
\end{figure*}

\begin{figure*}
\centering
\includegraphics[width=\textwidth]{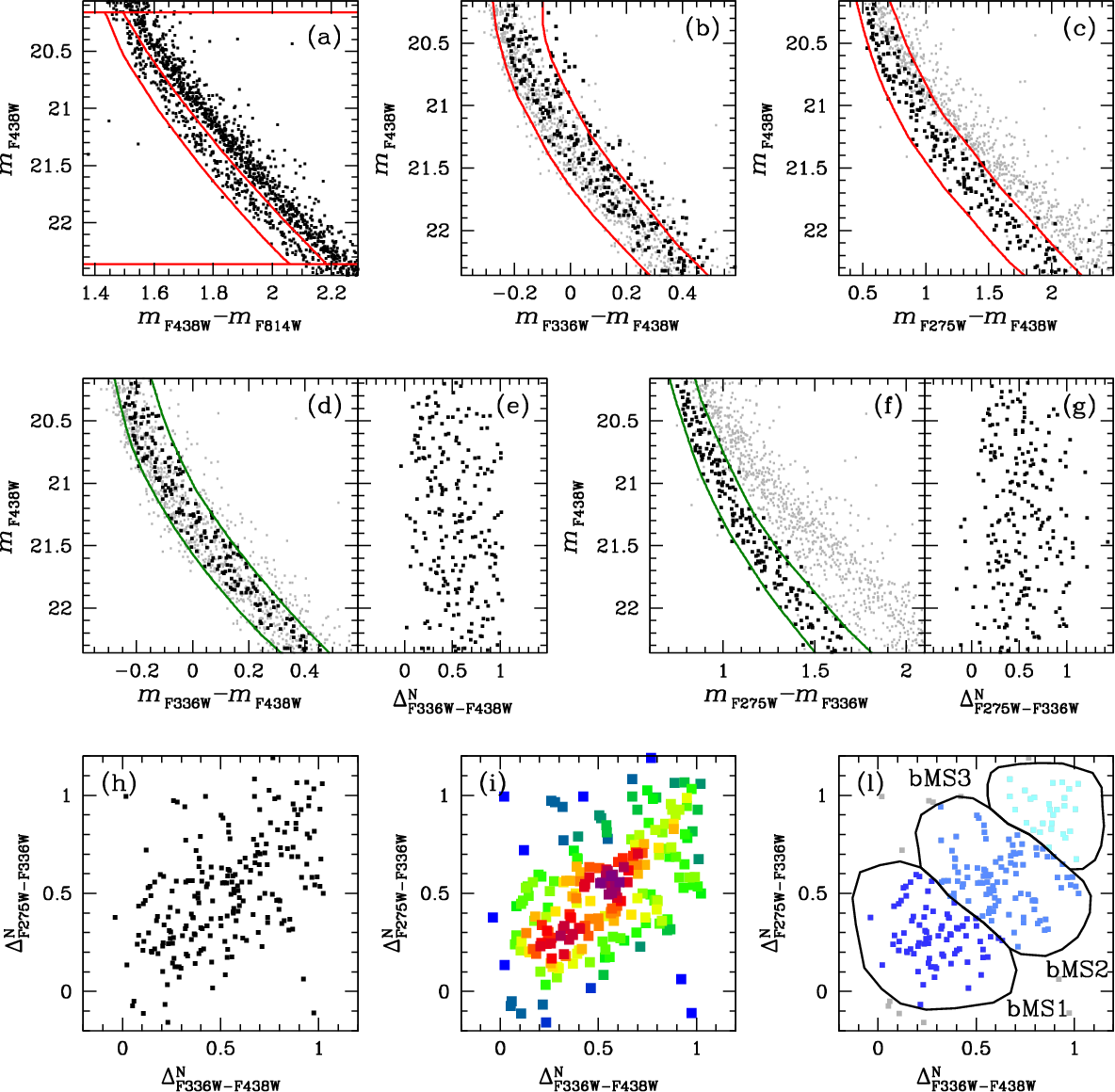}
\caption{Same as Fig.\,\ref{bMS} but for the F2 fields.}
\label{bMS_2}
\end{figure*}

\begin{figure*}
\centering
\includegraphics[width=\textwidth]{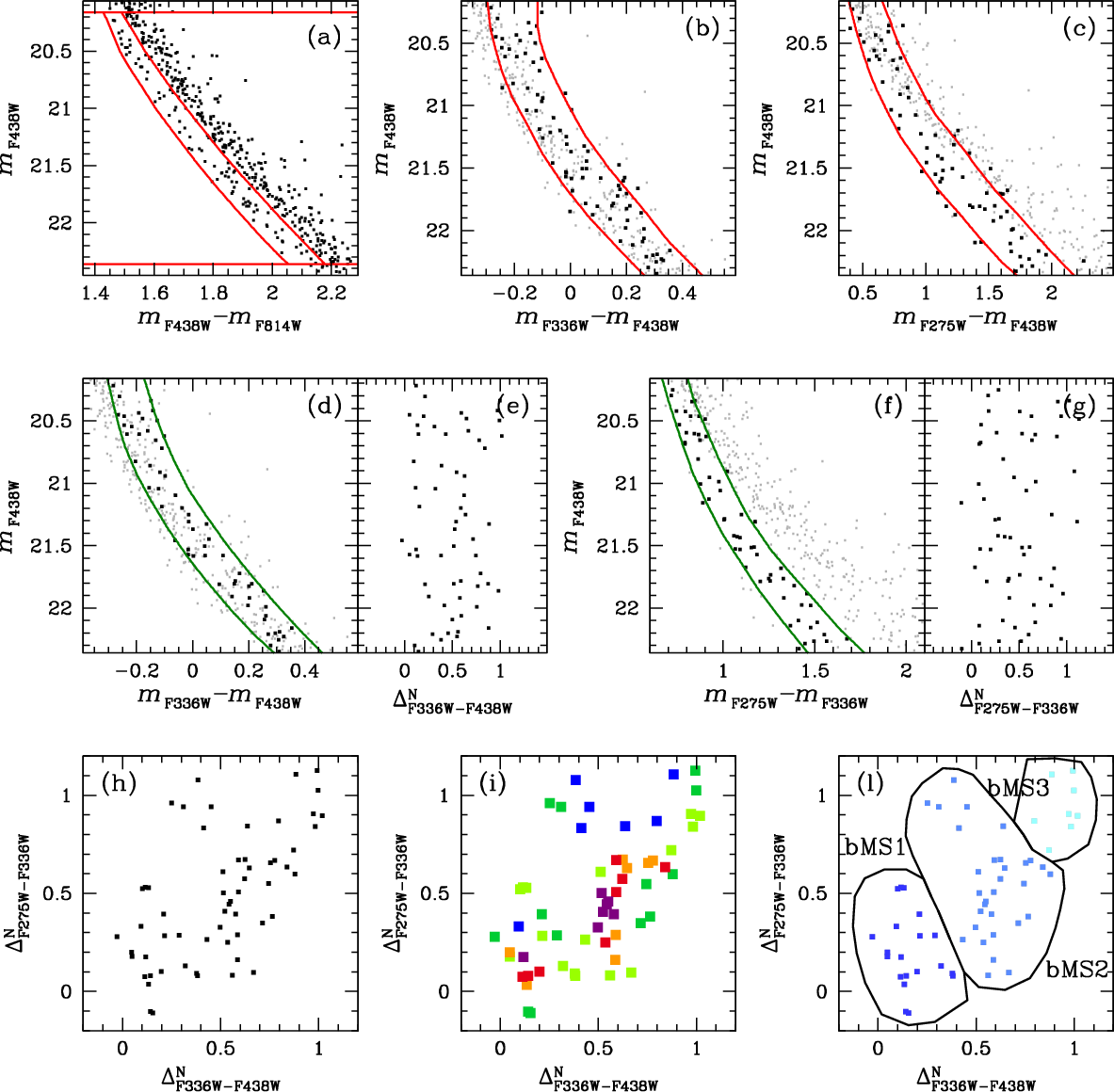}
\caption{Same as Fig.\,\ref{bMS} but for the F1 fields.}
\label{bMS_1}
\end{figure*}

\begin{figure*}
\centering
\includegraphics[width=\textwidth]{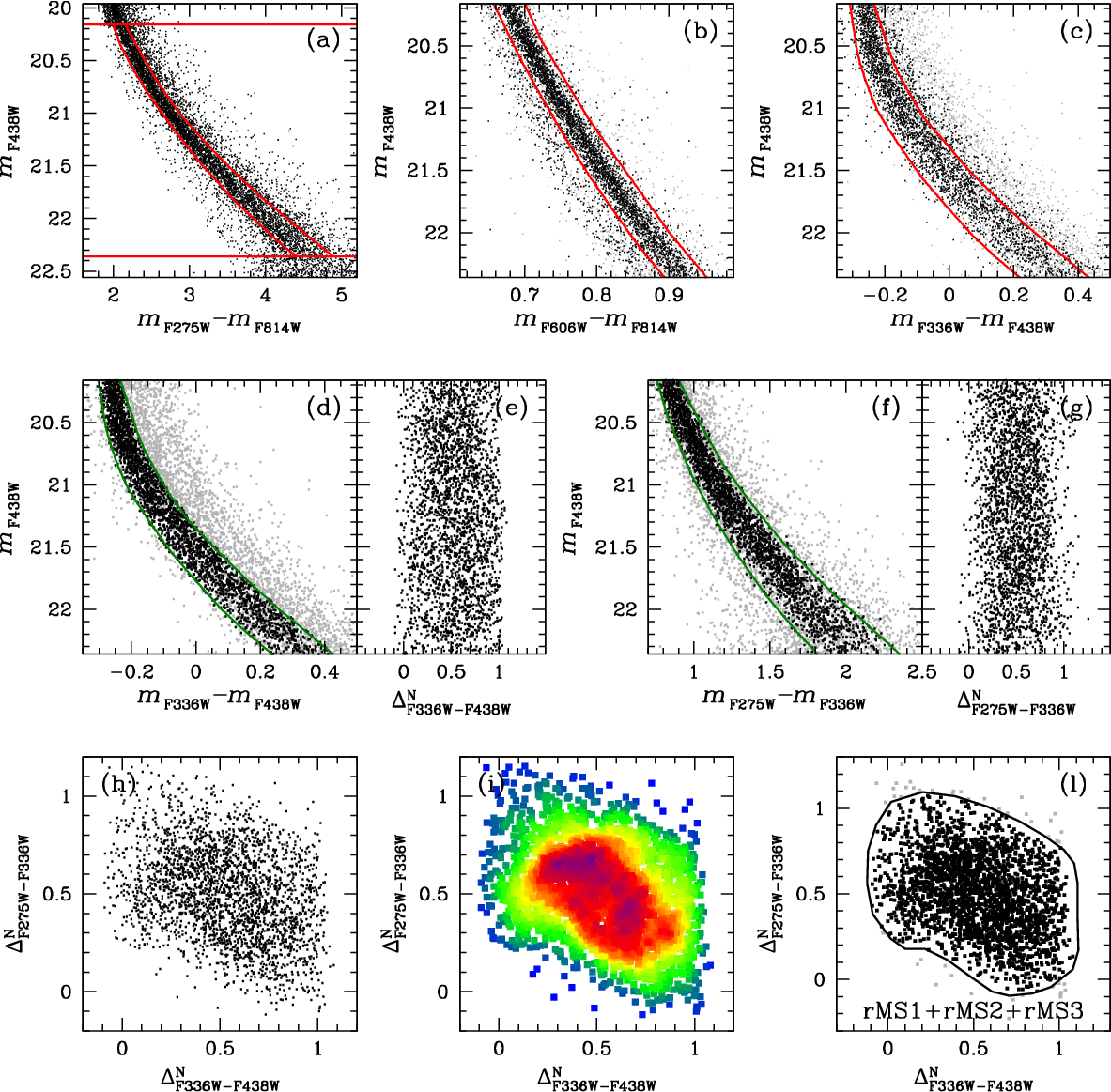}
\caption{Same as Fig.\,\ref{rMS} but for the F5 fields.}
\label{rMS_5}
\end{figure*}

\begin{figure*}
\centering
\includegraphics[width=\textwidth]{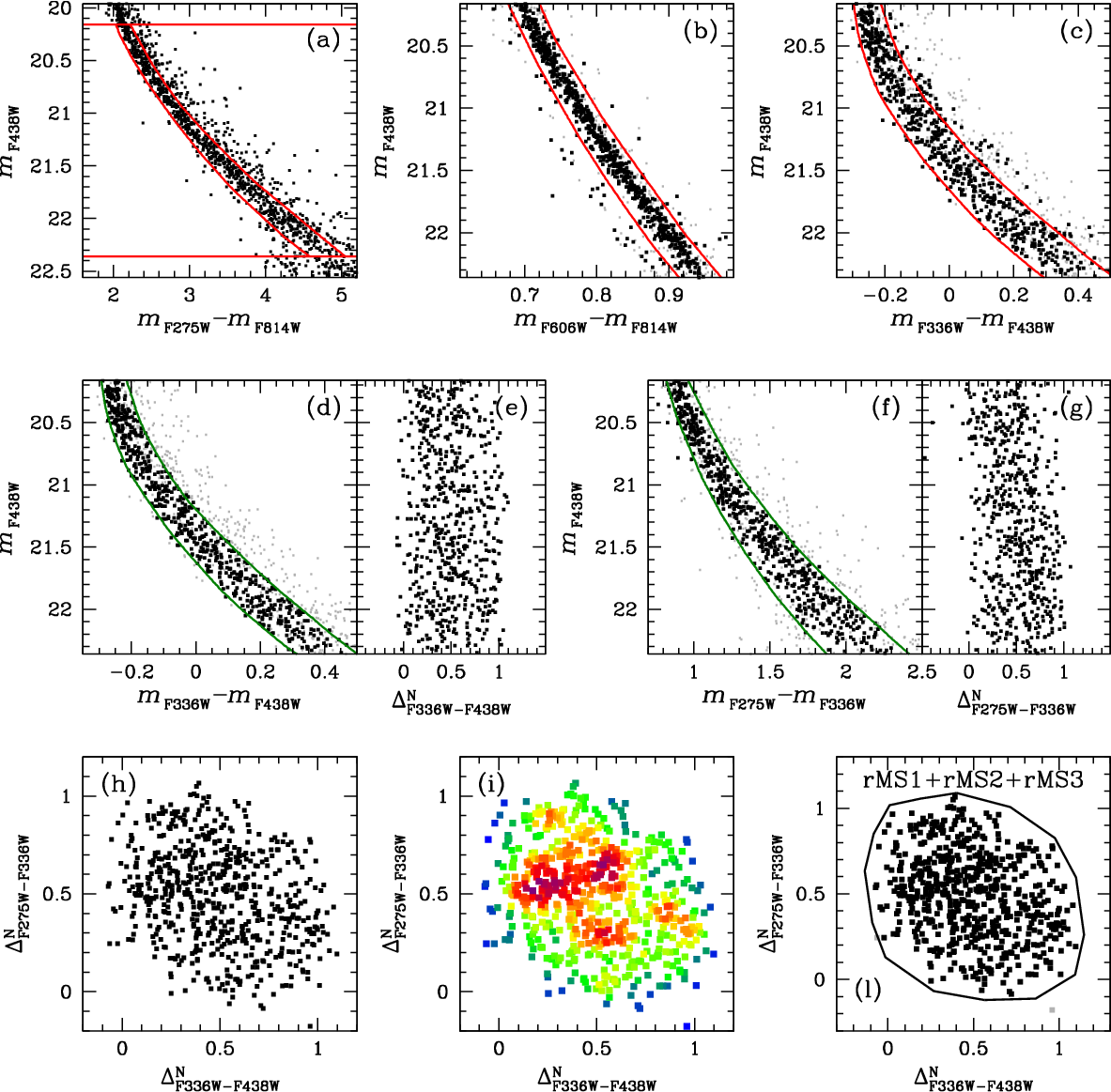}
\caption{Same as Fig.\,\ref{rMS} but for the F3 fields.}
\label{rMS_3}
\end{figure*}

\begin{figure*}
\centering
\includegraphics[width=\textwidth]{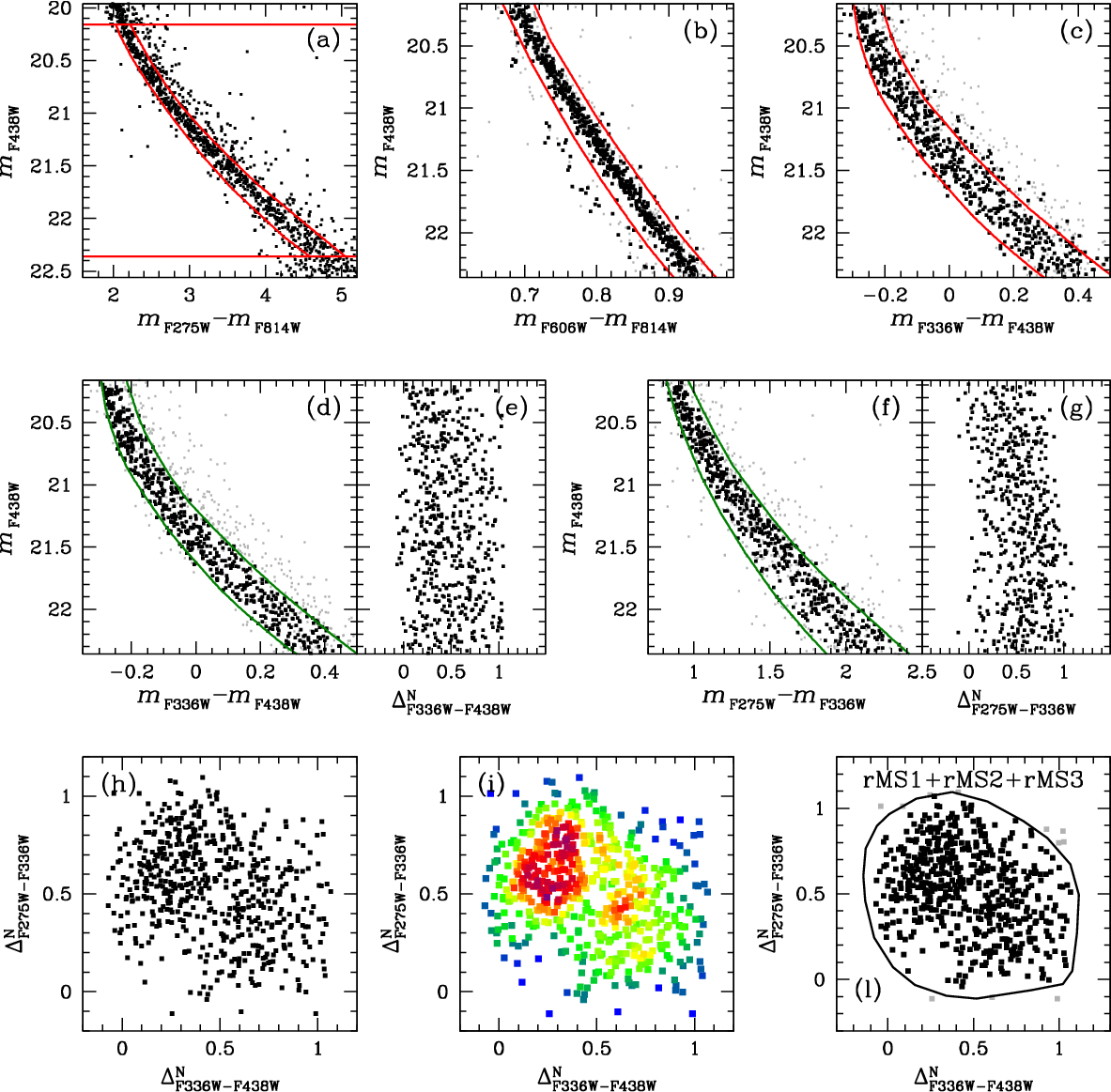}
\caption{Same as Fig.\,\ref{rMS} but for the F2 fields.}
\label{rMS_2}
\end{figure*}

\begin{figure*}
\centering
\includegraphics[width=\textwidth]{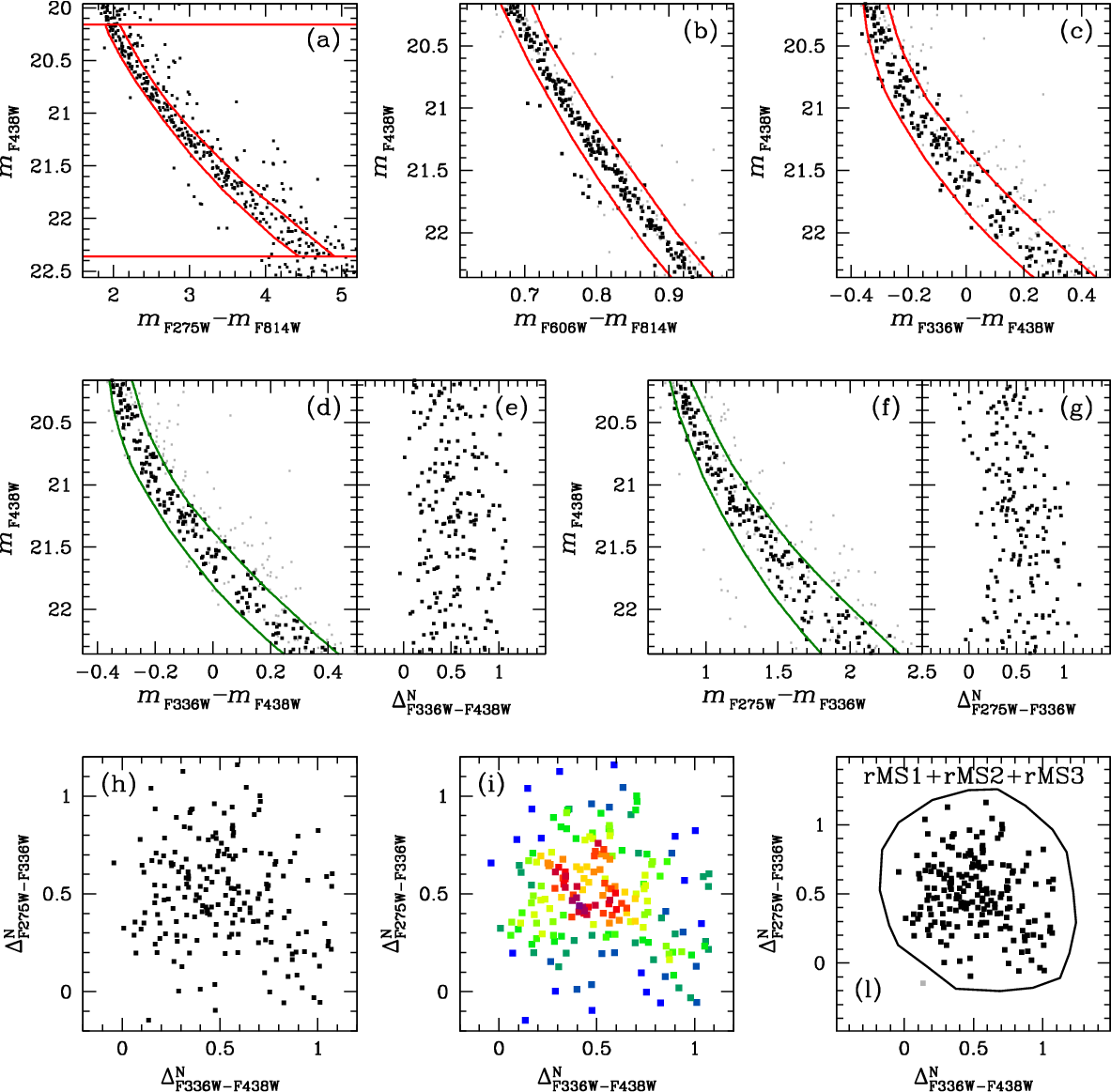}
\caption{Same as Fig.\,\ref{rMS} but for the F1 fields.}
\label{rMS_1}
\end{figure*}

\begin{figure*}
\centering
\includegraphics[width=\textwidth]{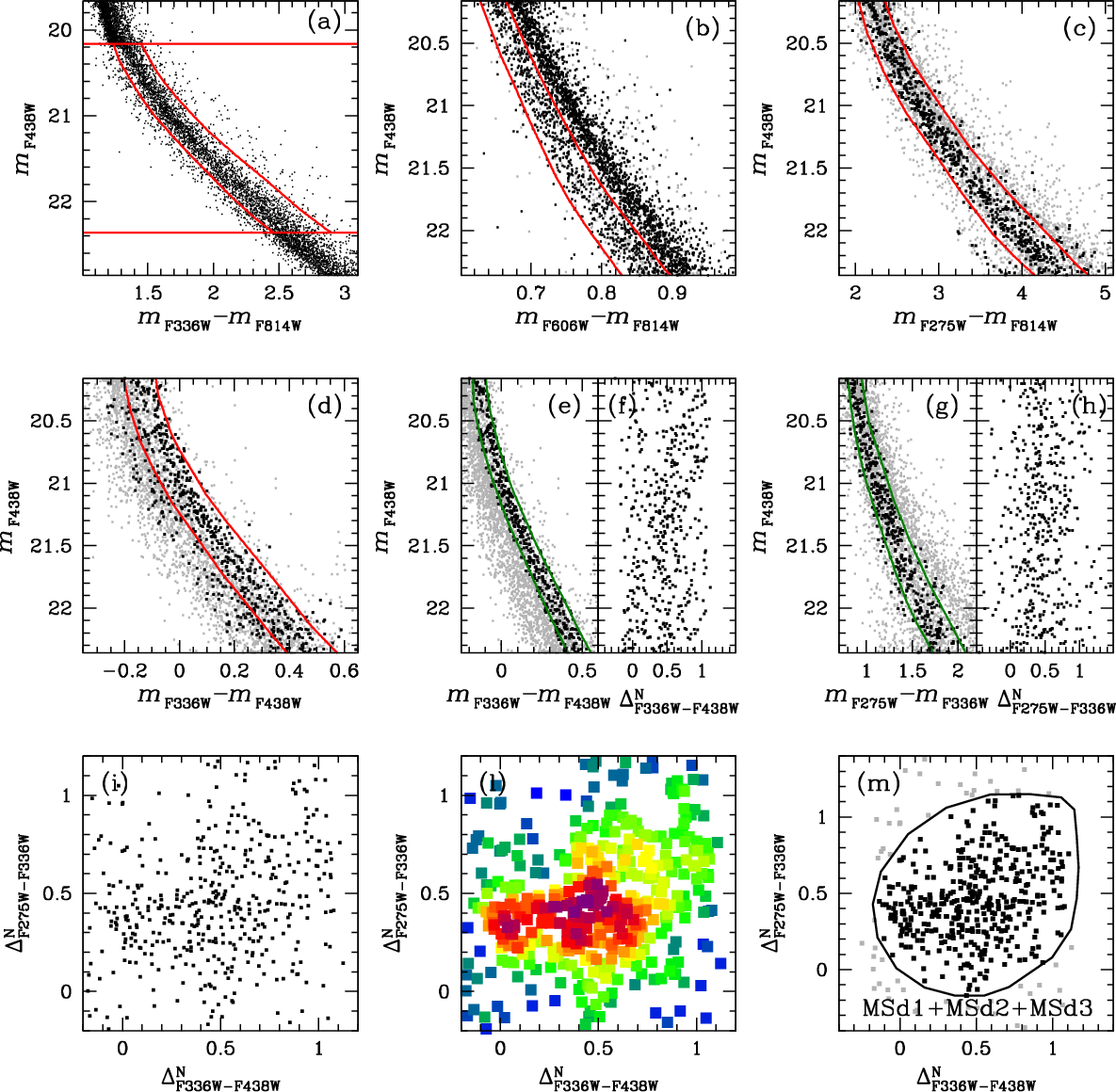}
\caption{Same as Fig.\,\ref{MSd} but for the F5 fields.}
\label{MSd_5}
\end{figure*}

\begin{figure*}
\centering
\includegraphics[width=\textwidth]{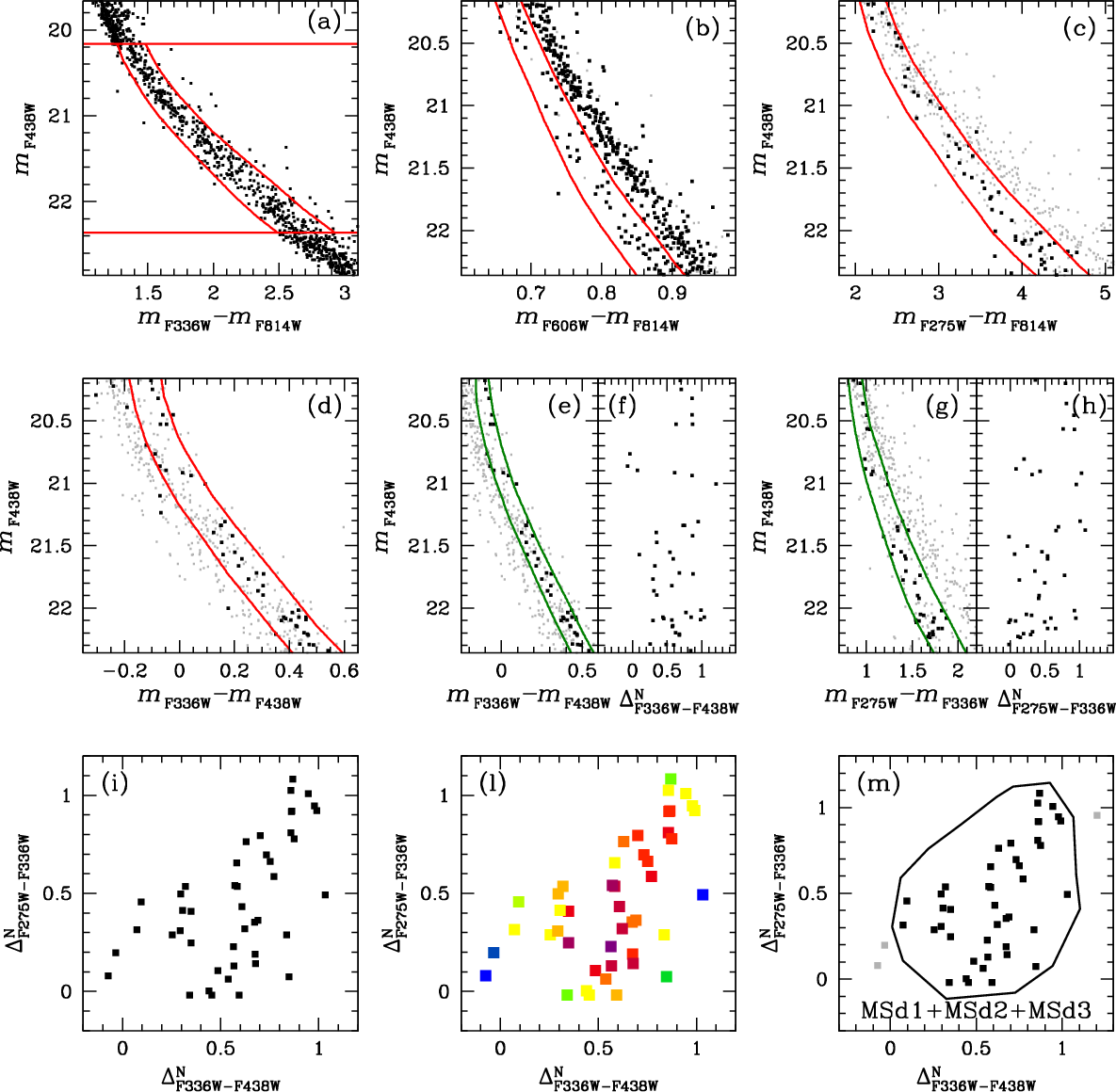}
\caption{Same as Fig.\,\ref{MSd} but for the F3 fields.}
\label{MSd_3}
\end{figure*}

\begin{figure*}
\centering
\includegraphics[width=\textwidth]{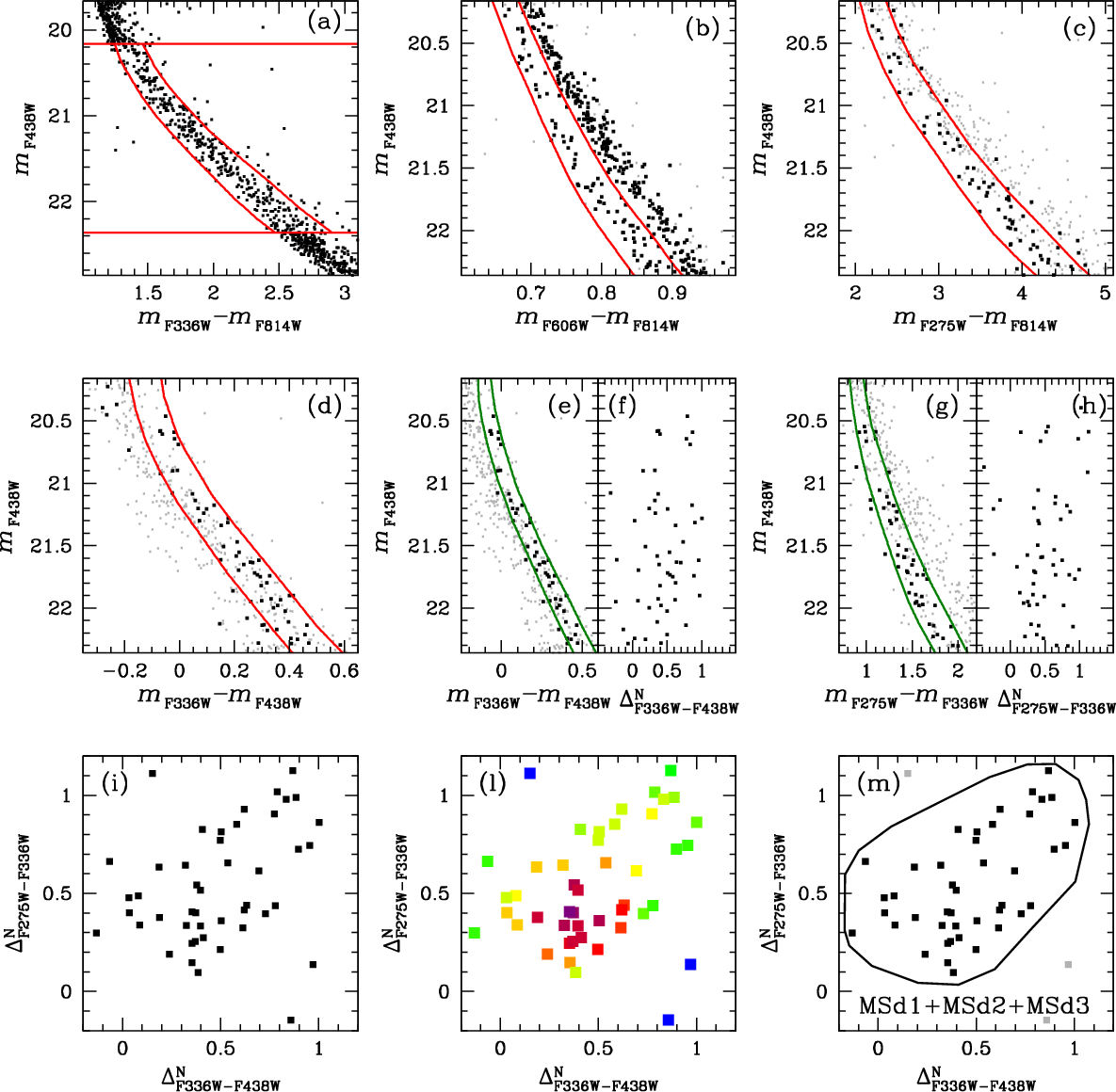}
\caption{Same as Fig.\,\ref{MSd} but for the F2 fields.}
\label{MSd_2}
\end{figure*}

\begin{figure*}
\centering
\includegraphics[width=\textwidth]{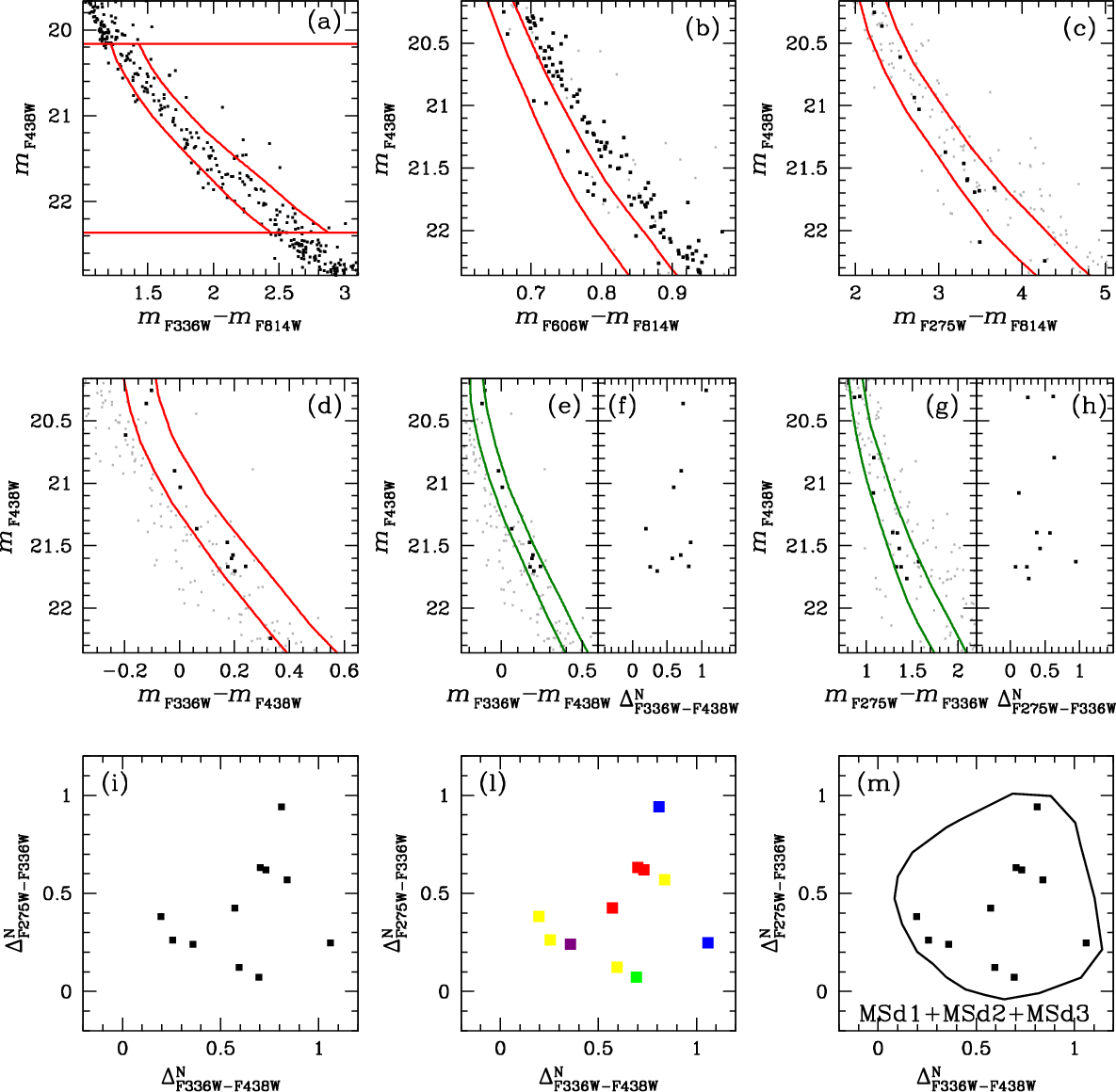}
\caption{Same as Fig.\,\ref{MSd} but for the F1 fields.}
\label{MSd_1}
\end{figure*}

\begin{figure*}
\centering
\includegraphics[width=\textwidth]{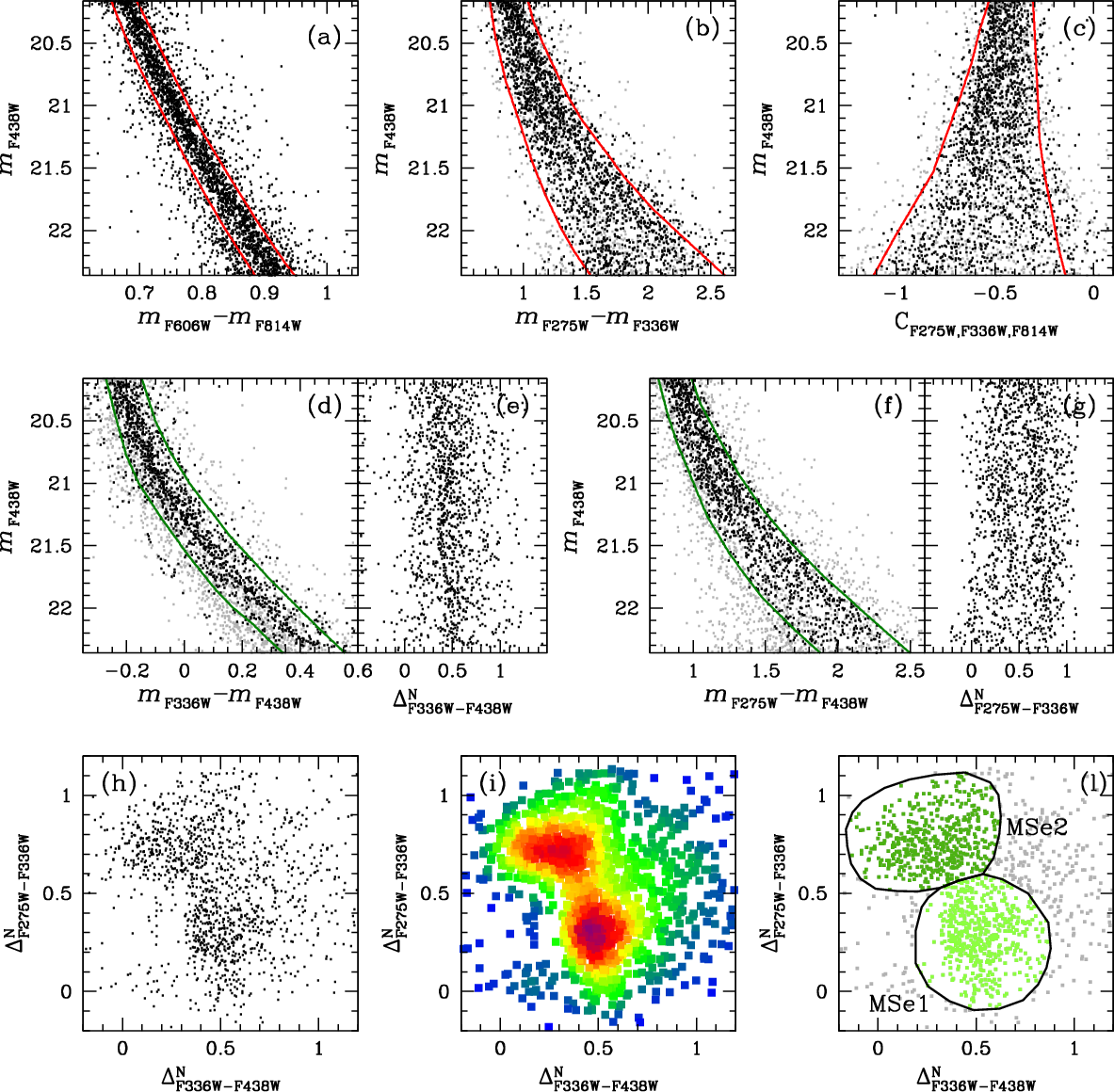}
\caption{Same as Fig.\,\ref{MSe} but for the F5 fields.}
\label{MSe_5}
\end{figure*}

\begin{figure*}
\centering
\includegraphics[width=\textwidth]{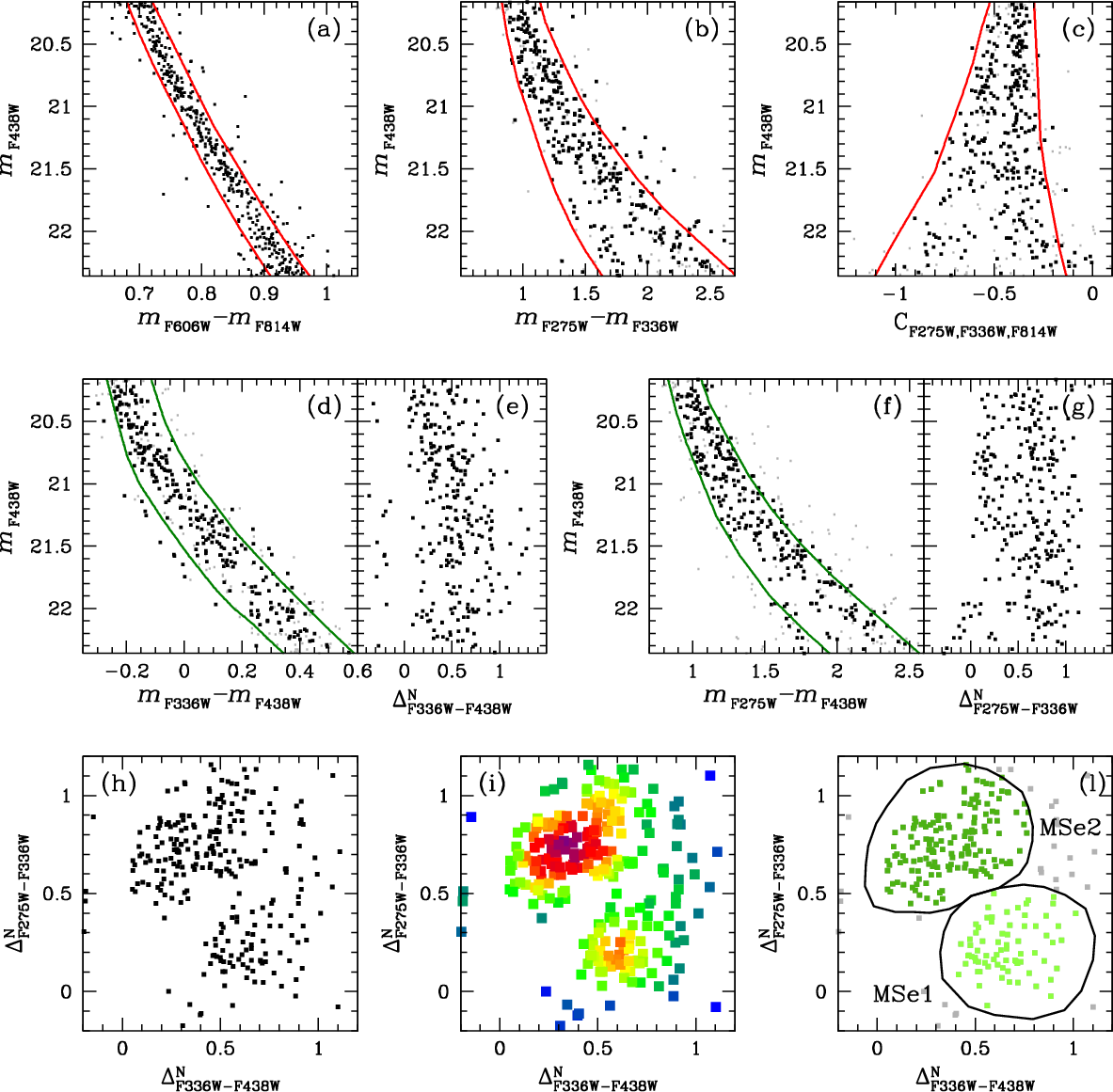}
\caption{Same as Fig.\,\ref{MSe} but for the F3 fields.}
\label{MSe_3}
\end{figure*}

\begin{figure*}
\centering
\includegraphics[width=\textwidth]{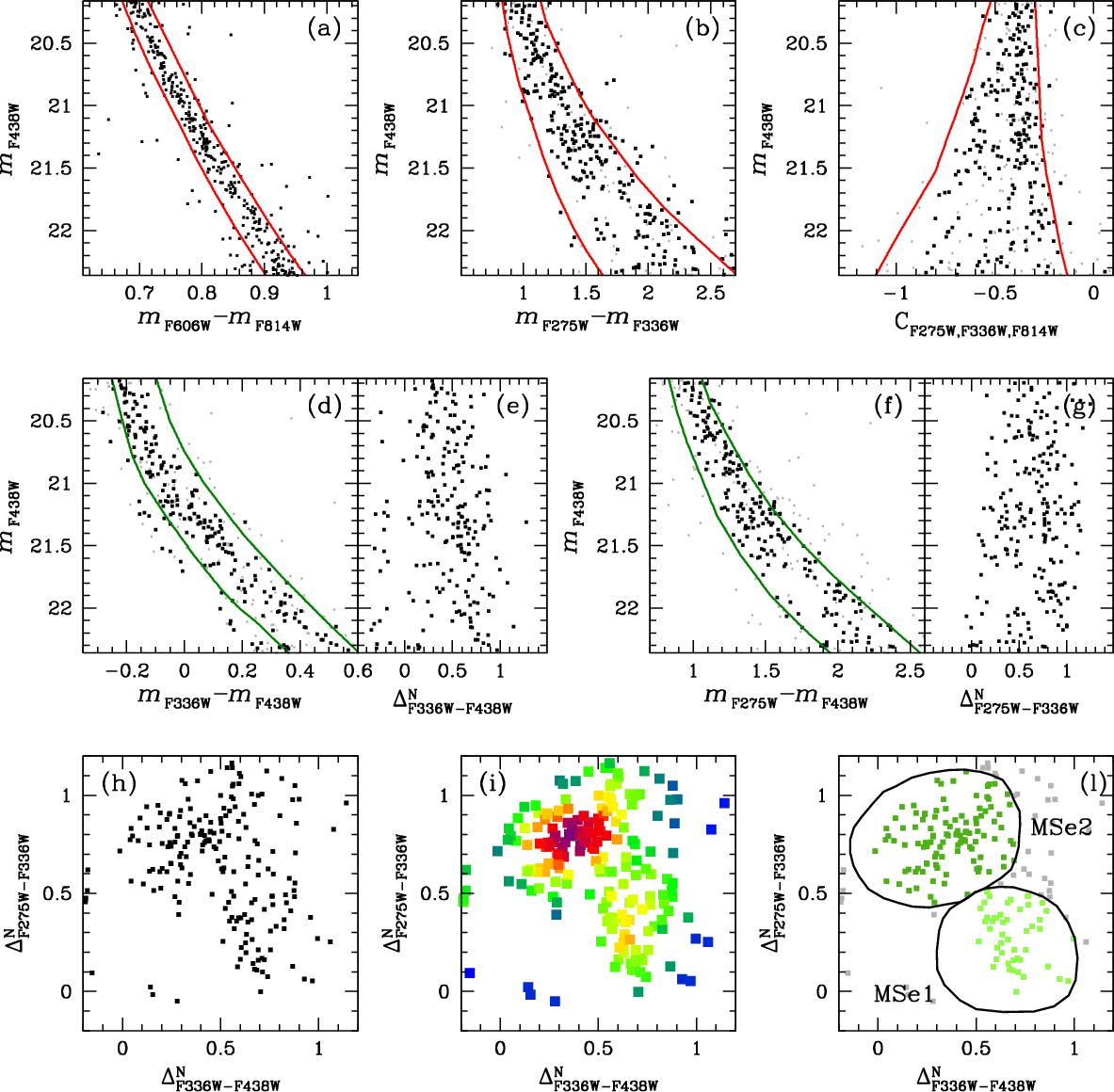}
\caption{Same as Fig.\,\ref{MSe} but for the F2 fields.}
\label{MSe_2}
\end{figure*}

\begin{figure*}
\centering
\includegraphics[width=\textwidth]{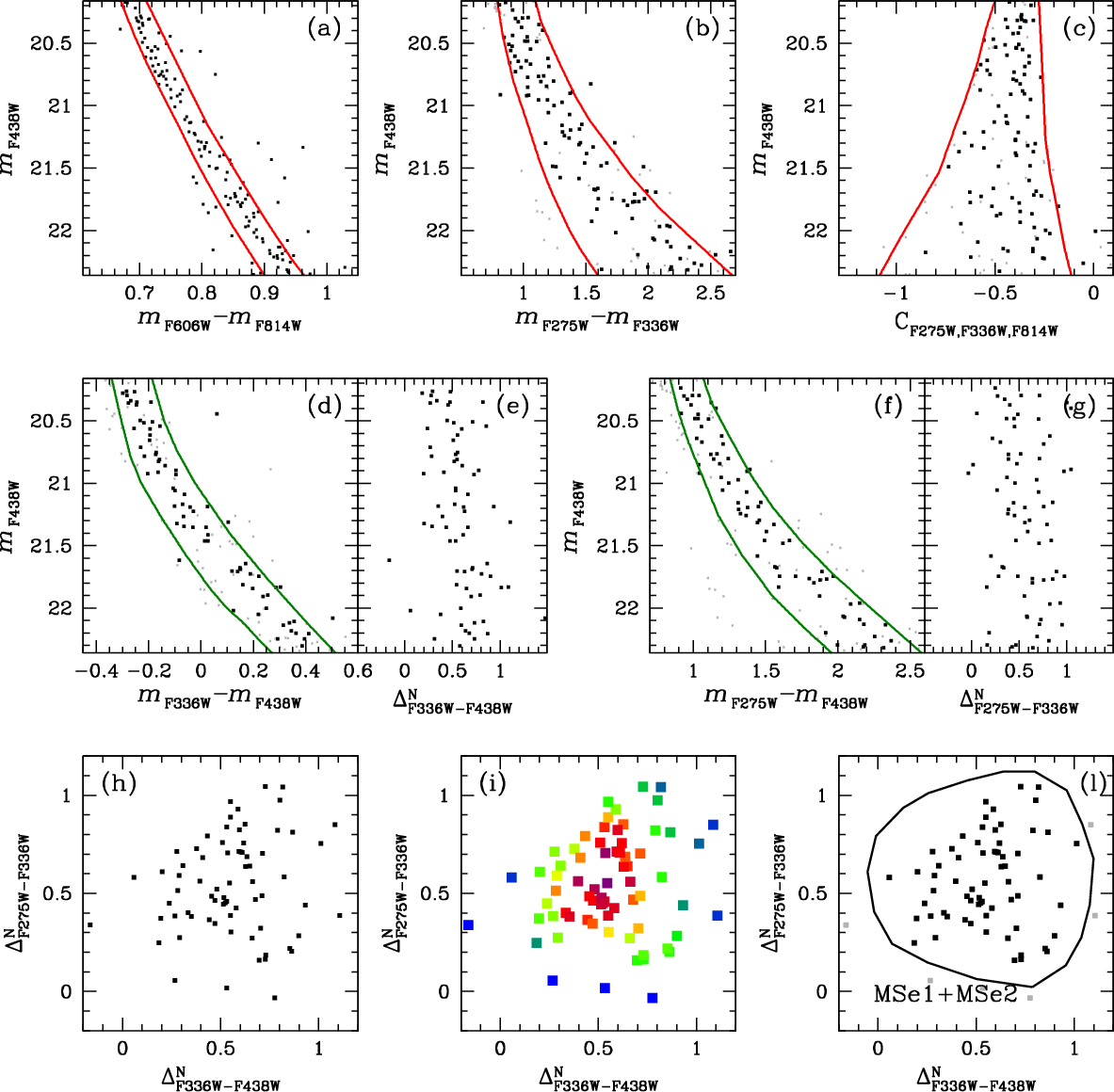}
\caption{Same as Fig.\,\ref{MSe} but for the F1 fields.}
\label{MSe_1}
\end{figure*}

\end{appendix}

\end{document}